\documentclass[
 twocolumn,
 amsmath,amssymb,
 aps,
 superscriptaddress, floatfix
]{revtex4-1}
\makeatletter
\let\MYcaption\@makecaption
\makeatother
\usepackage{subcaption}
\makeatletter
\let\@makecaption\MYcaption
\makeatother
\usepackage{dcolumn}
\usepackage{bm}
\usepackage{amsmath}
\usepackage{physics}
\usepackage{multirow}
\usepackage{scrextend}
\usepackage[pdftex]{graphicx}
\usepackage[T1]{fontenc}
\usepackage{algorithm}
\usepackage{algorithmicx}
\usepackage{algpseudocode}
\usepackage{blkarray}
\usepackage{amssymb}
\usepackage{xspace}
\usepackage{tabularx}
\usepackage{qcircuit}
\usepackage{sansmathaccent} 
\usepackage{color}
\usepackage{url}
\usepackage{tikz}
\usetikzlibrary{backgrounds,fit,decorations.pathreplacing}

\newcommand{\cnx}{\textsc{$C^{\otimes n}X$}\xspace}

\newcommand{\ux}{\textsc{$U3$}\xspace}
\newcommand{\uh}{\textsc{$U2$}\xspace}
\newcommand{\uz}{\textsc{$U1$}\xspace}
\newcommand{\RTX}{\textsc{$RTOF_{iX}$}\xspace}
\newcommand{\RTM}{\textsc{$RTOF_{M}$}\xspace}

\tikzset{meter/.append style={draw, inner sep=10, rectangle, font=\vphantom{A}, minimum width=30, line width=.8,
 path picture={\draw[black] ([shift={(.1,.3)}]path picture bounding box.south west) to[bend left=50] ([shift={(-.1,.3)}]path picture bounding box.south east);\draw[black,-latex] ([shift={(0,.1)}]path picture bounding box.south) -- ([shift={(.3,-.1)}]path picture bounding box.north);}}}
 
\usetikzlibrary{fit,shapes}

\protected\def\vvv#1{\leavevmode\bgroup\vbox\bgroup\xvvv#1\relax}

\def\xvvv{\afterassignment\xxvvv\let\tmp= }

\def\xxvvv{%
\ifx\tmp\@sptoken\egroup\ \vbox\bgroup\let\next\xvvv
\else\ifx\tmp\relax\egroup\egroup\let\next\relax
\else
\hbox to 1.1em{\hfill\tmp\hfill}
\let\next\xvvv\fi\fi
\next}

\begin{document}

\preprint{APS/123-QED}

\title{Subdivided Phase Oracle for NISQ Search Algorithms}

\author{Takahiko Satoh}
\affiliation{Keio Quantum Computing Center}
\affiliation{Keio University Shonan Fujisawa Campus}
\author{Yasuhiro Ohkura}
\affiliation{Keio University Shonan Fujisawa Campus}
\author{Rodney Van Meter}
\email{\{satoh,rum,rdv\}@sfc.wide.ad.jp}
\affiliation{Keio Quantum Computing Center}
\affiliation{Keio University Shonan Fujisawa Campus}

\date{\today}

\begin{abstract}
Because noisy, intermediate-scale quantum (NISQ) machines
accumulate errors quickly, we need new approaches to designing
NISQ-aware algorithms and assessing their performance. Algorithms with
characteristics that appear less desirable under ideal circumstances,
such as lower success probability, may in fact outperform their ideal
counterparts on existing hardware. We propose an adaptation of
Grover's algorithm, subdividing the phase flip into segments to
replace a digital counter and complex phase flip decision logic.  We
applied this approach to obtaining the best solution of the MAX-CUT problem in
sparse graphs, utilizing multi-control, Toffoli-like gates with
residual phase shifts. We implemented this algorithm on IBM Q
processors and succeeded in solving a 5-node MAX-CUT problem,
demonstrating amplitude amplification on four qubits. This approach
will be useful for a range of problems, and may shorten the time to
reaching quantum advantage.

\end{abstract}

\maketitle

\section{Introduction}
\label{sec:Intro}
With the advent of NISQ~(Noisy Intermediate-Scale Quantum~\cite{preskill2018quantum}) processors, implementation of various NISQ-friendly algorithms, such as VQE~\cite{peruzzo2014variational}, is in progress.
On the other hand, many algorithms whose theoretical computational complexity guarantees quantum acceleration require large-scale quantum circuits.
Practical scale implementation of these algorithms will be difficult with NISQ devices, and future quantum computers with error correction capabilities will be needed.

Cross \emph{et al.} proposed Quantum Volume (QV) as a quantitative indicator of the computing power of quantum processors~\cite{PhysRevA.100.032328}.
QV might double every year due to improvements in quantum processor performance~\cite{QV2019}.
Determining the relationship between the QV of a processor and the size of the quantum circuit it can perform is essential in determining when a future quantum processor can solve a particular problem.

\begin{figure}[htb]
    \begin{center}
        \includegraphics[width=85mm]{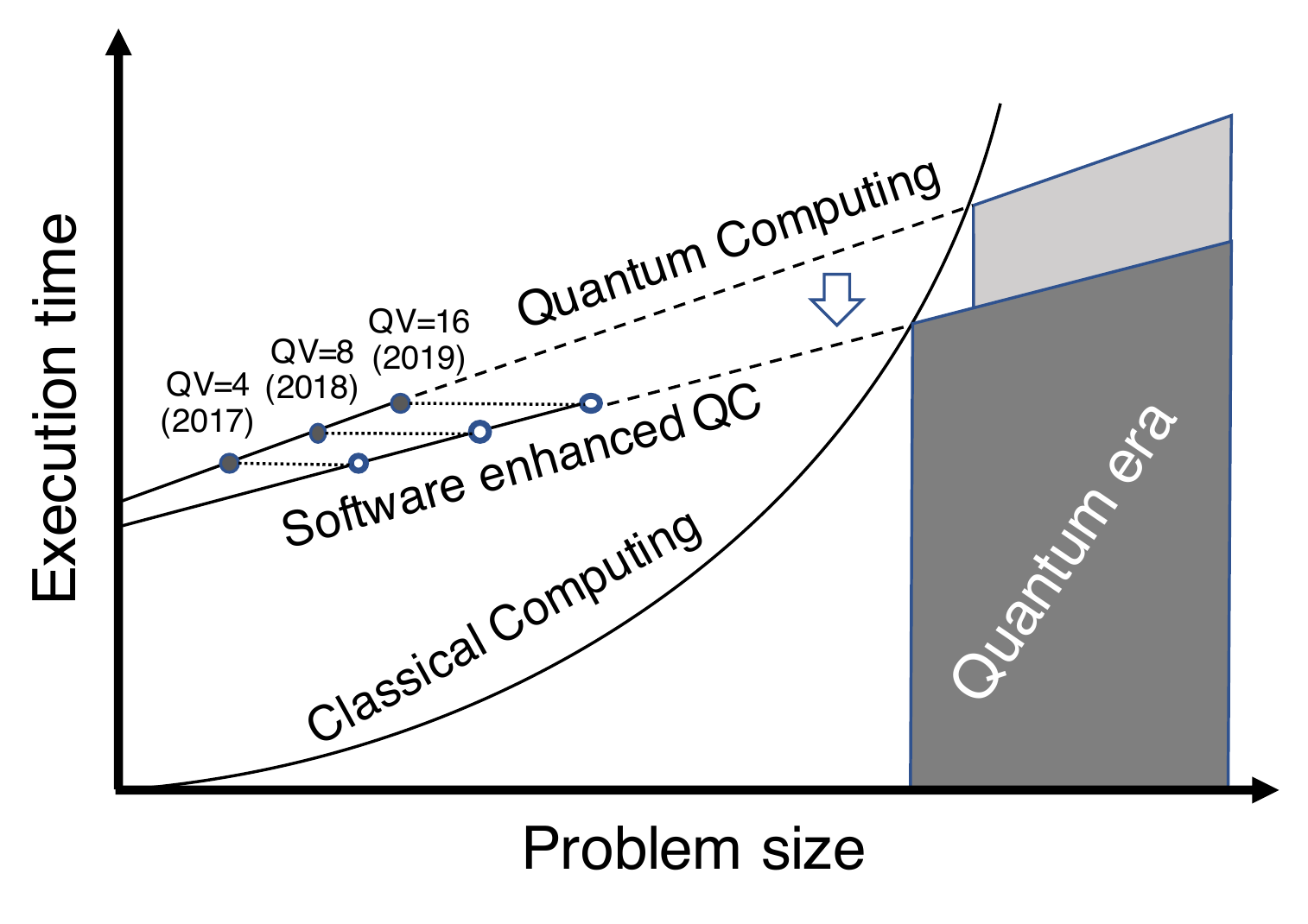}
        \caption{{\bf The significance of software development.} The solid, straight lines indicate the quantum computing power achieved to date, and the dashed line is the performance that will be realized assuming continuing increases in QV. Through the combined improvement of software and hardware, the aim is to reach the intersection with the curve of the ability of classical computers. Thus, software advances have the potential to shorten the time to the achievement of quantum advantage.} 
        \label{fig:qvcdiagram}
    \end{center}
\end{figure}

FIG.~\ref{fig:qvcdiagram} shows an abstract diagram of the relationship between classical and quantum computers.
Hardware improvements and error mitigation reduce the effect of decoherence.
The increased QV due to their contribution allows us to move to the upper right along this line.
Improvements in algorithm, compilation, and structural connectivity both move down and change the slope of this line.

Focusing on the algorithm aspect, we describe the following contributions in this paper: 1) replacing the combination of the digital accumulator plus the binary ($0$ or $\pi$) phase flip with the subdivided oracle phase, and 2) an implementation method for $n$-controlled Toffoli gate suitable for processors with low connectivity.
As an application of the first technique, we present an implementation for the MAX-CUT problem. The second technique addresses a fundamental need and may become an essential component of many algorithms.

Using these approaches, we have attempted to clarify the relationship between Grover's algorithm~\cite{grover1996fast} (Sec.~\ref{subsec:Grover}) and QV.
As a preliminary step, we designed an algorithm to obtain an exact solution in the MAX-CUT problem (Sec.~\ref{subsec:maxcut} and~\ref{sec:Exact}).
In this algorithm, when the input length exceeds 4 qubits, the total number of Controlled-NOT~($CX$) gates exceeds 100, and present-day quantum processors cannot obtain a useful answer.
To miniaturize the algorithm as much as possible, we reduced the weight of the \cnx gate used in the diffusion operator (Sec.~\ref{subsec:impdif}) and adapted the phase information fragmentation in the oracle (Sec.~\ref{subsec:impora}).
Although this makes it possible to realize a smaller quantum circuit than the above algorithm, it is not possible to transform a given problem into a decision problem, so we cannot call our solution NP-Complete. The correctness of the solution obtained depends on the average degree of the graph.

We executed our proposed algorithm on two IBM transmon systems, {\bf ibm\_ourense} with QV $=8$ and {\bf ibm\_valencia} with QV $=16$, and evaluated the success probability and KL divergence. 
The 3-data qubit Grover algorithm for the $K_{1,3}$ MAX-CUT found the correct answer over $29$\% (theoretical $34.7$\%) of the time on both processors (Sec.~\ref{subsec:exponq}).
The 4-data qubit Grover algorithm for the $K_{1,4}$ MAX-CUT found the correct answer more than $11$\% (theoretical $21.2$\%) of the time on both processors.
In the second experiment, the average KL divergence value of {\bf ibm\_valencia} was $0.457$, while that of {\bf ibm\_ourense} was $0.831$, substantially better than completely mixed state values of $1.149$.

These results indicate that probability amplification using Grover on a 4-qubit problem, which has conventionally been considered difficult~\cite{stromberg20184,mandviwalla2018implementing}, is possible using current processors. For this particular problem, differences in the decoherence characteristics of the two processors result in the off-answer elements of the superposition decaying more rapidly than the correct answer, resulting in an unexpectedly small decrease in overall success probability in the processor with the smaller QV. However, we expect that in more general cases, the success probability will more closely track the KL divergence.
Also, our algorithm scales reasonably well on processor topologies with degree 3 qubits.
Therefore, as processors with higher QV appear in the future, we can benchmark the maximum executable size of the Grover algorithm using our algorithm.

\section{Background}
\label{sec:BG}
\subsection{Grover's Algorithm}
\label{subsec:Grover}
Grover's algorithm is a quantum search algorithm to find the index of the target element $x \in \{0, 1, ...2^n-1\}$ s.t. $f(x) = y$, given $f$ and $y$, in $\mathcal{O} (\sqrt{N})$ operations with high probability, where $n$ is the number of qubits and $N = 2^n$ is the size of the list~\cite{grover1996fast}.
The feature of this algorithm is that even if the database is disordered, the square root acceleration is guaranteed with respect to the classical search, which requires an average of $\frac{N}{2}$ operations~\cite{NC}.

\subsubsection{Procedure}
The procedure of Grover's algorithm is as follows: 

\begin{enumerate}
	\item Initialization
	
		Prepare $\ket{0}^{\otimes n}$ and apply Hadamard gates $H^{\otimes n}$ to create a superposition of $2^n$ states. All states have the same amplitude $\frac{1}{\sqrt{N}}$.
	
	\item Oracle
	
		Apply the oracle operator $O$ to invert the sign of target element(s):
		\begin{align}
			O\ket{x} \xrightarrow{}  (-1)^{f(x)}\ket{x}.
		\end{align}
		Here, $f(x)=1$ if  $x$ is the target element, otherwise $0$.

	\item Diffusion
	
		Apply the diffusion operator $D$ to amplify the probability amplitude of the target element:
		\begin{align}
		\label{eq:diff}
			D&=H^{\otimes n}(2\ket{00..0} \bra{00..0} - I)H^{\otimes n} \\
			&=H^{\otimes n}X^{\otimes n}H_{T}C^{\otimes n-1}XH_{T}X^{\otimes n}H^{\otimes n}.
		\end{align}
		Here, $C^{\otimes n-1}X$ and $H_{T}$ denote $n$-controlled $X$ gate and $H$ to the target qubit of $C^{\otimes n-1}X$. $H^{\otimes n}$  corresponds to the gates for initialization.

	\item Iteration
	
		Repeat $O$ and $D$. The optimal number of iterations is $\frac{4}{\pi}\sqrt{N}$ when the number of targets is 1.
	\item Measurements
	
		Measure all qubits to  read the target data.	
\end{enumerate}

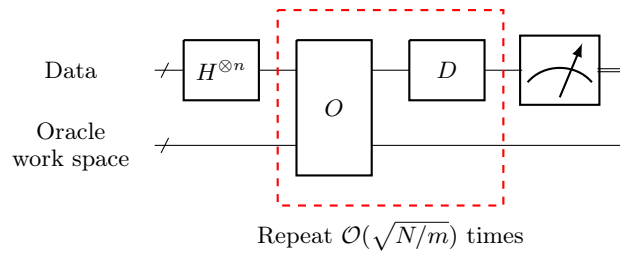
\begin{figure}[htb]
    \begin{center}
        \begin{tikzpicture}[background rectangle/.style={fill=white},
                show background rectangle]
                
%
\tikzstyle{operator} = [draw, fill=white,minimum size=1.0em] 
\tikzstyle{ctrl} = [fill,shape=circle,minimum size=5pt,inner sep=0pt]
\node at (-1, -0.2) {Data};
\node at (-1, -1) {Oracle};
\node at (-1, -1.4) {work space};
%
\node at (0, -0.2) (d2) {};
\node at (0,-1.2) (w2) {};

\draw (0.2, -0.3) -- (0.3, -0.1);
\draw (0.2, -1.3) -- (0.3, -1.1);
%
\draw (d2) -- (0.5, -0.2);

\draw[line width=0.8] (0.5, 0.2) -- (1.5, 0.2) -- (1.5, -0.6) -- (0.5, -0.6) -- (0.5, 0.2);
\node at (1, -0.2) {$H^{\otimes n}$};
%
\draw (1.5, -0.2) -- (2, -0.2); 
\draw (w2) -- (2, -1.2); 

\draw[line width=0.8] (2, 0.2) -- (3, 0.2) -- (3, -1.6) -- (2, -1.6) -- (2, 0.2);
\node at (2.5, -0.7) {$O$};
%
\draw (3, -0.2) -- (3.5, -0.2); 
    
\draw[line width=0.8] (3.5, 0.2) -- (4.5, 0.2) -- (4.5, -0.6) -- (3.5, -0.6) -- (3.5, 0.2);
\node at (4, -0.2) {$D$};

\draw (4.5, -0.2) -- (5, -0.2);

\node [meter] (meter)  at (5.5, -0.2) {};
%
\draw (6, -0.18) -- (6.3, -0.18); 
\draw (6, -0.22) -- (6.3, -0.22);

\draw (3, -1.2) -- (6.3, -1.2); 

%
\draw[red,thick,dashed] (1.75, 0.6) -- (4.75, 0.6) -- (4.75, -2) -- (1.75, -2) -- (1.75, 0.6);
\node at (3.25, -2.4) {Repeat $\mathcal{O}(\sqrt{N/m})$~times};

\end{tikzpicture}
        \caption{{\bf General circuit for Grover's algorithm} Grover's algorithm consists of data space and oracle working space. First, initialize all data qubits, then repeat Grover's operator (dashed box), which consists of oracle $O$ and diffusion operator $D$,  $\mathcal{O}(\sqrt{N/m})$ times when the number of target states is $m$ and the search space size is $N$.} 
        \label{fig:grovercircuit}
    \end{center}
\end{figure}
In general, Grover's algorithm uses an $n$-qubit data register and work space qubits for oracle execution, as in FIG.~\ref{fig:grovercircuit}.

\subsection{The MAX-CUT problem}
\label{subsec:maxcut}
MAX-CUT is the graph theory problem of finding the maximum cut of given graph $G(V,E)$.
MAX-CUT can be considered to be a vertex coloring problem using two colors that involves filling in some of the vertices with one color, and the rest of vertices with another color.
Then we count the edges that exist between vertices of different colors as if they were cut.
To solve this puzzle, we need to find a coloring combination which contains the highest number of edges connecting different color of vertices from $2^{\vert V \vert -1}$ possible colorings.
On a general graph, MAX-CUT is known to be an NP-hard class problem~\cite{garey1979guide}.

\subsection{Current quantum processors}
In recent years, NISQ (Noisy Intermediate Scale Quantum~\cite{preskill2018quantum}) devices that can perform quantum computation with a short circuit length have appeared, although the scale and accuracy are insufficient to perform continuous, effective error correction.
Various physical systems such as superconductors, ion traps, quantum dots, NV centres, and optics are used in NISQ devices~\cite{ladd2010quantum, van2016local}.

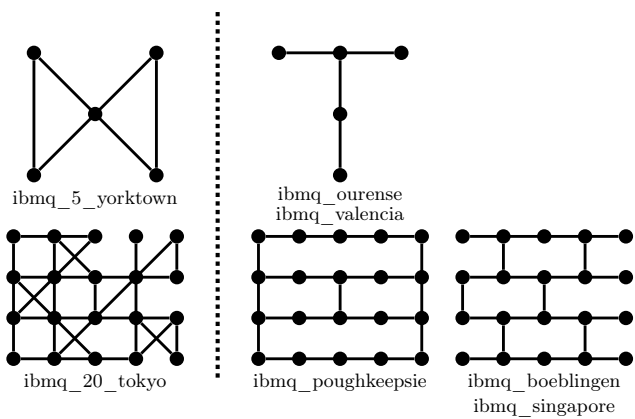
\begin{figure}[htb]
    \begin{center}
        \resizebox{0.48\textwidth}{!}{
    \begin{tikzpicture}[background rectangle/.style={fill=white},
                    show background rectangle]

    \tikzstyle{qubit} = [fill,shape=circle,minimum size=10pt,inner sep=0pt]
    %
    %
    
    \node [qubit] (x1) at (0.5, -0.5) {};
    \node [qubit] (x2) at (0.5, -3.5) {};
    \node [qubit] (x3) at (2, -2) {};
    \node [qubit] (x4) at (3.5, -0.5) {};
    \node [qubit] (x5) at (3.5, -3.5) {};
    \draw[line width = 0.7mm](x1)--(x2);
    \draw[line width = 0.7mm](x1)--(x5);
    \draw[line width = 0.7mm](x2)--(x4);
    \draw[line width = 0.7mm](x4)--(x5);
    
    \node[scale=1.5] at (2, -4.1) {{\rm ibmq\_5\_yorktown}};

    \node [qubit] (x11) at (0, -5) {};
    \node [qubit] (x12) at (1, -5) {};
    \node [qubit] (x13) at (2, -5) {};
    \node [qubit] (x14) at (3, -5) {};
    \node [qubit] (x15) at (4, -5) {};
    
    \node [qubit] (x21) at (0, -6) {};
    \node [qubit] (x22) at (1, -6) {};
    \node [qubit] (x23) at (2, -6) {};
    \node [qubit] (x24) at (3, -6) {};
    \node [qubit] (x25) at (4, -6) {};
    
    \node [qubit] (x31) at (0, -7) {};
    \node [qubit] (x32) at (1, -7) {};
    \node [qubit] (x33) at (2, -7) {};
    \node [qubit] (x34) at (3, -7) {};
    \node [qubit] (x35) at (4, -7) {};
    
    \node [qubit] (x41) at (0, -8) {};
    \node [qubit] (x42) at (1, -8) {};
    \node [qubit] (x43) at (2, -8) {};
    \node [qubit] (x44) at (3, -8) {};
    \node [qubit] (x45) at (4, -8) {};
    
    \draw[line width = 0.7mm](x11)--(x41);
    \draw[line width = 0.7mm](x12)--(x42);
    \draw[line width = 0.7mm](x23)--(x33);
    \draw[line width = 0.7mm](x14)--(x44);
    \draw[line width = 0.7mm](x15)--(x25); \draw[line width = 0.7mm](x35)--(x45);
    
    \draw[line width = 0.7mm](x11)--(x13);
    \draw[line width = 0.7mm](x21)--(x25);
    \draw[line width = 0.7mm](x31)--(x35);
    \draw[line width = 0.7mm](x41)--(x44);
    
    \draw[line width = 0.7mm](x21)--(x43);
    \draw[line width = 0.7mm](x12)--(x23); \draw[line width = 0.7mm](x34)--(x45);
    
    \draw[line width = 0.7mm](x13)--(x31);
    \draw[line width = 0.7mm](x15)--(x42);
    \draw[line width = 0.7mm](x35)--(x44);
    
    \node[scale=1.5] at (2, -8.6) {{\rm ibmq\_20\_tokyo}};
    
    \draw[line width=1mm, style=dashed](5, 0.5) -- (5, -8.5);
    
    \node [qubit] (x1) at (6.5, -0.5) {};
    \node [qubit] (x2) at (8, -0.5) {};
    \node [qubit] (x3) at (9.5, -0.5) {};
    \node [qubit] (x4) at (8, -2) {};
    \node [qubit] (x5) at (8, -3.5) {};
    \draw[line width = 0.7mm](x1)--(x3);
    \draw[line width = 0.7mm](x2)--(x5);
    
    \node[scale=1.5] at (8, -4) {{\rm ibmq\_ourense}};
    \node[scale=1.5] at (8, -4.5) {{\rm ibmq\_valencia}};

    \node [qubit] (x11) at (6, -5) {};
    \node [qubit] (x12) at (7, -5) {};
    \node [qubit] (x13) at (8, -5) {};
    \node [qubit] (x14) at (9, -5) {};
    \node [qubit] (x15) at (10, -5) {};
    
    \node [qubit] (x21) at (6, -6) {};
    \node [qubit] (x22) at (7, -6) {};
    \node [qubit] (x23) at (8, -6) {};
    \node [qubit] (x24) at (9, -6) {};
    \node [qubit] (x25) at (10, -6) {};
    
    \node [qubit] (x31) at (6, -7) {};
    \node [qubit] (x32) at (7, -7) {};
    \node [qubit] (x33) at (8, -7) {};
    \node [qubit] (x34) at (9, -7) {};
    \node [qubit] (x35) at (10, -7) {};
    
    \node [qubit] (x41) at (6, -8) {};
    \node [qubit] (x42) at (7, -8) {};
    \node [qubit] (x43) at (8, -8) {};
    \node [qubit] (x44) at (9, -8) {};
    \node [qubit] (x45) at (10, -8) {};
    
    \draw[line width = 0.7mm](x11)--(x41);
    \draw[line width = 0.7mm](x23)--(x33);
    \draw[line width = 0.7mm](x15)--(x45);
    
    \draw[line width = 0.7mm](x11)--(x15);
    \draw[line width = 0.7mm](x21)--(x25);
    \draw[line width = 0.7mm](x31)--(x35);
    \draw[line width = 0.7mm](x41)--(x45);
    
    \node[scale=1.5] at (8, -8.6) {{\rm ibmq\_poughkeepsie}};
    
    \node [qubit] (x11) at (11, -5) {};
    \node [qubit] (x12) at (12, -5) {};
    \node [qubit] (x13) at (13, -5) {};
    \node [qubit] (x14) at (14, -5) {};
    \node [qubit] (x15) at (15, -5) {};
    
    \node [qubit] (x21) at (11, -6) {};
    \node [qubit] (x22) at (12, -6) {};
    \node [qubit] (x23) at (13, -6) {};
    \node [qubit] (x24) at (14, -6) {};
    \node [qubit] (x25) at (15, -6) {};
    
    \node [qubit] (x31) at (11, -7) {};
    \node [qubit] (x32) at (12, -7) {};
    \node [qubit] (x33) at (13, -7) {};
    \node [qubit] (x34) at (14, -7) {};
    \node [qubit] (x35) at (15, -7) {};
    
    \node [qubit] (x41) at (11, -8) {};
    \node [qubit] (x42) at (12, -8) {};
    \node [qubit] (x43) at (13, -8) {};
    \node [qubit] (x44) at (14, -8) {};
    \node [qubit] (x45) at (15, -8) {};
    
    \draw[line width = 0.7mm](x12)--(x22); \draw[line width = 0.7mm](x14)--(x24);
    
    \draw[line width = 0.7mm](x21)--(x31); \draw[line width = 0.7mm](x23)--(x33); width = 0.7mm](x25)--(x35);
    
    \draw[line width = 0.7mm](x32)--(x42); \draw[line width = 0.7mm](x34)--(x44);
    
    \draw[line width = 0.7mm](x11)--(x15);
    \draw[line width = 0.7mm](x21)--(x25);
    \draw[line width = 0.7mm](x31)--(x35);
    \draw[line width = 0.7mm](x41)--(x45);
    
    \node[scale=1.5] at (13, -8.6) {{\rm ibmq\_boeblingen}};
    \node[scale=1.5] at (13, -9.2) {{\rm ibmq\_singapore}};
    
    \end{tikzpicture}
}
        \caption{{\bf Qubit topology of IBM Q processors} Early devices (left side) had a dense structure, while the recent devices (right side) are composed of relatively sparse qubit connections.} 
        \label{fig:processors}
    \end{center}
\end{figure}

The early 20-qubit superconducting processors from IBM had high connectivity and the maximum degree was 6, while the latest processors have a high gate accuracy but the maximum degree is 3 (FIG.~\ref{fig:processors}).

\subsubsection*{Quantum Volume and KQ}
Quantum Volume~(QV) is a measure proposed by IBM that shows the performance of NISQ~\cite{PhysRevA.100.032328}.
Quantum Volume QV is defined as 
\begin{align}
    QV=2^{{\rm min}(m,d)},
\end{align}
where $m$ denotes circuit width~(number of qubits) and $d$ denotes circuit $SU(4)$ depth.
The QV for each processor is calculated from single and two-qubits gate errors, connectivity, measurement errors, etc.
The computation fails with high probability when a given circuit satisfies
\begin{align}
\label{eq:epeff}
    md \simeq \frac{1}{\epsilon_{\textrm{eff}}}.
\end{align}
Here, ${\epsilon_{\textrm{eff}}}$ is an effective $CX$ gate error value that gradually increases with connectivity.

In this paper, we experimented with two 5-qubit processors, {\bf ibmq\_ourence} with QV$ = 8$ and {\bf ibmq\_valencia} with QV$ = 16$.

KQ is a measure of the capabilities of the machine, independent of the algorithm.  In 2003, Steane proposed a similar measure focusing on the algorithm’s needs and on error correction~\cite{steane2003overhead}. For an algorithm using $Q$ qubits and requiring $K$ time steps on those qubits (in suitable units), the space-time product $KQ$ is a guideline to the required error rate, which should be below $1/(KQ)$.

\subsubsection*{Open Quantum Assembly Language (QASM)}
The IBM Q processors accept gates written in the QASM language~\cite{cross2017open}.
All circuits are decomposed into four types of gate. 
We describe those gates and the required pulses in the IBM Q superconducting processors in Tab.~\ref{tab:gates}.
\begin{table}[htb]
\begin{tabular}{c|c}
gate type &  remarks \\ \hline
$U1(\lambda)$      & No pulse.     Rotation $Z$~($R_Z$) gate.   \\
$U2(\phi,\lambda)$    & One $\frac{\pi}{2}$ pulse.     $H$ gate is $U2(0,\pi)$. \\
$U3(\theta, \phi, \lambda)$     & Two  $\frac{\pi}{2}$ pulses.   $R_Y(\theta)$ gate is $U3(\theta, 0, 0)$. \\
$CX$ &  Cross-resonance pulses and One $\frac{\pi}{2}$ pulse.
\end{tabular}
\caption{Gate set for QASM}
\label{tab:gates}
\end{table}
Since no pulse is required, we can perform \uz~with zero cost.
The error level of \ux~is twice \uh~and approximately an order of magnitude less than the $CX$ gate~\cite{QV2019}.
The performance of {\bf ibmq\_ourense} and {\bf ibmq\_valencia} is shown in Tab.~\ref{table:device_qubit} and~\ref{table:device_cx} in the appendix.

\section{Grover algorithm to solve MAX-CUT problem}
\label{sec:Exact}
\begin{figure}[hbt]
        \includegraphics[width=60mm]{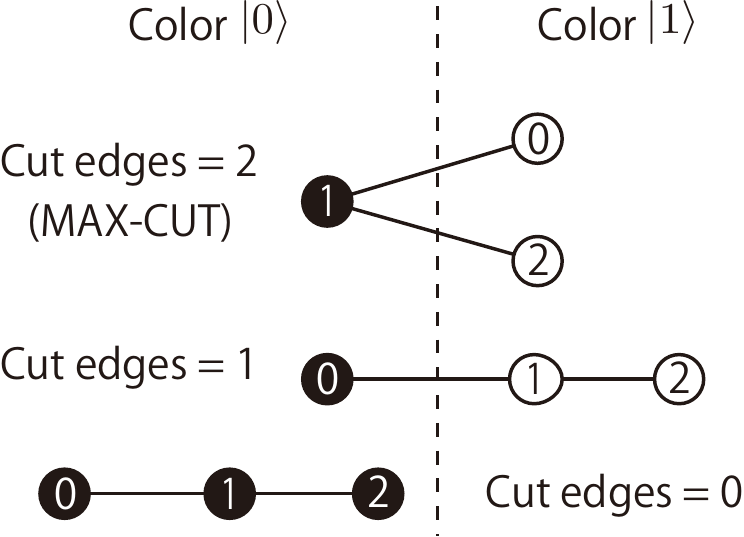}
        \caption{{\bf Data structure for MAX-CUT.} We can find MAX-CUT $\vert 010 \rangle_{012}$~(or $\vert 101 \rangle_{012}$) by counting the cases where the states of the qubits corresponding to both ends of the edge are different.} 
        \label{fig:cutstar}
\end{figure}

We propose Grover's algorithm for solving the MAX-CUT of a given graph $G$.
The following simple coloring approach is an exhaustive classical search:
\begin{itemize}
    \item[Step 1.] Color all vertices black or white.
    \item[Step 2.] Count the number of edges with different color vertices at both ends.
    \item[Step 3.] Color the vertices with a different pattern from the existing one and return to Step 2.
    \item[Step 4.] After testing all possible coloring patterns, the pattern with the largest number of edges counted corresponds to the MAX-CUT.
\end{itemize}
We can apply Grover's algorithm by assigning black to $\vert 0 \rangle$ and white to $\vert 1 \rangle$ in this procedure~\cite{QC2019_3}.
To illustrate this correspondence, we show a simple example using a star graph $K_{1,2}$ in FIG.~\ref{fig:cutstar}.
The MAX-CUT for a graph with $m$ edges and $n$ vertices can be found by the following procedure.
\begin{itemize}
    \item[Step 0.] Set threshold value $t~(\leq m)$.
    \item[Step 1.] Initialize all $n$ qubits to $\vert + \rangle$.
    \item[Step 2.] Flip the sign of the input where the number of edges to be cut exceeds $t$.~(the oracle)
    \item[Step 3.] Amplify the probability of any input whose sign is inverted.~(diffusion)
    \item[Step 4.] Repeat Steps 1-3 $\mathcal{O}(\sqrt{2^n})$ times.
    \item[Step 5.] Increase $t$ if the output is legal for the graph, decrease if the output is illegal. If $t$ returns to a value taken in a prior iteration, it is MAX-CUT, and the algorithm ends. Otherwise, the process returns to Step 1.
\end{itemize}

The number of iterations can be optimized by the quantum counting algorithm~\cite{brassard1998quantum}. In addition, if an excessively low value $t$ is set such that the sign of the majority of inputs is inverted, the probability of the input with the sign not inverted is amplified.
Since a binary search can be done by appropriately increasing and decreasing $t$, we can get accurate MAX-CUT by $\log_2{m}$ iterations.

The most straightforward way to implement an oracle for a counting problem is by using a binary accumulator register.
We describe the oracle's construction  below.

\subsection{Oracle circuit design}
\begin{figure}[htb]
    \begin{minipage}[t]{.13\textwidth}
    \centering
    \resizebox{1\textwidth}{!}{
    \begin{tikzpicture}[every fit/.style={ellipse,
                        draw,
                        inner sep=-2pt,
                        text width=2cm, 
                        line width=1pt
                        }]
        \tikzstyle{circle} = [draw,
                            shape=circle,
                            minimum size=15pt,
                            inner sep=0pt]
        
        \node[circle](v0) at(0, -1) {$0$};
        \node[circle](v1) at(1.8, 0) {$1$} edge [-] (v0);
        \node[circle](v2) at(1.8, -0.65) {$2$} edge [-] (v0);
        \node[circle](v3) at(1.8, -1.35) {$3$} edge [-] (v0);
        \node[circle](v4) at(1.8, -2) {$4$} edge [-] (v0);
        
    \end{tikzpicture}   
}
    \subcaption{{\bf $K_{1,4}$}}
    \label{fig:k_14}
  \end{minipage} 
  \begin{minipage}[t]{.33\textwidth}
    \centering    
    \resizebox{1\textwidth}{!}{
    \begin{tikzpicture}[background rectangle/.style={fill=white}, show background rectangle, 
    scale=1]
    
    \tikzstyle{operator} = [draw, fill=white,minimum size=1.5em] 
    \tikzstyle{ctrl} = [fill,shape=circle,minimum size=5pt,inner sep=0pt]
     \tikzstyle{cross} = [cross/.style={path picture={ \draw[black] (path picture bounding box.east) -- (path picture bounding box.west) (path picture bounding box.south) -- (path picture bounding box.north);}}]
    
    \node at(-1, 0) {};
    
    \node[scale=1.5] at (-0.1, 0) (q1) {$\ket{\psi_A}$};
    \node[scale=1.5] at (-0.1,-1) (q2) {$\ket{\psi_B}$};
    \node[scale=1.5] at (-0.1,-2) (q3) {$\ket{\psi_S}$};
    
    \draw (0.5, 0) -- (3, 0);
    \draw (0.5, -1) -- (1.5, -1);
    \draw (0.5, -2) -- (1.5, -2);
    \draw (0.7, -2.2) -- (0.9, -1.8);
    %
    %
    \node [ctrl] (ctrl01) at (1.0, 0) {};
    \node (targ03) at (1.0, -1) {$\bigoplus$};
    \draw (ctrl01) -- (1.0, -1);
    %
    %
    \draw (1.5, -0.8) -- (2, -0.8) -- (2, -2.2) -- (1.5, -2.2) -- (1.5, -0.8);
    \node [rotate=90] at (1.75, -1.5) {Incr};
    %
    %
    \node [ctrl] (ctrl04) at (2.5, 0) {};
    \node (targ42) at (2.5 ,-1) {$\bigoplus$};
    \draw (ctrl04) -- (2.5, -1);
    %
    %
    \draw (2, -1) -- (3, -1);
    \draw (2, -2) -- (3, -2);

    \node[scale=1.5] at (3.5, 0) (q1) {$\ket{\psi_A}$};
    \node[scale=1.5] at (3.5,-1) (q2) {$\ket{\psi_B}$};
    \node[scale=1.5] at (5,-2) (q3) {$\ket{\psi_S + (\psi_A \oplus \psi_B)}$};
    
    \end{tikzpicture}
}
    \subcaption{{\bf Sub-oracle $O_{s(A,B)}$.}}
    \label{fig:suboracle}
  \end{minipage}
  \caption{(a) A star graph $K_{1,4}$. Each node number denotes the corresponding data qubit. (b)  If the states of qubit $A$ and $B$ are different, the accumulator register $\vert \psi_{s} \rangle$ becomes $\vert \psi_{s} + 1\rangle$ .}
  \label{fig:exbasic}
\end{figure}
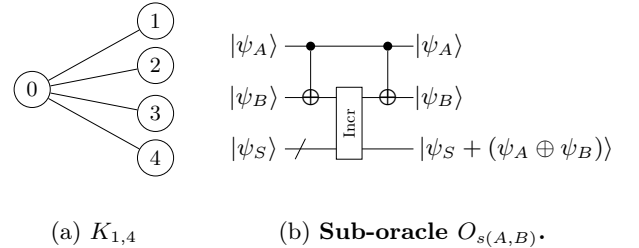

We discuss how to apply the above procedure when given a star graph $K_{1,4}$ (FIG.~\ref{fig:k_14}).
First, we prepare 5 data qubits to describe the state of nodes.
When there is an edge between node $A$ and $B$, as a cut checker for each edge, we introduce the following sub-oracle $O_{S(A,B)}$~\cite{QC2019_3}:
\begin{align}
    O_{s(A,B)}\vert \psi_{A}\psi_{B}\rangle\vert \psi_{S}\rangle \rightarrow \vert \psi_{A}\psi_{B}\rangle\vert \psi_{S}\! +\! (\psi_{A}\!\oplus \! \psi_{B})\rangle.
\end{align}
Here, $S$ is an accumulator register large enough to  store the number of cut edges.
For this problem, $ \lceil \log (\vert E \vert + 1 ) \rceil =3$ qubits are enough. 
When the states of $A$ and $B$ are different, the edge between $A,B$ is cut, and the information of cut edges on $S$ is updated.
We can implement $O_{s(A,B)}$ using a quantum increment circuit as shown in FIG.~\ref{fig:suboracle}.

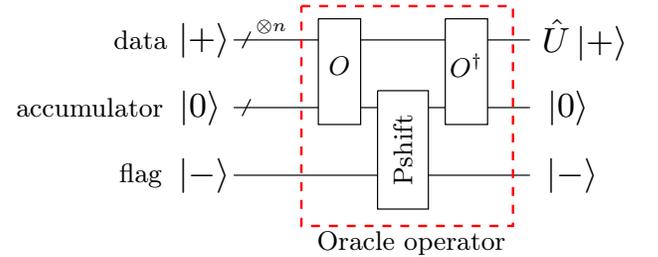
\begin{figure}[htbp]
        \resizebox{0.5\textwidth}{!}{
    \begin{tikzpicture}[background rectangle/.style={fill=white},
                    show background rectangle]
                    
    \tikzstyle{operator} = [draw, fill=white,minimum size=1.0em] 
    \tikzstyle{ctrl} = [fill,shape=circle,minimum size=5pt,inner sep=0pt]
    \node at (-1.1, 0) {data};
    \node at (-1.7, -0.8) {accumulator};
    \node at (-1.1, -1.6) {flag};
    %
    \node[scale=1.3] at (-0.35, 0) (q1) {$\ket{+}$};
    \node[scale=1.3] at (-0.4,-0.8) (q4) {$\ket{0}$};
    \node[scale=1.3] at (-0.35,-1.6) (q5) {$\ket{-}$};

    \draw (0, 0) -- (1, 0);
    \draw (0.1, -0.1) -- (0.2, 0.1);
    \node at (0.45, 0.1){$^{\otimes n}$};
    
    \draw (0, -0.8) -- (1, -0.8);
    \draw (0.1, -0.9) -- (0.2, -0.7);
    
    %
    \draw (1, 0.25) -- (1.5, 0.25) -- (1.5, -1) -- (1, -1) -- (1, 0.25);
    \node at (1.25, -0.3) {$O$};
    %
    \draw (1.5, 0) -- (2.5, 0);
    \draw (1.5, -0.8) -- (1.7, -0.8);
    \draw (0, -1.6) -- (1.7, -1.6);
    
    \draw (1.7, -0.6) -- (2.3, -0.6) -- (2.3, -2) -- (1.7, -2) -- (1.7, -0.6);
    \node[rotate=90] at (2, -1.3) {Pshift};
    
    \draw (2.3, -0.8) -- (2.5, -0.8);
    \draw (2.3, -1.6) -- (3.5, -1.6);
    
    
    \draw (2.5, 0.25) -- (3, 0.25) -- (3, -1) -- (2.5, -1) -- (2.5, 0.25);
    \node at (2.75, -0.3) {$O^\dagger$};
    
    \draw (3, 0) -- (3.5, 0);
    \draw (3, -0.8) -- (3.5, -0.8);

    \draw[red,thick,dashed] (0.8, 0.4) -- (3.3, 0.4) -- (3.3, -2.2) -- (0.8, -2.2) -- (0.8, 0.4);
    \node at (2.1, -2.4) {Oracle operator};
    %
    %
    \node[scale=1.3] at (4.15, 0) (q1) {$\hat{U}\ket{+}$};
    \node[scale=1.3] at (3.95,-0.8) (q4) {$\ket{0}$};
    \node[scale=1.3] at (4,-1.6) (q5) {$\ket{-}$};

    \end{tikzpicture}
}
        \caption{{\bf Oracle circuit.} $O$ denotes the sequence of all sub-oracles $O_S$. After the execution of Pshift, we have to uncompute $O^\dagger$ to propagate sign reversal for inputs equal to or exceeding the threshold value $t$. } 
        \label{fig:oracles}
\end{figure}

After the execution of $O_S$ for all edges, we set the threshold value $t$ and perform the phase inversion operation for inputs that equal or exceed $t$ using the flag qubit.
(In this problem, $t$ corresponding to MAX-CUT is obviously $4$.)
We show the circuit corresponding to these operations in FIG.~\ref{fig:oracles}.
We also show in detail how to configure phase shift~(Pshift) operation in Appendix~\ref{pshift}.

\subsection{Complete circuit implementation}
\label{comp_exact}
When $t = 4$, we can get $\vert 01111 \rangle$ and $\vert 10000 \rangle$ as solutions by combining the above oracle and diffusion and repeating those the appropriate number of times.
When implementing on a processor with the current QV, the proposed circuit is too large in both number of qubits and depth.

For example, the half adder contains a Toffoli gate that requires 6~$CX$ gates on IBM Q devices.
From the discussion in Sec.~\ref{sec:BG}, the upper limit of $CX$~ gates that can be used to obtain valid results is understood to be around 10.
Taking into account the need to uncompute portions of the circuit, we will not be able to include multiple sub-oracles and anticipate successful execution.

We have already proposed a method to reduce $CX$~ gates by eliminating adders and increasing ancilla qubits~\cite{QC2019_3}.
This implementation requires $\vert V \vert + \vert E \vert$ qubits and $2 \vert E \vert$ Toffoli gates, plus some $CX$~ gates for Pshift$(t)$ for one iteration.
In summary, we still need more than 36~$CX$~per iteration to solve MAX-CUT in the smaller graph $K_{1,3}$.
Needless to say, there is room for improvement in our proposed oracles.
However, in order to solve MAX-CUT with Grover's algorithm on a real processor in the near future, drastic improvement is necessary.
Therefore, we next propose a new data structure that does not store the number of cut edges in binary data.

\section{Approximated Grover search for MAX-CUT}
\label{sec:Approx}
In this section, we describe Grover's algorithm using phase subdivided oracle operators  instead of the conventional $0$ and $\pi$.
By using this method, we can remove the adders used in the previous section and reduce the circuit length significantly.
We also propose a diffusion operator implementation that requires fewer $CX$ gates for an actual processor design by using relative phase Toffoli  gates~\cite{barenco1995elementary, maslov2016advantages}.
We describe those methods and the verification of the effectiveness for the MAX-CUT problem below.

\subsection{Oracle circuit using subdivided phases}
\label{subsec:oracle}
In Sec.~\ref{sec:Exact}, storage of the evaluation value $k$ (the number of cut edges) and its calculation using adders led to a large increase in the number of $CX$ gates and occupied the largest portion of the whole circuit.

Therefore, we propose a method to express the evaluation value by the number of subdivided phases.
In the MAX-CUT problem, we use the same data structure for node color as in Sec.~\ref{sec:Exact} and unit  phase
\begin{align}
\theta_0 = \frac{\pi}{\abs{E}}
\end{align}
where $\abs{E}$ denotes the number of edges in the graph $G$.

For the cut edge determination, we introduce the following sub-oracle $O'_s$ using sub-divided phase $\theta_0$.
If an input $\vert \psi_{a} \rangle$ has a cut edge between vertices A and B, then we add $\theta_0$ to the phase information:
\begin{align}
\label{eq:app_suboracle}
O'_{s(A,B)}\vert \psi_{a} \rangle \rightarrow e^{i\theta_0}\vert \psi_{a} \rangle.
\end{align}
Similarly, based on the whole oracle operation $O'$, the best answer input $\vert \psi_{b} \rangle$ becomes as follows, (for MAX-CUT value.):
\begin{align}
\label{eq:oracle}
O'\vert \psi_{b} \rangle \rightarrow e^{ik\theta_0}\vert \psi_{b} \rangle
\end{align}
where $k\theta_0$ does not exceed $\pi$.
We next discuss the performance of this oracle and the existence of the optimal subdivided phase $\theta_{opt}$.

\subsubsection*{Algorithm performance and optimal subdivided phase}
\label{subsec:opttheta}
When the given graph is a tree~($\vert V \vert = \vert E \vert+1$ for a connected graph), the average value of the added phase $\langle\alpha(\theta)\rangle$ after applying the above oracle $O'$ is:
\begin{align}
    \langle\alpha(\theta)\rangle=\frac{1}{2^{\vert E \vert}}\sum^{\vert E \vert}_{k=0}
    \left(
    \begin{array}{c}
      \vert E \vert  \\
      k  
    \end{array}
  \right)e^{ik\theta}.
\end{align}
From Eq.~(\ref{eq:diff}), the probability amplitude of each input with phase $k\theta$ after diffusion execution becomes:
\begin{align}
\label{eq:optp}
    \frac{1}{\sqrt{2^{\vert V \vert}}}(\vert 2\langle \alpha(\theta) \rangle - e^{ik\theta}\vert).
\end{align}

If $\vert V \vert=5$,
the oracle adds the phase $e^{i4\theta}$ to the input corresponding to the MAX-CUT.
There are two bit strings corresponding to the same MAX-CUT.
(e.g. As shown in Sec.~\ref{comp_exact}, $\vert 01111 \rangle$ and $\vert 10000 \rangle$ denote the same MAX-CUT of $K_{1,4}$.)
We describe how to eliminate this redundancy by using a virtual node in the final part of this section.

When $\theta = \theta_0$, the probability of finding either value of MAX-CUT $p(\theta)$ becomes:
\begin{align}
    p(\theta_0)=2\left(\frac{1}{\sqrt{32}}\vert 2\langle \alpha(\theta) \rangle - e^{i4\theta_0}\vert\right)^2\simeq 0.195.
\end{align}

If we have information about the phase each input is given by the Oracle,
we can maximize the amplification factor by adjusting the subdivided phase:
\begin{align}
    {\rm max}\{\vert 2\langle \alpha(\theta) \rangle - e^{i4\theta_0} \vert \}\simeq 1.84.
\end{align}
Then, maximized $p(\theta)$ and optimal subdivided phase are:
\begin{align}
    p(\theta_{opt}) \simeq 0.212, \\
    \theta_{opt} \simeq 0.323 \pi.
\end{align}
The amount of amplification depends on the difference between the average value of the added phase.
Therefore, the probability of the worst solution that does not cut any edges is amplified similarly to the proper MAX-CUT solution.

Using the exact solution in the Sec.~\ref{sec:Exact}, the average value of the added phase $\langle\alpha'\rangle$ after applying the above oracle $O$ with $t=4$ is:
\begin{align}
    \langle\alpha'\rangle=\frac{1}{2^{\vert E \vert}}
    \left(e^{i\pi} + 2^{\vert E \vert} -1
    \right).
\end{align}
Then, after performing oracle and diffusion only once, the probability of finding either of MAX-CUT $p'$ becomes:
\begin{align}
    p'=\frac{1}{16}(\vert 2\langle \alpha' \rangle - e^{i\pi}\vert)^2\simeq 0.473.
\end{align}

Thus, the performance of our method lies in between the random search and the standard Grover algorithm using $\pi$ for the phase shift.
Not only that, since the average value increases as the graph become denser, the worst-case probability becomes larger than MAX-CUT.
Despite such drawbacks, this algorithm requires many fewer gates than searching for an exact solution.

A naive implementation using QAOA, a useful NISQ algorithm for MAX-CUT, requires $\vert V \vert$ qubits to store data and $2\vert E \vert~CX$~ gates for each Cost Unitary~\cite{farhi2014quantum,otterbach2017unsupervised}.
As discussed below, this is on a scale comparable to the Oracle circuit in our proposal.
The circuit depth is equal to or greater than our circuit when the number of block repeats $p=2$ or more.
Since QAOA requires parameter changes and iterative execution, the number of CX gates needed to obtain a solution is an advantage of our method.

Besides, for problem sizes where the classical algorithm requires a few $milliseconds$~\cite{goemans1995improved,kugel2010improved,guerreschi2019qaoa}, it is difficult for our method to outperform the classical algorithm on the MAX-CUT problem, since it requires a huge number of trials (Eq.~(\ref{eq:optp})).
Next, we show a specific implementation method.

\subsubsection*{Implementation of oracle}
\label{subsec:impora}
When the $\theta$ is not $0$ or $\pi$, the sub-oracle in Eq.~\ref{eq:app_suboracle} consists of the following gate sequence:
\begin{align}
    O'_{s(A,B)}:=X_{B}CR^{B,A}_{Z(\theta)}X_{B}X_{A}CR^{A,B}_{Z(\theta)}X_{A}.
\end{align}
Due to the limitations of the current IBM Q processors within the framework of QASM~\cite{cross2017open}, we need two $CX$ gates and single-qubit gates to execute one $CR^{A,B}_{Z}(\theta)$ exactly.

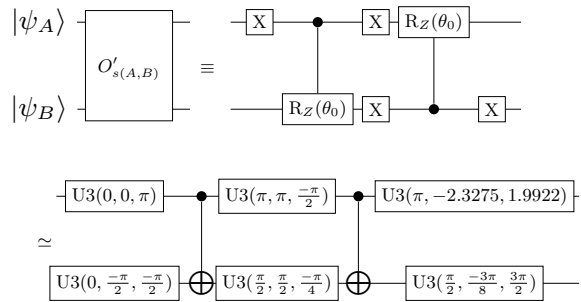
\begin{figure}[htb]
    \begin{center}
        \resizebox{0.45\textwidth}{!}{
    \begin{tikzpicture}[background rectangle/.style={fill=white},
                    show background rectangle]
        
    \tikzstyle{operator} = [draw, fill=white,minimum size=1.em] 
    \tikzstyle{ctrl} = [fill,shape=circle,minimum size=5pt,inner sep=0pt]
    \tikzstyle{targ} = [draw, shape=circle, cross]
    %
    %
    \node[scale=1.5] at (-0.3, 0) (q1) {$\ket{\psi_A}$};
    \node[scale=1.5] at (-0.3,-1.5) (q2) {$\ket{\psi_B}$};
    %
    %
    \draw (q1) -- (0.5, 0);
    \draw (q2) -- (0.5, -1.5);
    \draw (0.5, 0.2) -- (2, 0.2) -- (2, -1.7) -- (0.5, -1.7) -- (0.5, 0.2);
    \node at (1.25, -0.8) {$O'_{s(A, B)}$};
    \draw (2, 0) -- (2.3, 0);
    \draw (2, -1.5) -- (2.3, -1.5);

    \node at (2.6, -0.8){$\equiv$};
    \node at (-0.2, -3.8){$\simeq$};
    %
    %
    \draw(3, 0) -- (8, 0);
    \draw(3, -1.5) -- (8, -1.5);
    
    \draw(0, -3) -- (9, -3);
    \draw(0, -4.5) -- (9, -4.5);
    
    \node [operator] (R$_Z$2-1) at (3.5, 0) {X};
    \node [operator] (R$_Z$1) at (1, -3) {U3${(0, 0, \pi)}$};
    \node [operator] (R$_Z$2) at (1, -4.5) {U3${(0, \frac{-\pi}{2}, \frac{-\pi}{2})}$};
    %
    %
    \node [ctrl] (ctrl2-1) at (4.5, 0) {};
    \node [operator] (targ2-1) at (4.5, -1.5) {R$_Z(\theta_{0})$};
    \draw (ctrl2-1) -- (targ2-1);
    \node [ctrl] (ctrl1) at (2.5, -3) {};
    \node[scale=1.3] (targ1) at (2.5, -4.5) {$\bigoplus$};
    \draw (ctrl1) -- (2.5, -4.5);
    %
    %
    \node [operator] (R$_Z$2-3) at (5.5, 0) {X};
    \node [operator] (Rz2-4) at (5.5, -1.5) {X};
    \node [operator] (Rz3) at (3.8, -3) {U3${(\pi, \pi, \frac{-\pi}{2})}$};
    \node [operator] (Rz4) at (3.8 , -4.5) {U3${(\frac{\pi}{2}, \frac{\pi}{2}, \frac{-\pi}{4})}$};
    %
    %
    \node [ctrl] (ctrl2-2) at (6.5, -1.5) {};
    \node [operator] (targ2-2) at (6.5
    , 0) {R$_Z(\theta_{0})$} ;
    \draw (ctrl2-2) -- (targ2-2);
    \node [ctrl] (ctrl2) at (5.2, -3) {};
    \node[scale=1.3](targ2) at (5.2, -4.5) {$\bigoplus$};
    \draw (ctrl2) -- (5.2, -4.5);
    %
    %
    \node [operator] (Rz2-6) at (7.5, -1.5) {X};
    \node [operator] (Rz5) at (7.2, -3) {U3${(\pi, -2.3275, 1.9922)}$};
    \node [operator] (Rz6) at (7.2, -4.5) {U3${(\frac{\pi}{2}, \frac{-3\pi}{8}, \frac{3\pi}{2})}$};
    
    \end{tikzpicture}
}
        \caption{Approximation of sub-oracle circuit using KAK decomposition at $\theta_0 = \frac{\pi}{4}$. The approximation accuracy is over ${\bf 99}$\%, and the average error of the $CX$ gate of the Q processor as of January 2020 is about $1$\%. Until the $CX$ gate error is halved, the total error will be dominated by the two-qubit gates.}
        \label{fig:oracle_app}
    \end{center}
\end{figure}

Here, the error values on single-qubit gates are one order of magnitude smaller than that of two-qubit~($CX$) gates~\cite{QV2019}.
Therefore, we focused on reducing the number of $CX$ gates, and the number of single-qubit gates such as $U3$ gate is basically not a problem.
Hence we approximate the whole sub-oracle with two $CX$ gates and six $U3$ gates by KAK decomposition~\cite{bullock2003arbitrary,shende2004minimal} as shown in FIG.~\ref{fig:oracle_app}.
The error level of a $CX$ gate of the latest IBM Q processors used in this paper is about 1\% at best~\cite{QV2019}.
Hence, we approximate this oracle circuit with two $CX$ gates~\cite{PhysRevA.100.032328}.

\subsubsection*{Introduction of virtual vertex}
The output of the approach in Sec.~\ref{sec:Exact} has redundancy due to the symmetry of the problem.
In order to eliminate this and double the solution space in a given number of qubits, we introduce a virtual vertex  whose state is fixed at $\vert 0_V \rangle$.

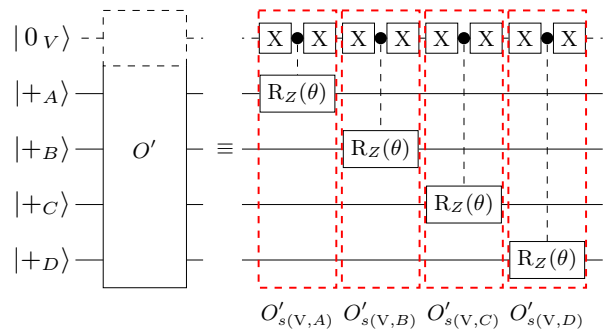
\begin{figure}[htb]
    \resizebox{0.47\textwidth}{!}{
    \begin{tikzpicture}[background rectangle/.style={fill=white},
                show background rectangle, scale=0.8]
    ]
    \tikzstyle{operator} = [draw, fill=white,minimum size=1.0em] 
    \tikzstyle{ctrl} = [fill,shape=circle,minimum size=5pt,inner sep=0pt]

    \node[scale=1.2] at (-0.6, 1) (v) {$\ket{\, 0\, _V}$};
    \node[scale=1.2] at (-0.6, 0) (q1) {$\ket{+_A}$};
    \node[scale=1.2] at (-0.6,-1) (q2) {$\ket{+_B}$};
    \node[scale=1.2] at (-0.6,-2) (q3) {$\ket{+_C}$};
    \node[scale=1.2] at (-0.6,-3) (q4) {$\ket{+_D}$};
    %
    \draw[dashed, thin] (v) -- (0.5, 1);
    \draw (0, 0) -- (0.5, 0);
    \draw (0, -1) -- (0.5, -1);
    \draw (0, -2) -- (0.5, -2);
    \draw (0, -3) -- (0.5, -3);
    \draw[dashed, thin] (0.5, 1.5) -- (2, 1.5) -- (2, 0.5) -- (0.5, 0.5) -- (0.5, 1.5);
    \draw (2, 0.5) -- (2, -3.5) -- (0.5, -3.5) -- (0.5, 0.5);
    \node at (1.25, -1) {$O'$};
    \draw[dashed] (2, 1) -- (2.3, 1);
    \draw (2, 0) -- (2.3, 0);
    \draw (2, -1) -- (2.3, -1);
    \draw (2, -2) -- (2.3, -2);
    \draw (2, -3) -- (2.3, -3);
    %
    \node at (2.7, -1){$\equiv$};
    %
    \node [operator] (x11) at (3.6, 1) {X} edge [dashed] (3, 1);
    \node [ctrl] (ctrl1) at (4, 1) {} edge [dashed] (x11);
    \node [operator] (x12) at (4.4, 1) {X} edge [dashed] (ctrl1);
    \node [operator] (Rz1) at (4, 0) {R$_Z(\theta)$} edge [-] (3, 0);
    \draw[dashed] (ctrl1) -- (Rz1);
    %
    \node [operator] (x21) at (5.1, 1) {X} edge [dashed] (x12);
    \node [ctrl] (ctrl2) at (5.5, 1) {} edge [dashed] (x21);
    \node [operator] (x22) at (5.9, 1) {X} edge [dashed] (ctrl2);
    \node [operator] (Rz2) at (5.5, -1) {R$_Z(\theta)$} edge [-] (3, -1);
    \draw[dashed, thin] (ctrl2) -- (Rz2);
    %
    \node [operator] (x31) at (6.6, 1) {X} edge [dashed] (x22);
    \node [ctrl] (ctrl3) at (7, 1) {} edge [dashed] (x31);
    \node [operator] (x32) at (7.4, 1) {X} edge [dashed] (ctrl3);
    \node [operator] (Rz3) at (7, -2) {R$_Z(\theta)$} edge [-] (3, -2);
    \draw[dashed, thin] (ctrl3) -- (Rz3);
    %
    \node [operator] (x41) at (8.1, 1) {X} edge [dashed] (x32);
    \node [ctrl] (ctrl4) at (8.5, 1) {} edge [dashed] (x41);
    \node [operator] (x42) at (8.9, 1) {X} edge [dashed] (ctrl4);
    \node [operator] (Rz4) at (8.5, -3) {R$_Z(\theta)$} edge [-] (3, -3);
    \draw[dashed, thin] (ctrl4) -- (Rz4);
    %
    \draw[dashed, thin] (x42) -- (9.5, 1);
    \draw (Rz1) -- (9.5, 0);
    \draw (Rz2) -- (9.5, -1);
    \draw (Rz3) -- (9.5, -2);
    \draw (Rz4) -- (9.5, -3);
    
    \draw[red, thick, dashed] (3.3, 1.5) -- (4.7, 1.5) -- (4.7, -3.5) -- (3.3, -3.5) -- (3.3, 1.5);
    \node (vs1) at (4, -4) {$O'_{s({\rm V}, A)}$};
    
    \draw[red, thick, dashed] (4.8, 1.5) -- (6.2, 1.5) -- (6.2, -3.5) -- (4.8, -3.5) -- (4.8, 1.5);
    \node (vs1) at (5.5, -4) {$O'_{s({\rm V}, B)}$};
    
    \draw[red, thick, dashed] (6.3, 1.5) -- (7.7, 1.5) -- (7.7, -3.5) -- (6.3, -3.5) -- (6.3, 1.5);
    \node (vs1) at (7, -4) {$O'_{s({\rm V}, C)}$};    
    
    \draw[red, thick, dashed] (7.8, 1.5) -- (9.2, 1.5) -- (9.2, -3.5) -- (7.8, -3.5) -- (7.8, 1.5);
    \node (vs1) at (8.5, -4) {$O'_{s({\rm V}, D)}$};
    
    \end{tikzpicture}
}
    \caption{Implementation of oracle circuit $O'$ using sub-divided phase for the star graph $K_{1,4}$.
    All sub-oracles $O'_{s(V,k)}$ can be replaced with $R_Z(\theta)$ by assigning the highest degree vertex to the virtual qubit.}
    \label{fig:oracle_virtual}
\end{figure}

The oracle for the edge connected to this virtual vertex can be replaced by a single qubit operation  $R_Z(\theta_0)$ on the other vertex.
In order to reduce the number of $CX$ gates in the oracle part, it is effective to virtualize the highest degree vertex.
For example, when the given graph is $K_{1,4}$, we can perform the oracle circuit without using $CX$ gates, as shown in Fig.~\ref{fig:oracle_virtual}.

\subsection{Implementation of diffusion}
\label{subsec:impdif}
After executing the oracle in Sec.~\ref{subsec:oracle}, we perform the normal diffusion operator for Grover's algorithm.
As described in Sec.~\ref{sec:BG}, the diffusion circuit for $n+1$ data qubits require one $n$-controlled NOT (\cnx) gate.

We can implement the (three-qubit) Toffoli gate with a well-known circuit using 6 $CX$, Hadamard gates and T gates~\cite{NC}. DiVincenzo and Smolin found a Toffoli gate decomposition using 5 controlled unitary gates~\cite{divincenzo1994results}, and this number of two-qubit operations was later shown to be optimal~\cite{yu2013five}.
If we allow imperfect phases, Margolus created a Toffoli gate using 3 $CX$ gates~\cite{margolus1994simple,divincenzo1998quantum}.
Maslov showed other configurations for imperfect Toffoli gates, such as based on $controlled-controlled-iX$~\cite{PhysRevA.87.042302}, and collectively called them relative-phase Toffoli gates ($RTOF$)~\cite{maslov2016advantages}.

Barenco {\it et al.} described the decomposition of an $n$-control qubit Toffoli gate (\textsc{$C^{\otimes n}X$} gate) using $2n-3$~-Toffoli gates and $n-2$ ancillary qubits~\cite{barenco1995elementary}.
To decrease the number of gates, Maslov replaced all but one Toffoli with $RTOF$ and composed \cnx with $6n-6$~-CX gates~\cite{maslov2016advantages}.

On the other hand, executing these Toffoli gate circuits requires a fully-connected three-qubit structure, which we cannot directly implement on the actual quantum processors used in this paper.  Thus, it becomes important to relax the semantic constraints, including both phase and variable qubit placement.
Therefore, we discuss how to implement a \cnx~gate under the constraints of the IBM Q processors.

\subsubsection*{\cnx~gate implementation}
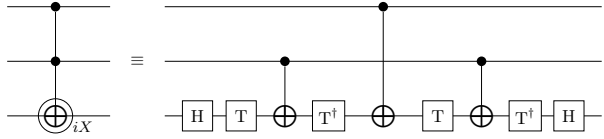
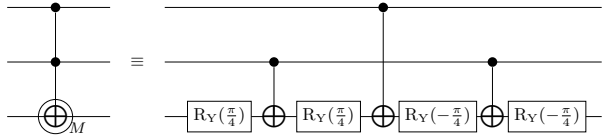
\begin{figure}[htb]
    \begin{minipage}[t]{.47\textwidth}
    \centering
    \resizebox{1\textwidth}{!}{
    \begin{tikzpicture}[background rectangle/.style={fill=white},
                    show background rectangle]
    
    \tikzstyle{operator} = [draw, fill=white,minimum size=1.7em] 
    \tikzstyle{ctrl} = [fill,shape=circle,minimum size=5pt,inner sep=0pt]
    \node at (0, 0) (c0) {};
    \node at (0,-1) (c1) {};
    \node at (0, -2) (t) {};
    \node [ctrl] (ctrl11) at (1, 0) {};
    \node [ctrl] (ctrl21) at (1, -1) {};
    \node [scale=1.3, draw=black, circle, inner sep=0.03pt] at (1,-2) {$\bigoplus$};
    \node at (1.5,-2.2) {{\footnotesize $iX$}};
    \draw[line width=0.1mm] (ctrl11) -- (1, -2);
    
    \draw[line width=0.1mm] (c0) -- (2,  0);
    \draw[line width=0.1mm] (c1) -- (2, -1);
    \draw[line width=0.1mm] (t) -- (2, -2);
    
    \node at (2.5, -1) {$\equiv$};
    \draw[line width=0.1mm](3, -2) -- (11, -2);
    %
    %
    \node [operator] at (3.6, -2) {H};
    \node [operator] at (4.4, -2) {T};
    
    %
    %
    \node[ctrl] (ctrl22) at (5.2, -1){};
    \node[scale=1.3] at (5.2, -2) {$\bigoplus$};
    \draw[line width=0.1mm] (ctrl22) -- (5.2, -2);
    %
    %
    \node [operator] at (6, -2) {T$^\dagger$};
    %
    %
    \node[ctrl] (ctrl41) at (7, 0){};
    \node[scale=1.3] at (7, -2) {$\bigoplus$};
    \draw[line width=0.1mm] (ctrl41) -- (7, -2);
    %
    %
    \node [operator] at (8, -2) {T};
    %
    %
    \node[ctrl] at (8.8, -1){};
    \node[scale=1.3] at (8.8, -2) {$\bigoplus$};
    \draw[line width=0.1mm] (8.8, -1) -- (8.8, -2);
    %
    %
    \node [operator] at (9.6, -2) {T$^\dagger$};
    \node [operator] at (10.4, -2) {H};
    
    \draw[line width=0.1mm] (3, 0) -- (11, 0);
    \draw[line width=0.1mm] (3, -1) -- (11, -1);
    
    \end{tikzpicture}
}
    \subcaption{{\bf The controlled-controlled-iX gate~\cite{PhysRevA.87.042302}~(\RTX).} This gate uses four $U1$ gates and two $U2$ gates~(see Tab.~\ref{tab:gates}).}
    \label{fig:rccix}
  \end{minipage} \\
  \begin{minipage}[t]{.47\textwidth}
    \centering
    \resizebox{1\textwidth}{!}{ 
    \begin{tikzpicture}[background rectangle/.style={fill=white},
                    show background rectangle]
    
    \tikzstyle{operator} = [draw, fill=white,minimum size=1.7em] 
    \tikzstyle{ctrl} = [fill,shape=circle,minimum size=5pt,inner sep=0pt]
    \node at (0, 0) (c0) {};
    \node at (0,-1) (c1) {};
    \node at (0, -2) (t) {};
    \node [ctrl] (ctrl11) at (1, 0) {};
    \node [ctrl] (ctrl21) at (1, -1) {};
    \node [scale=1.3, draw=black, circle, inner sep=0.03pt] at (1,-2) {$\bigoplus$};
    \node [] at (1.4,-2.2) {{\footnotesize $M$}};
    \draw[line width=0.1mm] (ctrl11) -- (1, -2);
    
    \draw[line width=0.1mm] (c0) -- (2,  0);
    \draw[line width=0.1mm] (c1) -- (2, -1);
    \draw[line width=0.1mm] (t) -- (2, -2);
    
    \node at (2.5, -1) {$\equiv$};
    \draw[line width=0.1mm](3, -2) -- (11, -2);
    %
    %
    \node [operator] at (4, -2) {$\rm{R_{Y}(\frac{\pi}{4})}$};
    
    %
    %
    \node [ctrl] (ctrl22) at (5, -1){};
    \node [scale=1.3] at (5, -2) {$\bigoplus$};
    \draw[line width=0.1mm] (ctrl22) -- (5, -2);
    %
    %
    \node [operator] at (6, -2) {$\rm{R_{Y}(\frac{\pi}{4})}$};
    %
    %
    \node [ctrl] (ctrl41) at (7, 0){};
    \node [scale=1.3] at (7, -2) {$\bigoplus$};
    \draw[line width=0.1mm] (ctrl41) -- (7, -2);
    %
    %
    \node [operator] at (8, -2) {$\rm{R_{Y}(-\frac{\pi}{4})}$};
    %
    %
    \node [ctrl] (ctrl62) at (9, -1){};
    \node [scale=1.3] at (9, -2) {$\bigoplus$};
    \draw[line width=0.1mm] (ctrl62) -- (9, -2);
    %
    %
    \node [operator] at (10, -2) {$\rm{R_{Y}(-\frac{\pi}{4})}$};
    
    \draw[line width=0.1mm] (3, 0) -- (11, 0);
    \draw[line width=0.1mm] (3, -1) -- (11, -1);
    
    \end{tikzpicture}
}
    \subcaption{{\bf The Margolus gate~(\RTM).} In addition to the normal Toffoli operation, the sign of $\vert 101 \rangle$ is inverted. This gate uses four $U3$ gates.}
    \label{fig:rccxm}
  \end{minipage}
  \caption{Two $RTOF$ implementations adopted for \cnx.}
  \label{fig:rccx}
\end{figure}

To construct \cnx, 
we adopt two types of $RTOF$, shown in the FIG.~\ref{fig:rccx}.
Both of these $RTOF$ can be implemented on a system with only a one-dimensional qubit layout. Although the number of $CX$ gate is equal, $RTOF_{iX}$ does not require $U3$, which reduces single qubit rotation errors.
To implement a \cnx gate with ancillary qubits and $RTOF$~\cite{maslov2016advantages} while avoiding the above-mentioned connectivity problem, we also introduce the Toffoli gate with built-in SWAP operation.

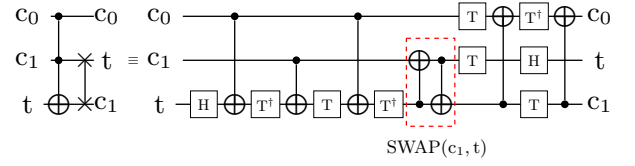
\begin{figure}[htb]
 \begin{center}
    \resizebox{0.47\textwidth}{!}{%
   \begin{tikzpicture}[background rectangle/.style={fill=white},
                    show background rectangle]

    \tikzstyle{operator} = [draw, fill=white,minimum size=1.7em] 
    \tikzstyle{ctrl} = [fill,shape=circle,minimum size=5pt,inner sep=0pt]
    \tikzstyle{targ} = [draw, shape=circle, cross]

    \node[scale=2] at (0, 0) (c0-0) {$\rm c_0$};
    \node[scale=2] at (0,-1) (c1-0) {$\rm c_1$};
    \node[scale=2] at (0, -2) (t-0) {$\rm t$};
    \draw[line width=0.3mm] (c0-0) -- (1.5,  0);
    \draw[line width=0.3mm] (c1-0) -- (1.5, -1);
    \draw[line width=0.3mm] (t-0) -- (1.5, -2);
    
    \node [ctrl] (ctrl00) at (0.7, 0) {};
    \node [ctrl] (ctrl01) at (0.7, -1) {};
    \node [scale=1.5] at (0.7, -2) {$\bigoplus$};
    \draw[line width=0.3mm][line width=0.3mm] (ctrl00) -- (0.7, -2);
    
    \node[scale=2] (ctrl10) at (1.3, -1) {$\times$};
    \node[scale=2] (ctrl11) at (1.3, -2) {$\times$};
    \draw[line width=0.3mm][line width=0.3mm] (1.3, -1) -- (1.3, -2);
    
    \node[scale=2] at (1.8, 0) (c0-0) {$\rm c_0$};
    \node[scale=2] at (1.8, -1) (t-0) {$\rm t$};
    \node[scale=2] at (1.8,-2) (c1-0) {$\rm c_1$};
    
    \node at (2.4, -1) {$\equiv$};

    \node[scale=2] at (3, 0) (c0) {$\rm c_0$};
    \node[scale=2] at (3,-1) (c1) {$\rm c_1$};
    \node[scale=2] at (3, -2) (t) {$\rm t$};
    
    \draw[line width=0.3mm][line width=0.3mm] (c0) -- (12.6,  0);
    \draw[line width=0.3mm][line width=0.3mm] (c1) -- (12.6, -1);
    \draw[line width=0.3mm][line width=0.3mm] (t) -- (12.6, -2);

    %
    %
    \node [operator, scale=1.1] at (4, -2) {H};
    %
    \node [ctrl] (ctrl22) at (4.7, 0) {};
    \node [scale=1.5] at (4.7, -2) {$\bigoplus$};
    \draw[line width=0.3mm] (ctrl22) -- (4.7, -2);
    %
    %
    \node [operator, scale=1.1] at (5.4, -2) {T$^\dagger$};
    %
    %
    \node [ctrl] (ctrl41) at (6.1, -1) {};
    \node [scale=1.5] at (6.1, -2) {$\bigoplus$};
    \draw[line width=0.3mm] (ctrl41) -- (6.1, -2);
    %
    %
    \node [operator, scale=1.1] at (6.8, -2) {T};
    %
    %
    \node [ctrl] (ctrl62) at (7.5, 0) {};
    \node [scale=1.5] at (7.5, -2) {$\bigoplus$};
    \draw[line width=0.3mm] (ctrl62) -- (7.5, -2);
    %
    %
    \node [operator, scale=1.1] at (8.2, -2) {T$^\dagger$};
    %
    %
    \node [ctrl] (ctrl82) at (8.9, -2) {};
    \node [scale=1.5] at (8.9, -1) {$\bigoplus$};
    \draw[line width=0.3mm] (ctrl82) -- (8.9, -1);
    %
    %
    \node [ctrl] (ctrl90) at (9.4, -1) {};
    \node [scale=1.5] at (9.4, -2) {$\bigoplus$};
    \draw[line width=0.3mm] (ctrl90) -- (9.4, -2);
    %
    %
    \node [operator, scale=1.1] at (10.1, 0) {T};
    \node [operator, scale=1.1] at (10.1, -1) {T};
    %
    %
    \node [ctrl] (ctrl112) at (10.8, -2) {};
    \node [scale=1.5] at (10.8, 0) {$\bigoplus$};
    \draw[line width=0.3mm] (ctrl112) -- (10.8, 0);
    %
    %
    \node [operator, scale=1.1] at (11.5, 0) {T$^\dagger$};
    \node [operator, scale=1.1] at (11.5, -1) {H};
    \node [operator, scale=1.1] at (11.5, -2) {T};
    %
    %
    \node [ctrl] (ctrl132) at (12.2, -2) {};
    \node [scale=1.5] at (12.2, 0) {$\bigoplus$};
    \draw[line width=0.3mm] (ctrl132) -- (12.2, 0);
    
    \node[scale=2] at (13, 0) (c0end) {$\rm c_0$};
    \node[scale=2] at (13,-1) (c1end) {$\rm t$};
    \node[scale=2] at (13, -2) (tend) {$\rm c_1$};
    
    \draw[line width=0.3mm][red, thick, dashed] (8.6, -0.5) -- (9.7, -0.5) -- (9.7, -2.5) -- (8.6, -2.5) -- (8.6, -0.5);
    \node[scale=1.3] (vs1) at (9.3, -3) {SWAP($\rm c_1, t$)};
    
    \end{tikzpicture}
}
    \caption{{\bf Toffoli with SWAP circuit.}  By adding the $CX$ gates surrounded by a broken line to the general Toffoli gate decomposition, SWAP is built in, and the circuit can be performed with qubits connected in a straight line.}
    \label{fig:TwithS}
 \end{center}
\end{figure}

A Toffoli gate implementation with the minimal 6 $CX$ gates requires three qubits interconnected in a triangle.
Recent IBM Q devices after {\bf ibm\_tokyo} do not have a structure that can embed triangles.
To deal with this situation, we propose a Toffoli  circuit suitable for a one-dimensional layout, as shown in FIG.~\ref{fig:TwithS}.
This circuit requires one additional $CX$, the minimum overhead.
However, since SWAP is built in, it is necessary to consider the location of qubits in the output state.

\begin{figure}[htb]
 \begin{center}
    \begin{tikzpicture}[scale=0.9]
    \tikzstyle{operator} = [draw, fill=white,minimum size=1.0em] 
    \tikzstyle{ctrl} = [fill,shape=circle,minimum size=5pt,inner sep=0pt]
     \tikzstyle{cross} = [cross/.style={path picture={ \draw[black] (path picture bounding box.east) -- (path picture bounding box.west) (path picture bounding box.south) -- (path picture bounding box.north);}}]
    \tikzstyle{targ} = [draw, shape=double circle, draw=black, size=1mm]

    
    \node at (0, 0) (c0) {$c_0$};
    \node at (0, -0.5) (c1) {$c_1$};
    \node at (0,-1) (a0) {$a_0$};
    \node at (0,-1.5) (c2) {$c_{2}$};
    \node at (0,-2) (a1) {$a_1$};
    \node at (0,-2.4) (dots) {$ \vdots $};
    \node at (0,-3) (an-2) {$a_{n-2}$};
    \node at (0,-3.5) (cn) {$c_n$};
    \node at (0,-4) (t) {$t\,\,\,$};
    %
    %
    \node [ctrl] (cnx0) at (0.8, 0) {} edge [-] (c0);
    \node [ctrl] (cnx1) at (0.8, -0.5) {} edge [-] (c1);
    \draw (cnx0) -- (cnx1);

    \node [ctrl] (cnx3) at (0.8, -1.5) {} edge [-] (c2);
    \draw (cnx1) -- (cnx3);
    \draw (cnx3) -- (0.8, -2.25);
    
    \node (dots) at (0.8, -2.4) {$\vdots$};
    
    \node [ctrl] (cnxn) at (0.8, -3.5) {} edge [-] (cn);
    \draw (0.8, -2.75) -- (cnxn);
    
    \node[] (target) at (0.8, -4) {$\bigoplus$};
    \draw (cnxn) -- (0.8, -4);
    \node at (1.3, -3.5) {$\times$};
    \node at (1.3, -4) {$\times$};
    \draw (1.3, -3.5) -- (1.3, -4);
    %
    %
    \draw (cnx0) -- (1.8, 0);
    \draw (cnx1) -- (1.8, -0.5);
    \draw (a0) -- (1.8, -1);
    \draw (cnx3) -- (1.8, -1.5);
    \draw (a1) -- (1.8, -2);
    \draw (an-2) -- (1.8, -3);
    \draw (cnxn) -- (1.8, -3.5);
    \draw (t) -- (1.8, -4);
    %
    %
    \node at (2.5, -2) {$\equiv$};
    %
    %
    \node [ctrl] (ctrl11) at (3.5, 0) {};
    \node [ctrl] (ctrl12) at (3.5, -0.5) {};
    \node [draw=black, circle, inner sep=0.2pt] at (3.5,-1) {$\bigoplus$};
    \draw (ctrl11) -- (3.5, -1);
    %
    %
    \node [ctrl] (ctrl21) at (4, -1) {};
    \node [ctrl] (ctrl22) at (4, -1.5) {};
    \node [draw=black, circle, inner sep=0.2pt] at (4,-2) {$\bigoplus$};
    \draw (ctrl21) -- (4, -2);
    %
    %
    \node at (4.5, -2.4) {$\ddots$};
    %
    %
    \node [draw=black, circle, inner sep=0.2pt] at (4.9,-3) {$\bigoplus$};
    \draw (4.9, -2.65) -- (4.9, -3);
    %
    %
    \node [ctrl] (ctrl51) at (5.5, -3) {};
    \node [ctrl] (ctrl52) at (5.5, -3.5) {};
    \node (targ) at (5.5, -4) {$\bigoplus$};
    \draw (ctrl51) -- (5.5, -4);
    %
    \node at (6, -3.5) {$\times$};
    \node at (6, -4) {$\times$};
    \draw (6, -3.5) -- (6, -4);
    %
    \node [draw=black, circle, inner sep=0.2pt] at (6.6,-3) {$\bigoplus$};
    \draw (6.6, -2.65) -- (6.6, -3);
    %
    %
    \node at (7, -2.4) {$\reflectbox{$\ddots$}$};
    %
    %
    \node [ctrl] (ctrl51) at (7.5, -1) {};
    \node [ctrl] (ctrl52) at (7.5, -1.5) {};
    \node [draw=black, circle, inner sep=0.2pt] at (7.5,-2) {$\bigoplus$};
    \draw (ctrl51) -- (7.5, -2);
    %
    %
    \node [ctrl] (ctrl61) at (8, 0) {};
    \node [ctrl] (ctrl62) at (8, -0.5) {};
    \node [draw=black, circle, inner sep=0.2pt] at (8,-1) {$\bigoplus$};
    \draw (ctrl61) -- (8, -1);
    %
    %
    \draw (3, 0) -- (8.5, 0);
    \draw (3, -0.5) -- (8.5, -0.5);
    \draw (3, -1) -- (8.5, -1);
    \draw (3, -1.5) -- (8.5, -1.5);
    \draw (3, -2) -- (8.5, -2);
    \draw (3, -3) -- (8.5, -3);
    \draw (3, -3.5) -- (8.5, -3.5);
    \draw (3, -4) -- (8.5, -4);
    
    \draw[red, thick, dashed] (5.2, -2.8) -- (5.2, -4.25) -- (6.3, -4.25) -- (6.3, -2.8) -- (5.2, -2.8);
\end{tikzpicture}
    \caption{{\bf Configuration of \cnx using $n-2$ ancillary qubits and $2n-4$ Toffoli gates.}  Toffoli gates other than the one enclosed in the dashed box can be replaced with the relative phase gate. Thereby, the number of $CX$ gates can be reduced.}
    \label{fig:cnx}
 \end{center}
\end{figure}
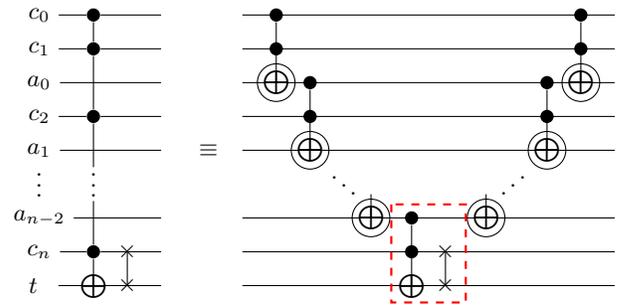

By using those components, we can configure a \cnx gate  for recent IBM Q devices using $6n-5$~$CX$  gates.
It is known that a \cnx~gate can consist of $2n-3$ Toffoli gates with $n-2$ ancillary qubits (initialized to $\vert 0 \rangle$)~\cite{barenco1995elementary}.
A Toffoli gate contains at least 6 $CX$ gates.
As shown in FIG.~\ref{fig:cnx}, Toffoli gates in \cnx can be replaced with $RTOF$ except for the central one.
To support additional implementations,
we show the procedure for \cnx in Algorithm~\ref{proc_cnx}.
\begin{algorithm}[H]
  \caption{\cnx~gate implementation}
  \label{proc_cnx}
  \begin{algorithmic}[1]
    \Statex {Input: $n+1$ data qubits $d$, $n-2$ ancillary qubits $a$.}
    \Statex {Output: Data qubits on which the \cnx~ gate is performed and SWAP gate between last two data qubits.}
    \Statex {}
    \Procedure{\cnx~gate with SWAP}{}
    \State {RTOF($d_0$, $d_1$, $a_0$)}
    \For {k=0; k<n-3; k++}
        \State {RTOF($a_k$, $d_{k+2}$, $a_{k+1}$)}
    \EndFor
    \State{TOF($a_n-3$, $d_n-1$, $d_n$) with SWAP($d_n-1$, $n$)}
    \For {k=n-4; k>-1; k-\,-}
        \State {RTOF($a_k$, $d_{k+2}$, $a_{k+1}$)}
    \EndFor
    \State {RTOF($d_0$, $d_1$, $a_0$)}
    \State {return Target states.}
    \EndProcedure
  \end{algorithmic}
\end{algorithm}

If the processor can embed the structure shown in FIG.~\ref{fig:qubits_arrange}, the procedure can be executed without additional SWAPs.
\begin{figure}[H]
 \begin{center}
        \resizebox{0.47\textwidth}{!}{%
\begin{tikzpicture}
    \tikzstyle{qubit} = [draw, shape=circle, fill=white,minimum size=3.0em]
    \tikzstyle{boldnode} = [draw, shape=circle,minimum size=4.0em]
    
    \node [qubit] at (0, 0) (d0) {\huge $d_0$};
    %
    \node [qubit] at (2, 0) (a0) {\huge $a_0$};
    \node [qubit] at (2, -2) (d1) {\huge $d_1$};
    \draw (d0) -- (a0);
    \draw (a0) -- (d1);
    %
    \node [qubit] at (4,0) (a1) {\huge $a_1$};
    \node [qubit] at (4,-2) (d2) {\huge $d_2$};
    \draw (a0) -- (a1);
    \draw (a1) -- (d2);
    %
    \node at (6,0) (dots) {$\cdots$};
    \draw (a1) -- (dots);
    %
    \node [qubit] at (8,0) (an-3) {\Large $a_{n-3}$};
    \node [qubit] at (8,-2) (dn-2) {\Large $d_{n-2}$};
    \draw (dots) -- (an-3);
    \draw (an-3) -- (dn-2);
    %
    \node [qubit] at (10,0) (dn-1) {\Large $d_{n-1}$};
    \draw (an-3) -- (dn-1);
    %
    \node [qubit] at (12,0) (dn) {\huge $d_n$};
    \draw (dn-1) -- (dn);
    
    \node[scale=1.5] at (6, -4) {(a)  Qubit connection for Algorithm 1.};

    \draw[line width = 0.4mm](0, -6)--(12, -6); \draw[line width = 0.4mm](0, -8)--(12, -8);
    
    \draw[line width = 0.4mm](0, -10)--(12, -10);
    \draw[line width = 0.4mm](0, -12)--(12, -12);
    
    \draw[line width = 0.4mm](3, -6)--(3, -8); \draw[line width = 0.4mm](9, -6)--(9, -8);
    
    \draw[line width = 0.4mm](0, -8)--(0, -10);
    \draw[line width = 0.4mm](6, -8)--(6, -10);
    \draw[line width = 0.4mm](12, -8)--(12, -10);
    
    \draw[line width = 0.4mm](3, -10)--(3, -12); \draw[line width = 0.4mm](9, -10)--(9, -12);

    \node [qubit] (x11) at (0, -6) {\huge $d_0$};
    \node [qubit] (x12) at (3, -6) {\huge $a_0$};
    \node [qubit] (x13) at (6, -6) {\huge $d_1$};
    \node [qubit] (x14) at (9, -6) {};
    \node [qubit] (x15) at (12, -6) {};

    \node [qubit] (x21) at (0, -8) {\huge $d_2$};
    \node [qubit] (x22) at (3, -8) {\huge $a_1$};
    \node [qubit] (x23) at (6, -8) {\huge $a_2$};
    \node [qubit] (x24) at (9, -8) {\huge $d_3$};
    \node [qubit] (x25) at (12, -8) {};

    \node [qubit] (x31) at (0, -10) {\huge $d_5$};
    \node [qubit] (x32) at (3, -10) {\huge $a_4$};
    \node [qubit] (x33) at (6, -10) {\huge $a_3$};
    \node [qubit] (x34) at (9, -10) {\huge $d_4$};
    \node [qubit] (x35) at (12, -10) {};

    \node [qubit] (x41) at (0, -12) {\huge $d_6$};
    \node [qubit] (x42) at (3, -12) {\huge $a_5$};
    \node [qubit] (x43) at (6, -12) {\huge $d_7$};
    \node [qubit] (x44) at (9, -12) {\huge $d_8$};
    \node [qubit] (x45) at (12, -12) {};

    \draw[line width = 1.7mm](x11)--(x12); \draw[line width = 1.7mm](x12)--(x13);
    
    \draw[line width = 1.7mm](x21)--(x22);
    \draw[line width = 1.7mm](x12)--(x22);
    \draw[line width = 1.7mm](x22)--(x23);
    \draw[line width = 1.7mm](x23)--(x24);
    
    \draw[line width = 1.7mm](x31)--(x32);
    \draw[line width = 1.7mm](x32)--(x33);
    \draw[line width = 1.7mm](x23)--(x33);
    \draw[line width = 1.7mm](x33)--(x34);
    
    \draw[line width = 1.7mm](x41)--(x42);
    \draw[line width = 1.7mm](x32)--(x42);
    \draw[line width = 1.7mm](x42)--(x43);
    \draw[line width = 1.7mm](x43)--(x44);

    \node[scale=1.5] at (6, -14) {(b) Example mapping for $n=8$};
    
\end{tikzpicture}
}
        \caption{Qubit connections for Algorithm~\ref{proc_cnx}. Data and ancilla qubits are denoted by $d_k$ and $a_k$, respectively. (a) shows the interactions required by the algorithm; (b) shows how they might map to one of the 20-qubit machines.}
        \label{fig:qubits_arrange}
    \end{center}
\end{figure}
When $n = 1$, we can embed this in all recent processors, including the 5-qubit processors {\bf ibmq\_vigo~(ourense)}. Similarly, when $n = 8$ or less, we can embed in the processors {\bf ibmq\_boeblingen~(singapore)}, which have 20 qubits.

\subsection{Experiments on IBM Q devices}
\label{subsec:exponq}
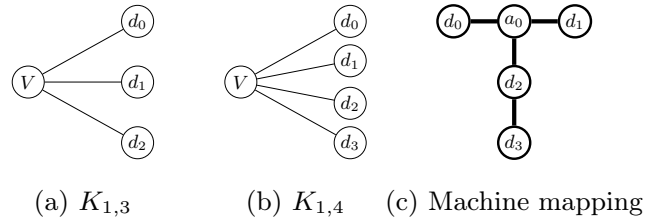
\begin{figure}[htb]
    \centering

    \resizebox{0.48\textwidth}{!}{%
\begin{tikzpicture}[every fit/.style={ellipse,
                    draw,
                    inner sep=-2pt,
                    text width=2cm, 
                    line width=1pt
                    }]
    \tikzstyle{circle} = [draw,
                        shape=circle,
                        minimum size=15pt,
                        inner sep=0pt]
                        
    \tikzstyle{qubit} = [draw=black,
    line width = 0.4mm, 
    shape=circle,
    minimum size=15pt,
    inner sep=0.2pt]
    
    \node[circle](v0) at(0, -1) {$V$};
    \node[circle](v1) at(1.8, 0) {$d_0$} edge [-] (v0);
    \node[circle](v2) at(1.8, -1) {$d_1$} edge [-] (v0);
    \node[circle](v3) at(1.8, -2) {$d_2$} edge [-] (v0);
    
    \node[scale=1.4] at (0.9, -3) {(a)~$K_{1, 3}$};
    
    \node[circle](v0) at(3.5, -1) {$V$};
    \node[circle](v1) at(5.3, 0) {$d_0$} edge [-] (v0);
    \node[circle](v2) at(5.3, -0.65) {$d_1$} edge [-] (v0);
    \node[circle](v3) at(5.3, -1.35) {$d_2$} edge [-] (v0);
    \node[circle](v4) at(5.3, -2) {$d_3$} edge [-] (v0);
    
    \node[scale=1.4] at (4.4, -3) {(b)~$K_{1, 4}$};
    
    \node [qubit] (x1) at (7, 0) {$d_0$};
    \node [qubit] (x2) at (8, 0) {$a_0$};
    \node [qubit] (x3) at (9, 0) {$d_1$};
    \node [qubit] (x4) at (8, -1) {$d_2$};
    \node [qubit] (x5) at (8, -2) {$d_3$};
    \draw[line width = 0.7mm](x1)--(x2);
    \draw[line width = 0.7mm](x2)--(x3);
    \draw[line width = 0.7mm](x2)--(x4);
    \draw[line width = 0.7mm](x4)--(x5);
    
    \node[scale=1.4] at (8, -3) {(c) Machine mapping};

\end{tikzpicture}
}
  \caption{
        {\bf Correspondence between qubits and star graphs $K_{1,n}$. }
        (a),~(b)~We assign the virtual qubit $V$ to the highest degree node and the other nodes to physical qubits. (c)~Mapping of variables to the machine for $K_{1,4}$ on both processors.
    }
  \label{fig:correspondence_gq}
\end{figure}

We evaluate our proposed algorithm by finding MAX-CUT of $K_{1,3}$ and $K_{1,4}$ on current processors. If the given graph is $K_{1,4}$, our algorithm requires 5 physical qubits (4 data, 1 ancillary) and 1 virtual qubit. FIG.~\ref{fig:correspondence_gq} illustrates the correspondence between the given graphs and qubits.

We investigate the performance of each component and the whole algorithm.

\subsubsection*{\cnx gate performance}
The $CX$ gate error rates of {\bf ibmq\_ourence} and {\bf ibmq\_valencia} are around 1~\%, an order of magnitude higher than errors of single-qubit gates~(see TAB.~\ref{table:device_qubit} and~\ref{table:device_cx} in Appendix~\ref{appendix:performance_device}).
We performed several experiments to verify the performance of $U3$ gates and measurement error mitigation~\cite{kandala2017hardware}. 
The results in FIG.~\ref{fig:single_rot} ~(Appendix~\ref{appendix:performance_ry}) show that the single-qubit gate error and the mitigated measurement error are much smaller than the $CX$ error.
 
\begin{figure}[htbp]
    \begin{center}
    \includegraphics[width=86mm]{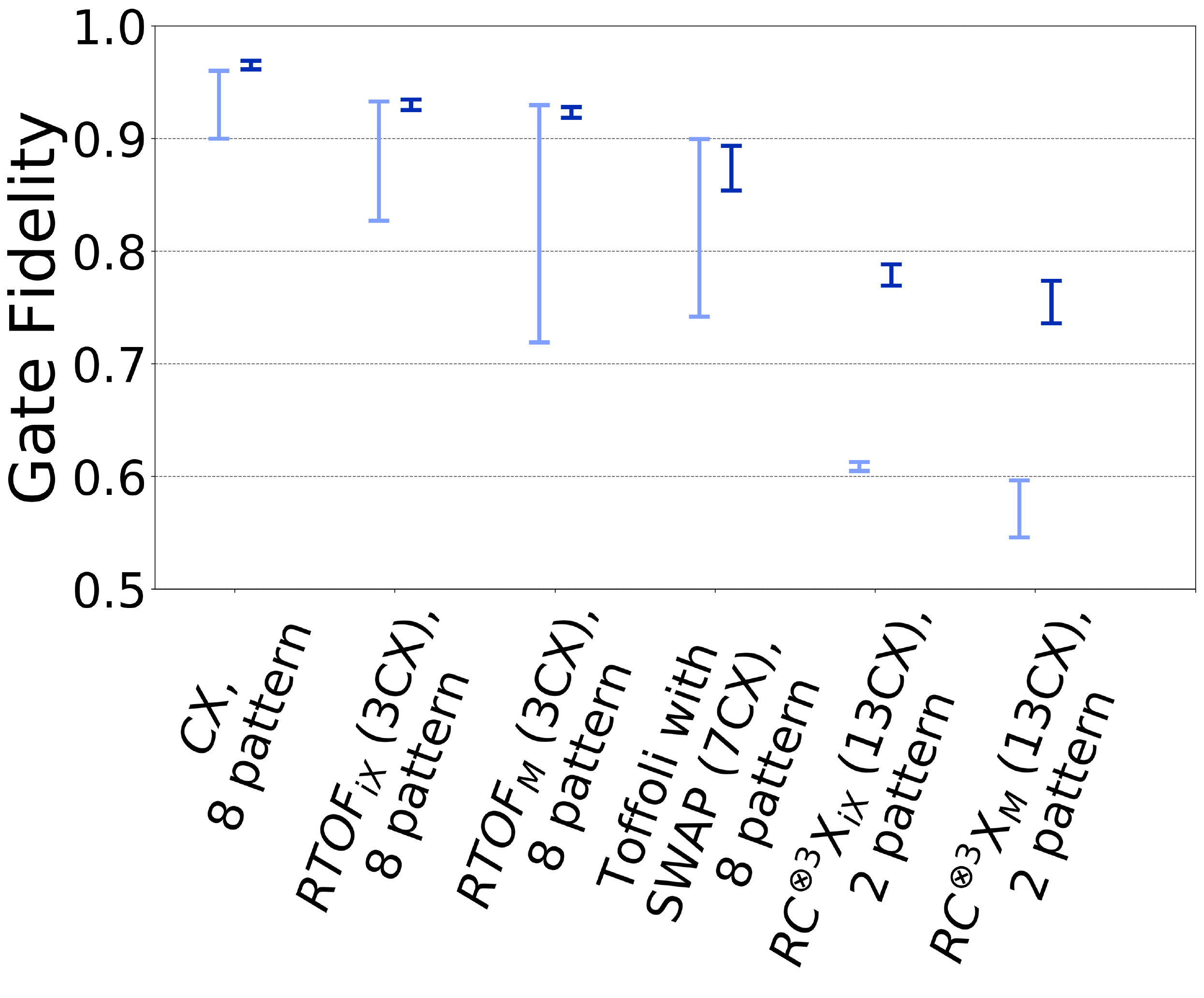}
    \caption{
        {\bf Gate fidelities of various Toffoli gates on real devices.} 
        Light blue points are the gate fidelity on {\bf ibmq\_ourense} with ${\rm QV} = 8$ and deep blue points  are the gate fidelity on {\bf ibmq\_valencia} with ${\rm QV} = 16$. For each gate type, we tested all possible mappings to the processor topology, collecting the results of $8192$ shots for each pattern. The top and bottom bar of each data bar are the maximum and minimum values of the experimental results.
    }
    \label{fig:tofs}
    \end{center}
\end{figure}

Using Algorithm~\ref{proc_cnx}, 
we can assemble a $C^{\otimes 3}X$ gate from a Toffoli with SWAP gate, and two types of $RTOF$ gates.
To evaluate these gate performances, we reconstructed output states.
We calculated fidelities of those states as shown in FIG.~\ref{fig:tofs}.
Additionally we also confirm the output of $RTOF$ gates $C^{\otimes 3}X$ gates in the computational basis in FIG.~\ref{fig:tof_prob} and~\ref{fig:c3x_prob} (Appendix~\ref{appendix:performance_cnx}).

These results show that {\bf ibmq\_valencia} is a better device than {\bf ibmq\_ourense} in accordance with their QV values in terms of average fidelity and variance.

\subsubsection*{Whole circuit performance on real processors}
\begin{figure}[htbp]
    \begin{minipage}[t]{.47\textwidth}
        \centering
        \includegraphics[width=86mm]{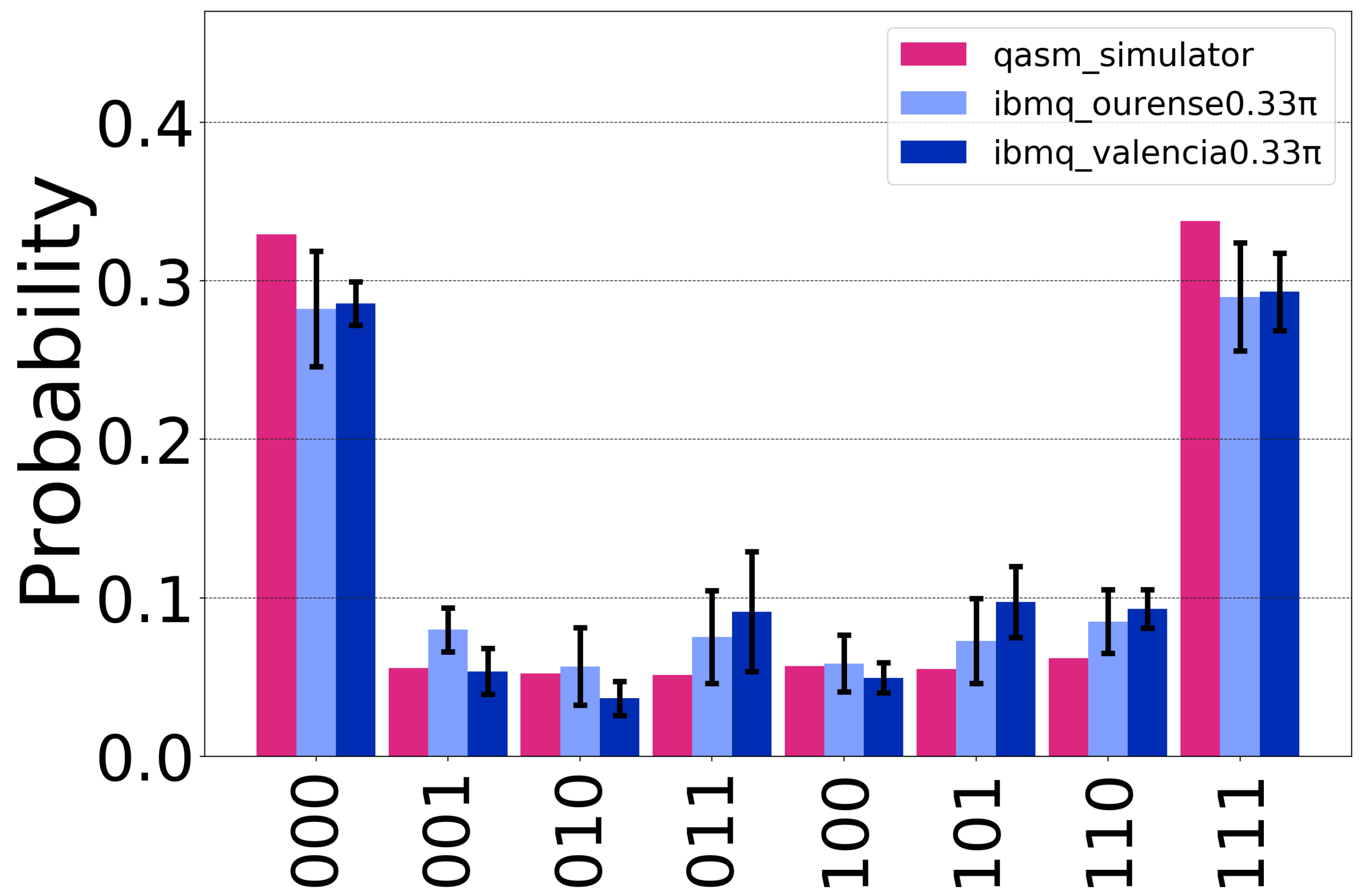}
        \subcaption{Subdivided phase angle $\theta = 0.333\pi$.}
    \end{minipage}
    \begin{minipage}[t]{.47\textwidth}
        \centering
        \includegraphics[width=86mm]{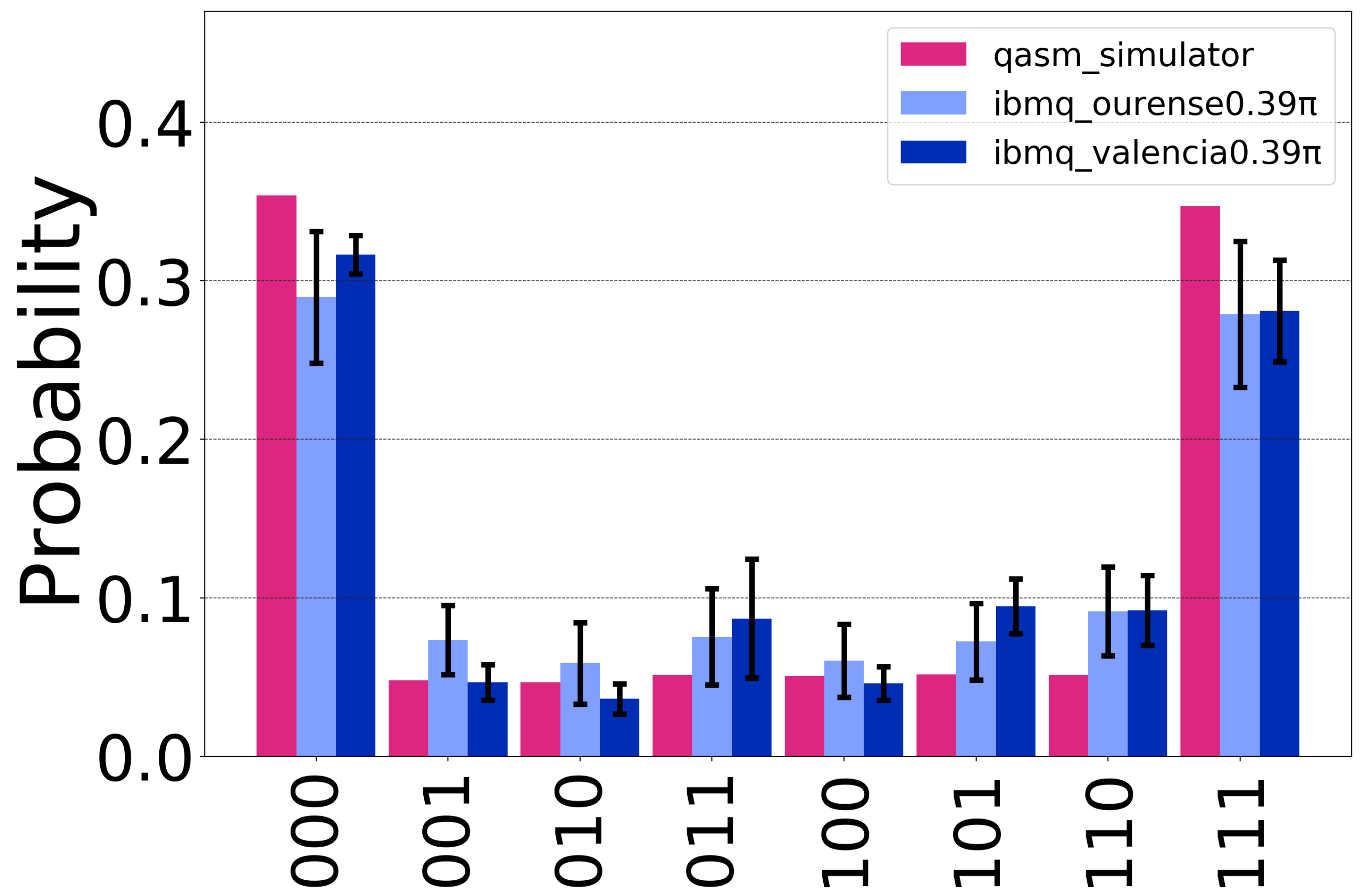}
        \subcaption{Subdivided phase angle $\theta = 0.392 \pi$ }
    \end{minipage}
    \caption{
    {\bf Results of the complete subdivided oracle search for $K_{1,3}$ MAX-CUT.} 
    On each processor, we created eight different qubit mappings, and executed each circuit $819200$ times with measurement error mitigation. Error bars represent the standard error $1 \sigma$.
    }
    \label{fig:3grover_result}
\end{figure}

\begin{figure}[htbp]
    \centering
    \includegraphics[width=80mm]{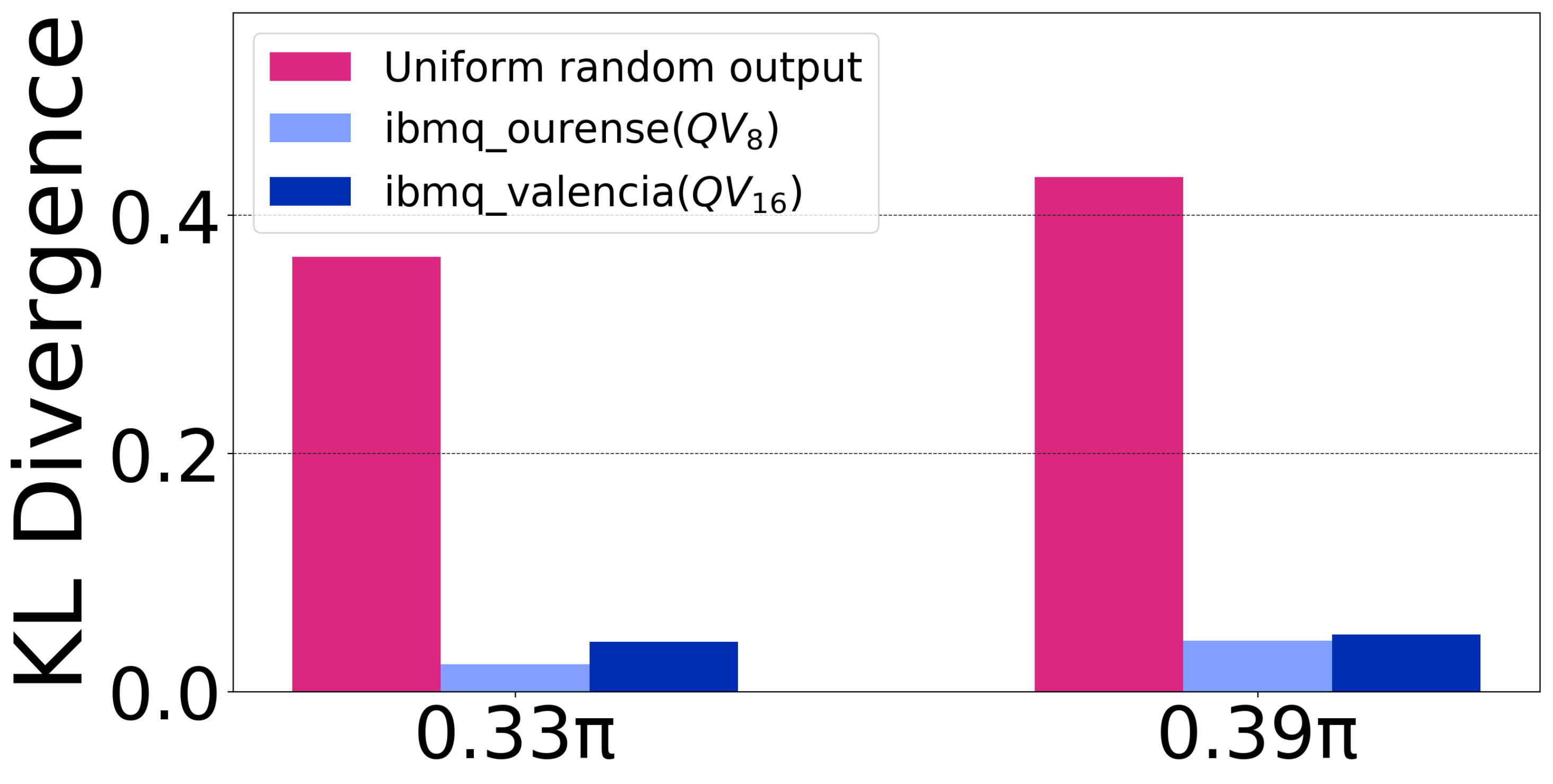}
    \caption{
        {\bf Kullback-Leibler divergence of 3 qubit subdivided Grover outputs } The KL  divergence of the data from FIG.~\ref{fig:kl-3grover} relative to the expected probability distribution (from pure state simulation) compares favorably to that of a uniform distribution with all output values having equal probability (as would be expected with high noise levels), showing that quantum algorithm performs well. Although the overall quality of valencia is superior to ourense, two output values (011 and 101) are more heavily weighted, giving a slightly worse KL divergence.
    }
    \label{fig:kl-3grover}
\end{figure}

To evaluate our algorithm performance, we first execute the whole circuit (see FIG.~\ref{fig:3grover_circuit}) with $7~CX$ for $K_{1,3}$.
In this experiment, we adopt two subdivided phases
\begin{align}
    \theta_0 = \frac{\pi}{3}
\end{align}
and
\begin{align}
    \theta_{opt} = 0.392\pi
\end{align}
obtained from Eq.~(\ref{eq:optp}).
FIG.~\ref{fig:3grover_result} shows the execution results using two processors.
In all experiments, the output probability of the correct answer $\vert 111 \rangle$ is about 28\%, which is a good result even when compared to the ideal value of 33.4\% with $\theta_{0}$ and 34.7\% with $\theta_{opt}$.
For a more quantitative evaluation we show the KL divergence in FIG.~\ref{fig:kl-3grover}.
A better value for {\bf ibm\_ourense} would suggest that the circuit is small enough for both processors.

\begin{figure*}[htbp]
\begin{tabular}{c}
    \begin{minipage}[t]{.47\textwidth}
        \centering
        \includegraphics[width=80mm]{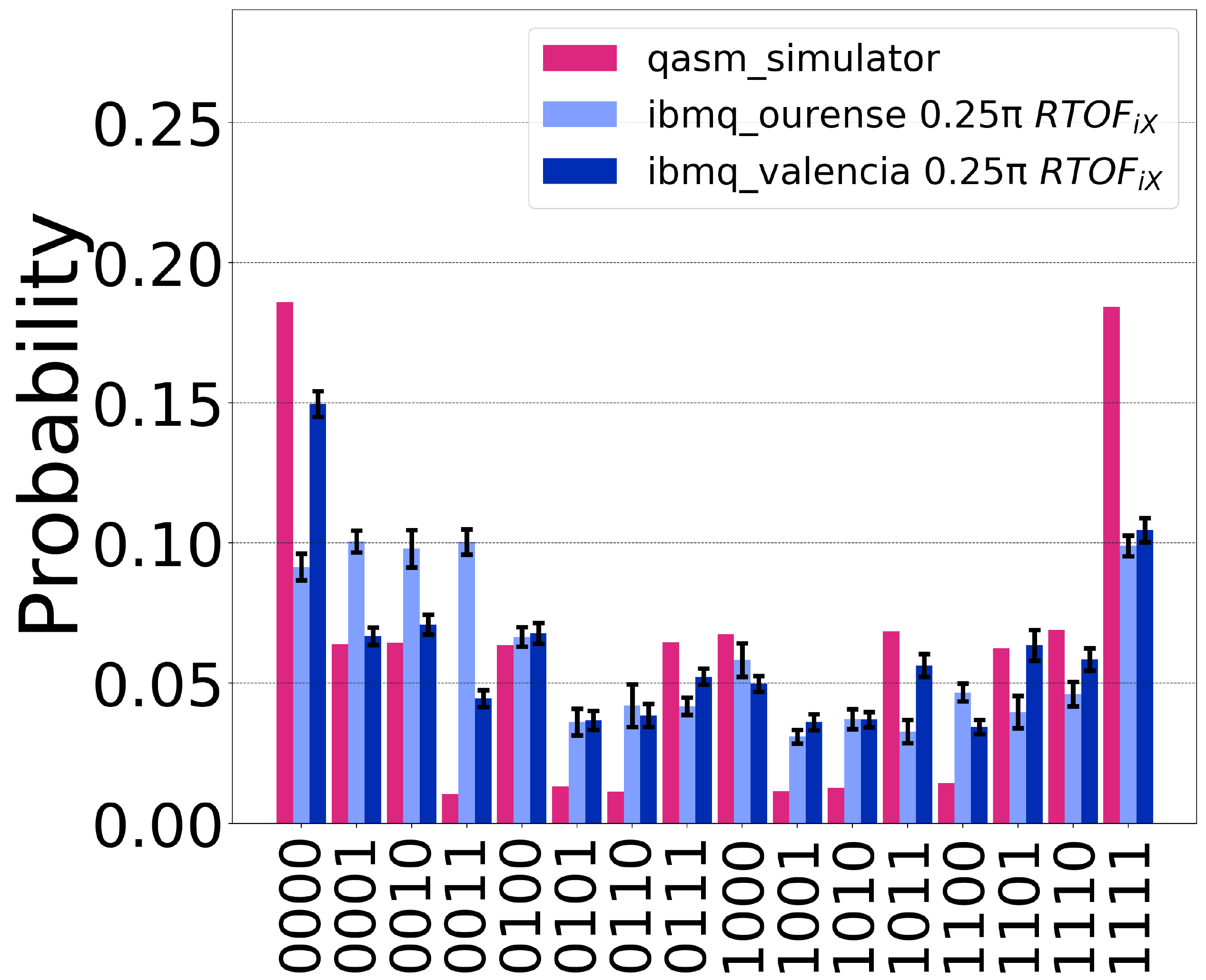}
        \subcaption{$\theta = 0.25\pi$, $C^{\otimes 3}X$ with $RTOF_{iX}$.}
        \label{fig:grover_u2_025}
    \end{minipage}
    \begin{minipage}[t]{.47\textwidth}
    \includegraphics[width=80mm]{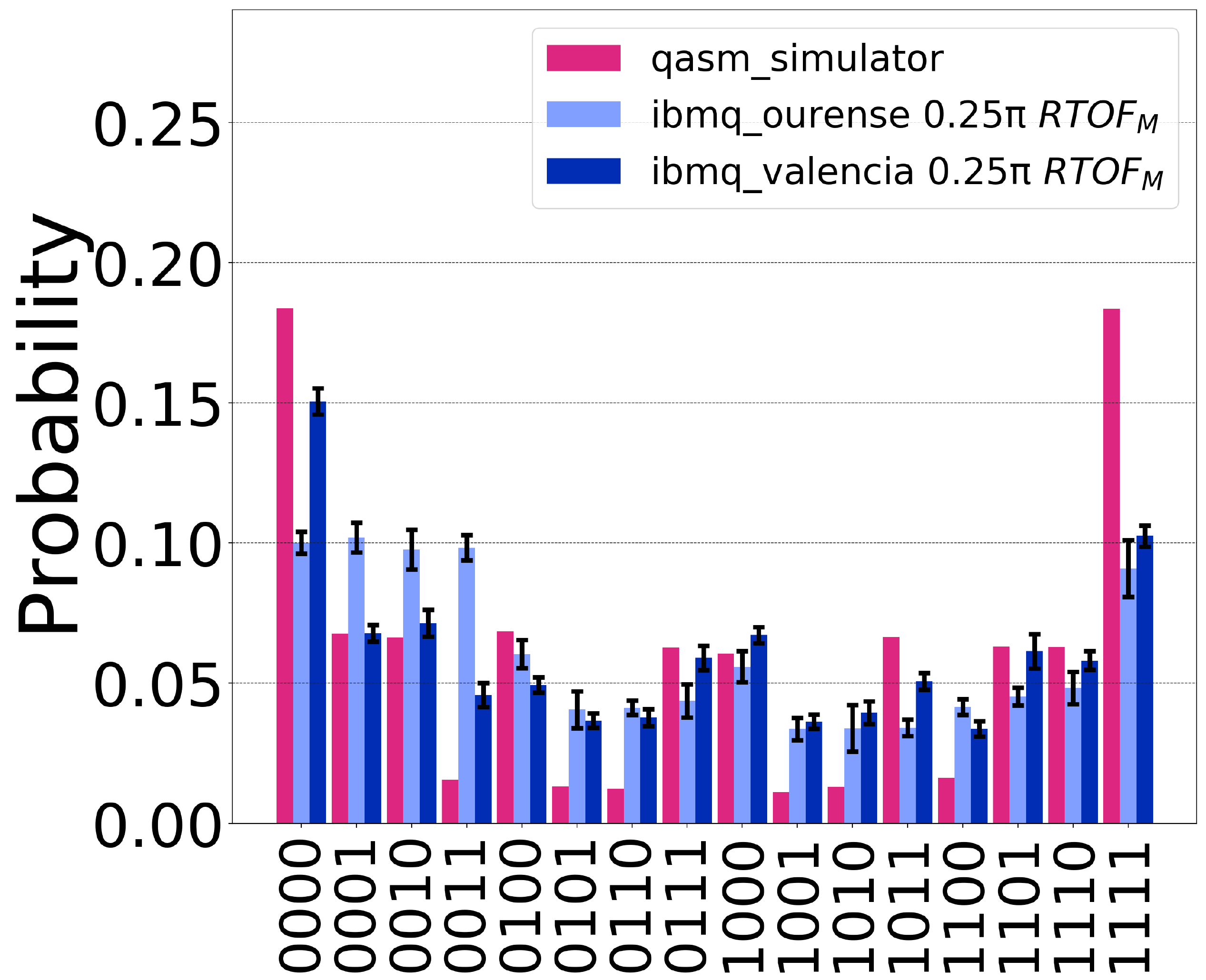}
        \subcaption{$\theta = 0.25\pi$, $C^{\otimes 3}X$ with $RTOF_{M}$.}
        \label{fig:grover_u3_025}
    \end{minipage}\\
    
    \begin{minipage}[t]{.47\textwidth}
        \centering
        \includegraphics[width=80mm]{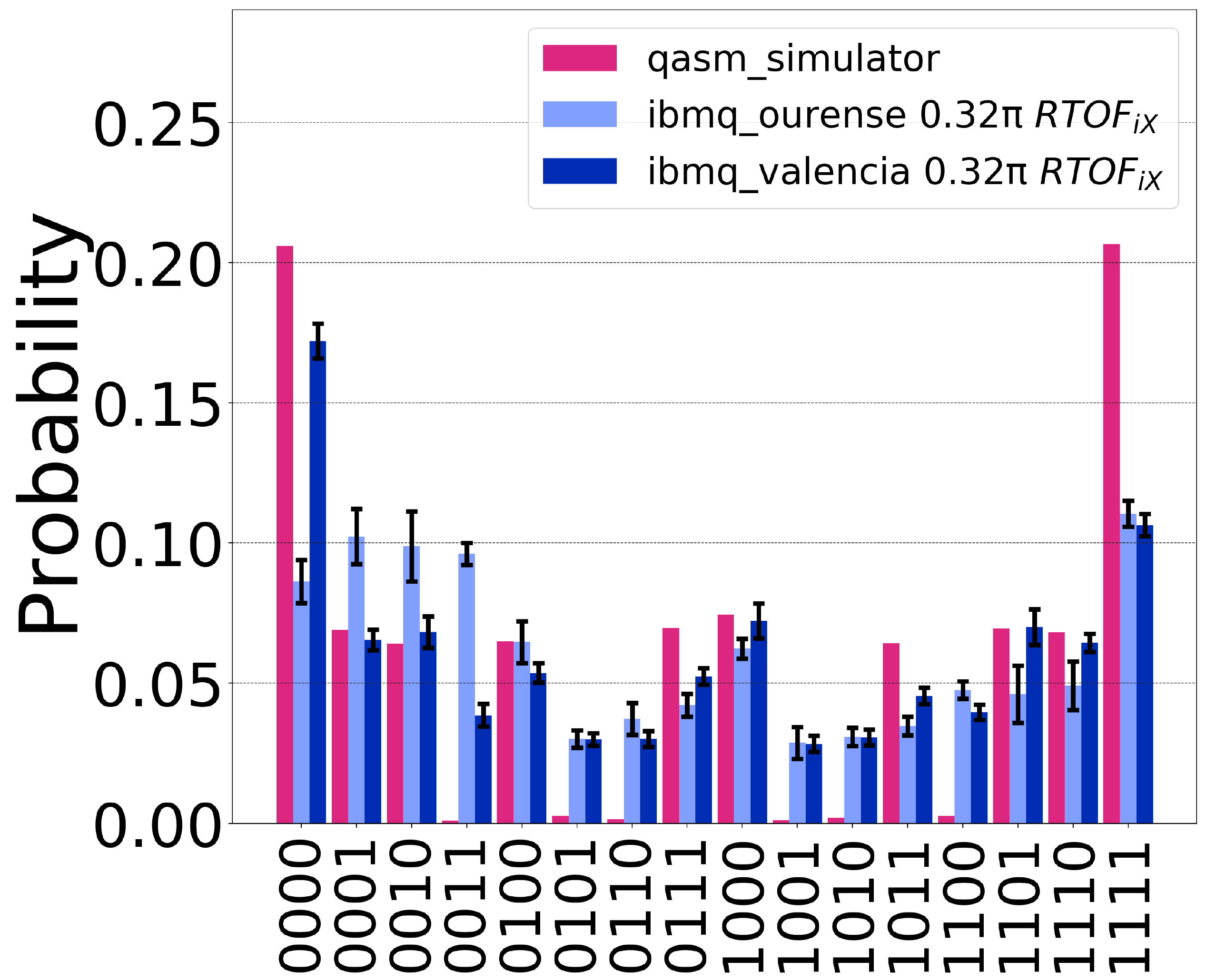}
        \subcaption{$\theta = 0.323\pi$, $C^{\otimes 3}X$ with $RTOF_{iX}$.}
        \label{fig:grover_u2_032}
    \end{minipage}
    \begin{minipage}[t]{.47\textwidth}
        \centering
        \includegraphics[width=80mm]{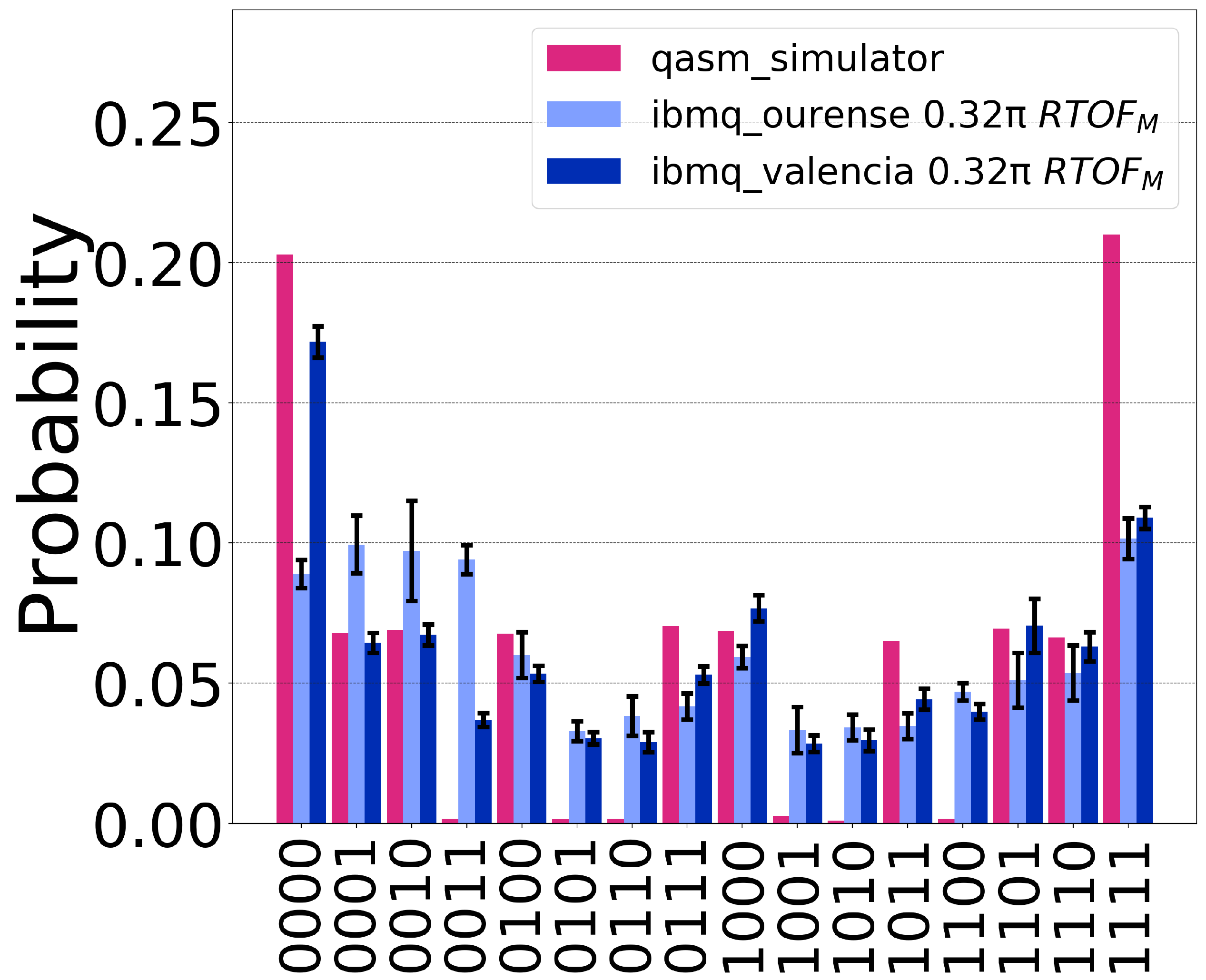}
        \subcaption{$\theta = 0.323\pi$, $C^{\otimes 3}X$ with $RTOF_{M}$.}
        \label{fig:grover_u3_032}
    \end{minipage} \\
    \end{tabular}
    \caption{
    {\bf Results from execution of the complete subdivided oracle search for $K_{1,4}$ MAX-CUT.} 
    Two qubit mappings were tested for each circuit. Each circuit is executed $819200$ times with measurement error mitigation. Error bars represent the standard error $1 \sigma$.
    }
    \label{fig:4grover_result}
\end{figure*}

\begin{figure}[htbp]
    \centering
    \includegraphics[width=86mm]{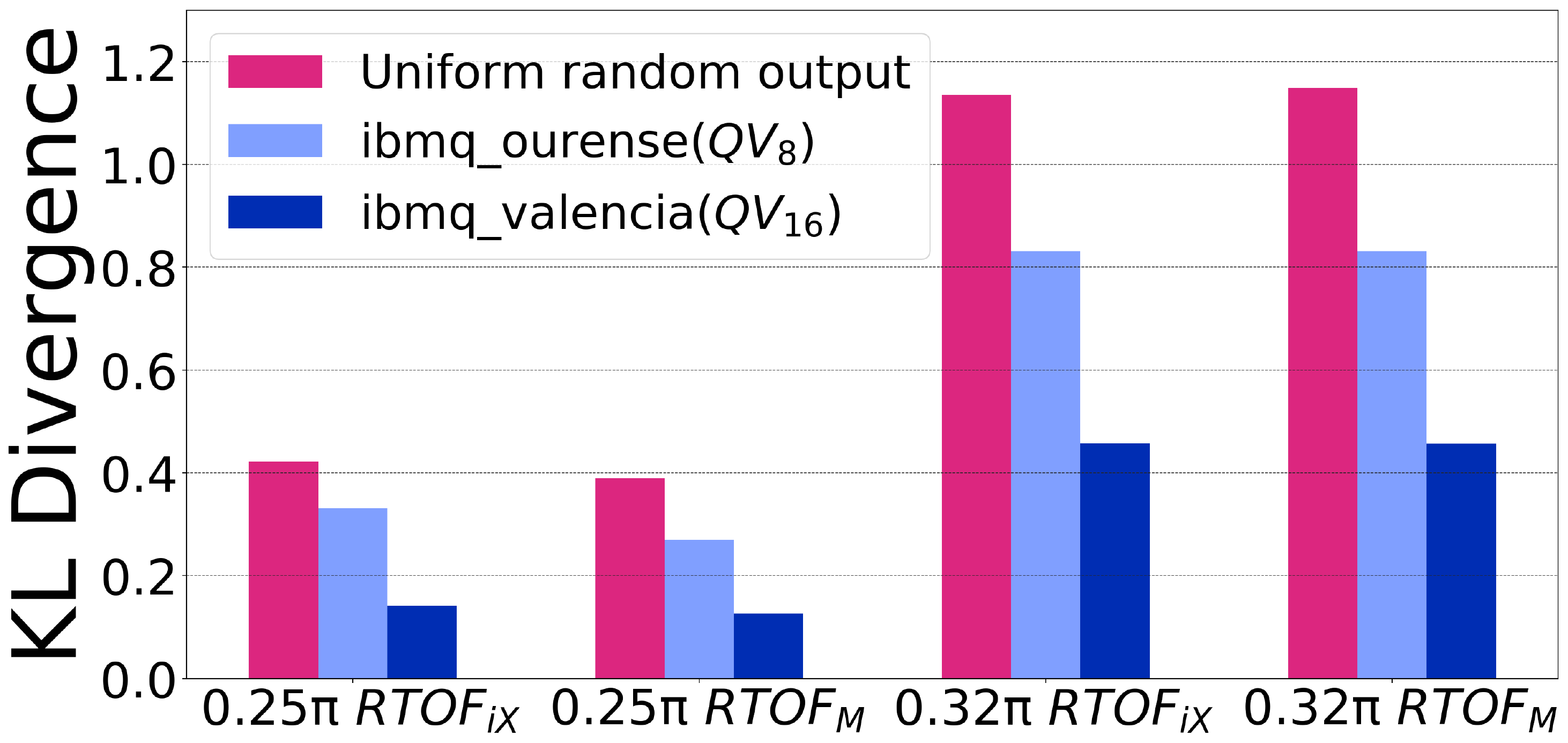}
    \caption{
        {\bf Kullback-Leibler divergence of 4-qubit subdivided oracle search outputs } The KL divergence values for the 4-qubit search are substantially higher than for the 3-qubit search, as expected, but still show a clear difference from the uniform distribution, evidence of the algorithm's effectiveness. {\bf ibmq\_valencia}'s higher QV is apparent here.
    }
    \label{fig:kl-4grover}
\end{figure}

We next execute the whole circuit (see FIG.~\ref{fig:4grover_circuit}) with $13~CX$ for $K_{1,4}$.
As discussed in Sec.~\ref{subsec:oracle}, we adopted both $0.25\pi$ and $0.323\pi$ for the angle of divided phase oracle.
We also adopt $RTOF_{iX}$ and $RTOF_{M}$ in $C^{\otimes 3}X$ gate.
We show results on two processors in FIG.~\ref{fig:4grover_result}.
In these experiments, the
probabilities of the correct answer $\vert 1111 \rangle$ are increasing.
Those probabilities are maximum when $\theta_{opt}$ is used in any processors, and is about 11\%, about half the theoretical probability of 21.2\%.
On the other hand, there is a significant difference between the processors in the probability amplification and suppression of incorrect answers.
This may be due in part to the $\vert 1111 \rangle$ output being susceptible to relaxation errors.
We show the difference in performance between the two processors using KL in FIG.~\ref{fig:kl-4grover}.
Due to the symmetry of the problem, the probability of $\vert 1111 \rangle$, which is the MAX-CUT value, and $\vert 0000 \rangle$, where no edge is cut, should be amplified the most.
Nevertheless, only one of the results is greatly amplified.
In the circuit used in this experiment, the oracle does not include $CX$, and diffusion includes the theoretically minimum number of $CX$ in current IBM Q processors.
The fact that we were unable to achieve the ideal probability amplification even when such a circuit was adopted seems to indicate that the number of qubits and circuit depth exceed the current processor capability.
Further, considering the effect of relaxation, an increase in the probability of a solution containing more $\vert 0 \rangle$ values seems natural. However, depending on the qubit mapping, the probability of solutions containing $\vert 1 \rangle$ clearly increases.
This may be due to an unknown difference between the data structure in the development environment Qiskit and the data structure on the actual IBM Q system.
\section{Conclusion}
\label{sec:Con}
As of this writing, there has been no report that any problem has been solved using 4-qubit unmodified Grover search on a solid-state quantum computer. As shown in Sec.~\ref{sec:Exact}, the scale of the circuit required for the algorithm exceeds the limit that existing quantum processors can handle. Thus, we investigated alternate solutions appropriate for the NISQ era, reducing the number of qubits and gates required by over one order of magnitude via the sub-divided phase oracle. This oracle, rather than the normal 0/$\pi$ phase flip of ordinary Grover, applies a smaller phase shift to less desirable outcomes and a larger phase shift to more desirable ones. While this initially appears less favorable, the dramatic reduction in required fidelity makes it a good tradeoff for small problems, as shown by our experimental results demonstrating effective amplitude amplification for 4-qubit search problems as exemplified by solving the MAX-CUT problem. Further work will help to determine the range of problem sizes and characteristics for which this technique can be applied.

With our current modest circuit depths, overall performance is still strongly affected by measurement errors, but it is worth comparing the KQ of our algorithms with the reported QV of the processors.  We found that the $K_{(1,3)}$ solution using 7 CNOTs on 3 qubits ($KQ=7\times 3=21$) works well on quantum volume QV=8, and very similarly on QV=16. The $K_{(1,4)}$ solution using 13 CNOTs on 4 qubits ($KQ=13\times 4=52$) works, although not well, on QV=8; it performs much better, but still with limited effectiveness, on QV=16.  This circuit is one of the largest KQ values reported to have been run successfully on a solid-state quantum computer to date.  KQ and QV are similar measures and it will be interesting to continue tracking their relationship and predictive value for execution success over the coming generations of computers.

In addition, we designed a diffusion operator using the minimum number of $CX$ gates within the constraints of recent IBM Q processors, by incorporating Toffoli gate variants with phase shifts that we compensate for later in the algorithm. This technique is exact, and will benefit a broad range of algorithms beyond the NISQ era.

\begin{acknowledgments}
This work was supported by MEXT Quantum Leap Flagship Program Grant Number JPMXS0118067285.
The results presented in this paper
were obtained in part using an IBM Q quantum computing system as part of the IBM Q Network. The views expressed are those of the authors and do not reflect the
official policy or position of IBM or the IBM Q team.
We thank Miguel Sozinho Ramalho and Lakshmi Prakash for working with TS and YO on the project that inspired this paper at Qiskit Camp Vermont 2019.
TS would like to thank Yuri Kobayashi, Atsushi Matsuo, and Shin Nishio for their collaborative activities for the Quantum Challenge, which helped refine the ideas in this paper.
We thank Ken M. Nakanishi at the University of Tokyo for a useful discussion on the optimal implementation of the Toffoli gate.
We are grateful for meaningful discussions with Shota Nagayama at Mercari, Inc.
\end{acknowledgments}

\bibliography{satoh_quantum}

\appendix
\onecolumngrid

\section{Multi-controlled $X$ gates for phase shift operator}
\label{pshift}
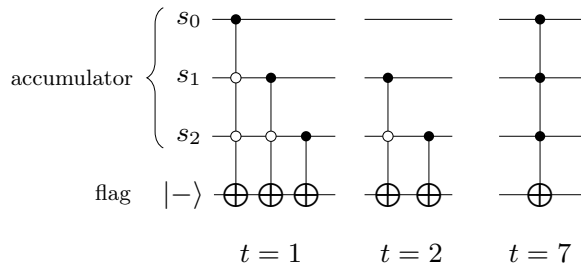
\begin{figure}[htbp]
    \begin{center}
        \resizebox{0.45\textwidth}{!}{
\begin{tikzpicture}[background rectangle/.style={fill=white}, show background rectangle]
    
    \tikzstyle{operator} = [draw, fill=white,minimum size=1.5em] 
    
    \tikzstyle{ctrl} = [fill,shape=circle,minimum size=5pt,inner sep=0pt]
    
    \tikzstyle{uctrl} = [draw, fill=white, shape=circle,minimum size=5pt,inner sep=0pt]
    
    \tikzstyle{cross} = [cross/.style={path picture={ \draw[black] (path picture bounding box.east) -- (path picture bounding box.west) (path picture bounding box.south) -- (path picture bounding box.north);}}]
    
    \tikzstyle{targ} = [draw, shape=circle, cross]
    \draw [decorate,decoration={brace,amplitude=8pt, mirror},xshift=-4pt,yshift=0pt] (-0.3,0.2) -- (-0.3,-2.2) node [black,midway,xshift=-0.6cm] {};
    \node[scale=1.2] at (-2, -1) {accumulator};
    
    \node[scale=1.2] at (-1.3, -3) {flag };
    %
    %
    \node[scale=1.5] at (0,0) (b0) {$s_0$};
    \node[scale=1.5] at (0,-1) (b1) {$s_1$};
    \node[scale=1.5] at (0,-2) (b3) {$s_2$};
    \node[scale=1.5] at (-0.1, -3) (c) {$\ket{-}$};
    %
    \node[scale=1.5] at (1.4, -4) {$t=1$};
    \draw (b0) -- (2.5, 0);
    \draw (b1) -- (2.5, -1);
    \draw (b3) -- (2.5, -2);
    \draw (c) -- (2.5, -3);
    
    \node [scale=1.3] (t1Z) at (0.8, -3) {$\bigoplus$};
    \draw (0.8, 0) -- (0.8, -3);
    \node [ctrl]  at (0.8, 0 ) {};
    \node [uctrl] at (0.8, -1) {};
    \node [uctrl] at (0.8, -2) {};
    
    \node [scale=1.3] (t1Z) at (1.4, -3) {$\bigoplus$};
    \draw (1.4, -1) -- (1.4, -3);
    \node [ctrl]  at (1.4, -1) {};
    \node [uctrl] at (1.4, -2) {};
    
    \node [scale=1.3] (t1Z) at (2, -3) {$\bigoplus$};
    \draw (2, -2) -- (2, -3);
    \node [ctrl]  at (2, -2) {};

    \node[scale=1.5] at (3.8, -4) {$t=2$};
    \draw (3, 0) -- (4.5, 0);
    \draw (3, -1) -- (4.5, -1);
    \draw (3, -2) -- (4.5, -2);
    \draw (3, -3) -- (4.5, -3);

    \node [scale=1.3] (t1Z) at (3.4, -3) {$\bigoplus$};
    \draw (3.4, -1) -- (3.4, -3);
    \node [ctrl]  at (3.4, -1) {};
    \node [uctrl] at (3.4, -2) {};
    
    \node [scale=1.3] (t1Z) at (4.1, -3) {$\bigoplus$};
    \draw (4.1, -2) -- (4.1, -3);
    \node [ctrl]  at (4.1, -2) {};
    
    \node[scale=1.5] at (6, -4) {$t=7$};
    \draw (5.3, 0) --  (6.7, 0);
    \draw (5.3, -1) -- (6.7, -1);
    \draw (5.3, -2) -- (6.7, -2);
    \draw (5.3, -3) -- (6.7, -3);

    \node [scale=1.3] (t1Z) at (6, -3) {$\bigoplus$};
    \draw (6, 0) -- (6, -3);
    \node [ctrl]  at (6, 0) {};
    \node [ctrl] at (6, -1) {};
    \node [ctrl] at (6, -2) {};
    
    \end{tikzpicture}
}
        \caption{{\bf Phase shift operators for given threshold value $t$.} The required number of $C^{\otimes n}X$ gates differs depending on the value of $t$. Here we show  implementation examples of the phase shift operator according to $t$.} 
        \label{fig:phase-flip}
    \end{center}
\end{figure}

To solve the MAX-CUT by combining the binary search and Grover's algorithm, we have to invert the sign of the input whose cut edges exceeds the  threshold value $t$.
With the proper $C^{\otimes n}X$ gate combination, the phase shift operator can distinguish whether the accumulated cut edges value exceeds $t$ or not.
We show the phase shift operator according to three different $t$ in Fig.~\ref{fig:phase-flip}.

\section{Results of supplemental experiments}
\subsection{Performance of $R_{Y}$ gate}
\label{appendix:performance_ry}

\begin{figure}
  \begin{minipage}[t]{.47\textwidth}
     \centering
    \includegraphics[width=80mm]{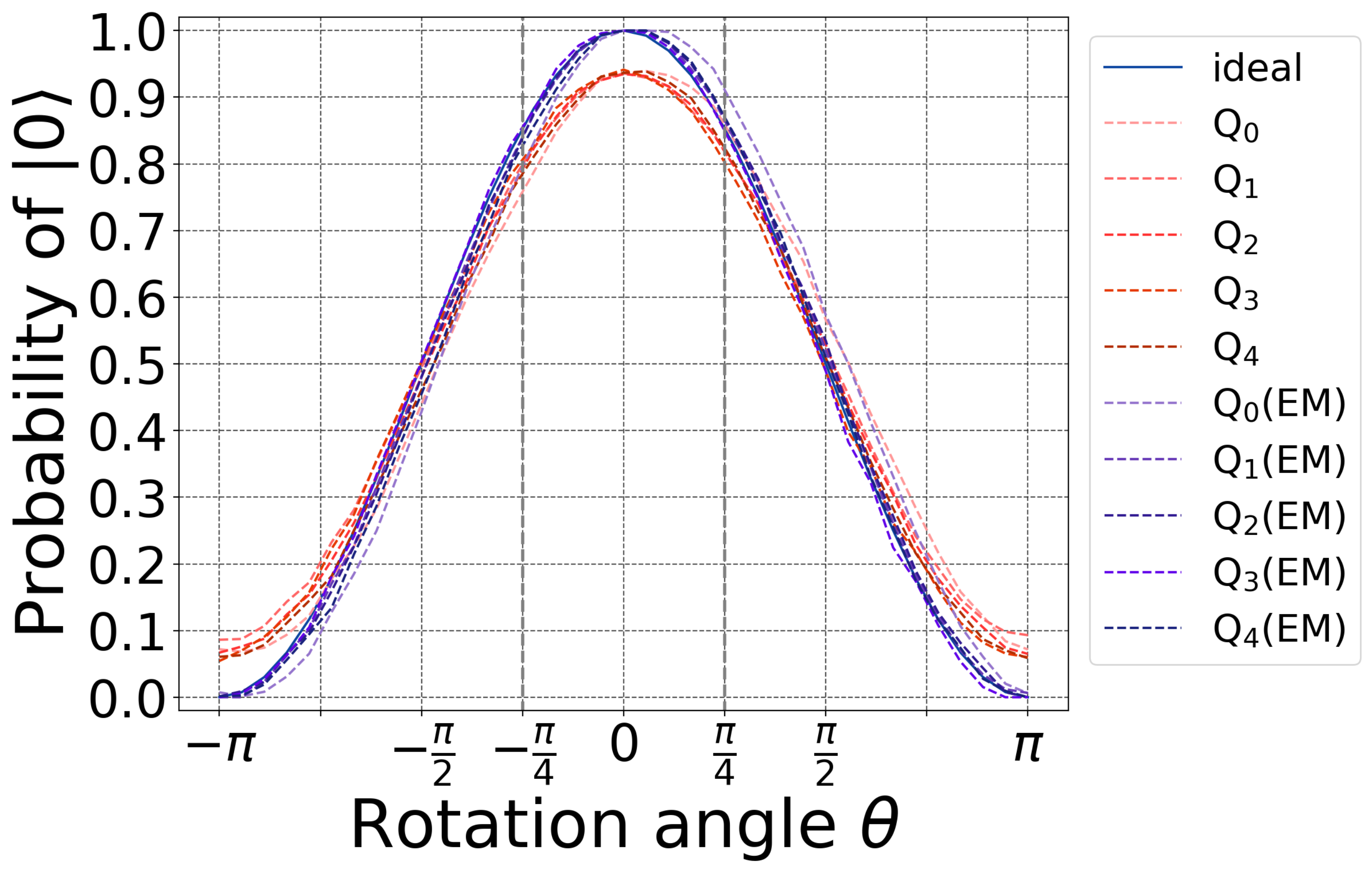}
        \subcaption{$R_{Y}(\theta)$ on {\bf ibmq\_ourense}~(${\rm QV}=8$).}
    \end{minipage} \\
    \begin{minipage}[t]{.47\textwidth}
        \centering
        \includegraphics[width=80mm]{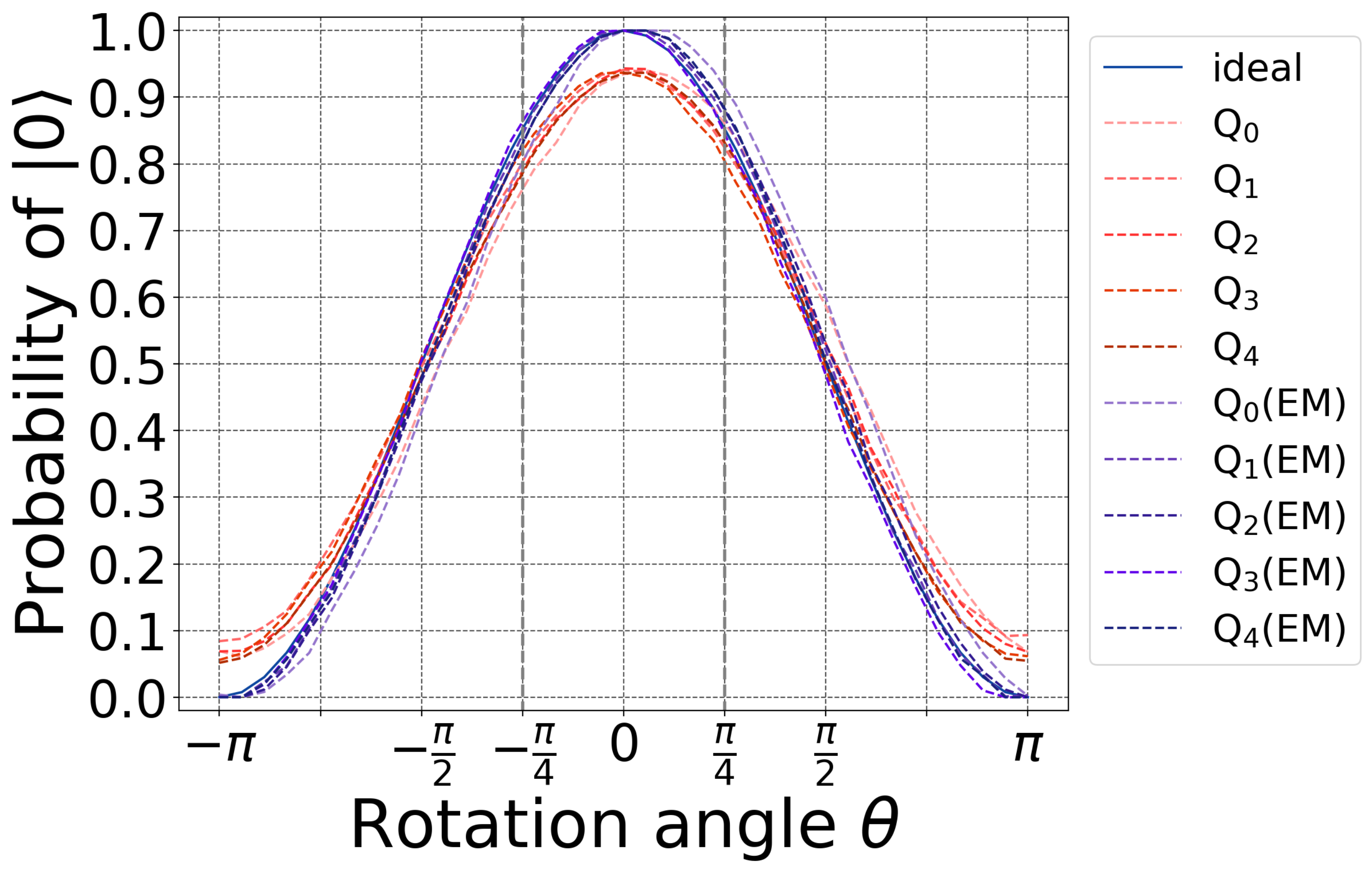}
        \subcaption{$R_{Y}(\theta)$  on {\bf ibmq\_valencia}~(${\rm QV}=16$).}
    \end{minipage} 
    \caption{Performance of $R_{Y}(\theta)$ gate as a probability of $\vert 0 \rangle$ for each rotation angle $\theta$. Dashed line correspond to $R_{Y}(\pm \frac{\pi}{4})$ in $RTOF_{M}$ gate. EM denotes result with measurement error mitigation.}
    \label{fig:single_rot}
\end{figure}

Fig.~\ref{fig:single_rot} shows the performance of $R_Y(\theta) \equiv U3(\theta,0,0)$, with and without measurement error mitigation. We show the probability of finding $\vert 0 \rangle$ when measuring in the computational basis with $-\pi \leq \theta \leq \pi$. The result after applying measurement error mitigation is close to the ideal value, regardless of the value of QV.

\begin{figure*}[tp]
    \begin{tabular}{c}
    \begin{minipage}[t]{.33\hsize}
        \includegraphics[keepaspectratio, width=60mm, angle=0]{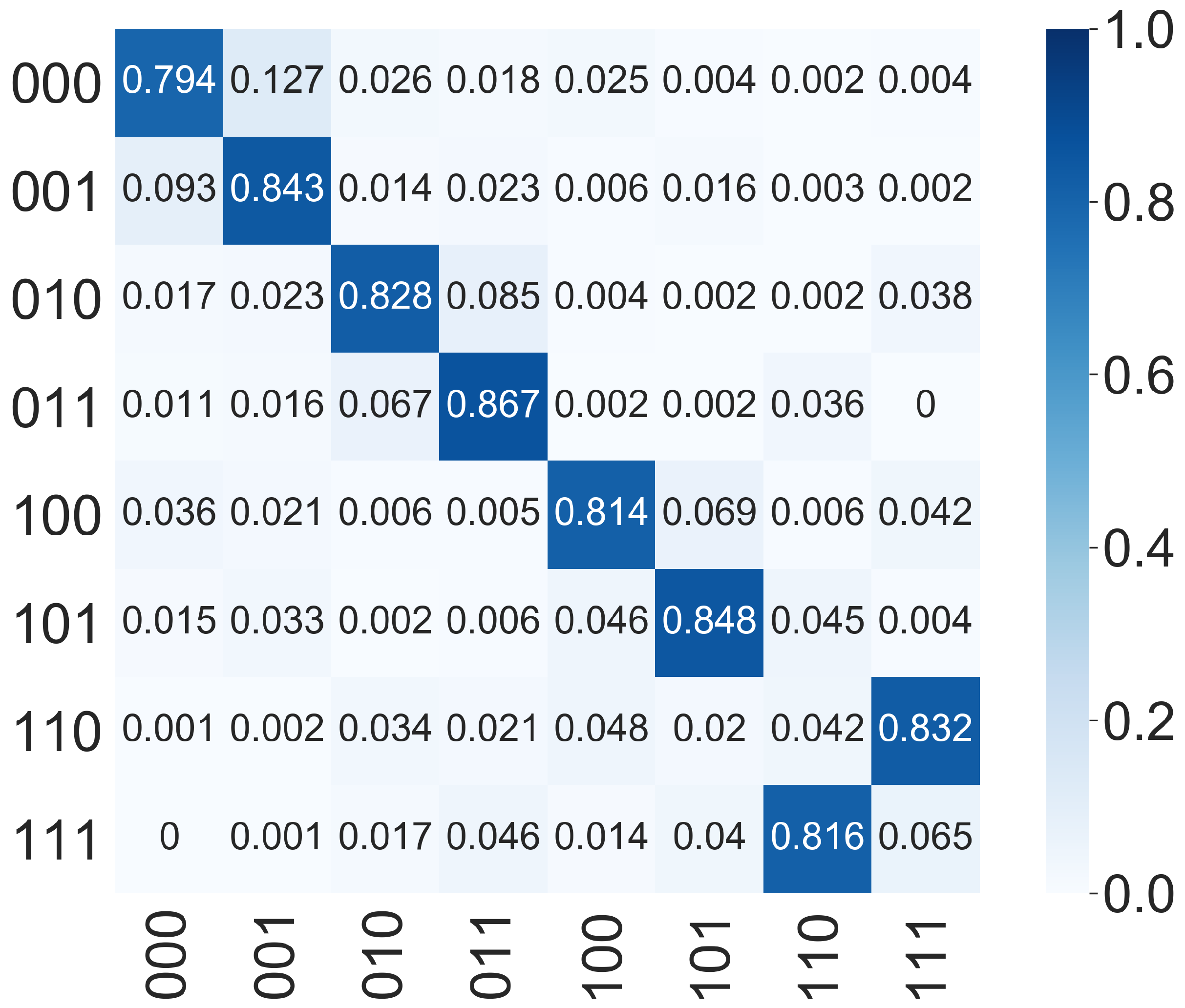}
        \subcaption{{\bf The performance of \RTX \\on {\bf ibmq\_ourense}.}}
        \label{fig:rtofix_ourense_prob}
    \end{minipage}
    
    \begin{minipage}[t]{.33\hsize}
        \centering
        \includegraphics[keepaspectratio, width=60mm, angle=0]{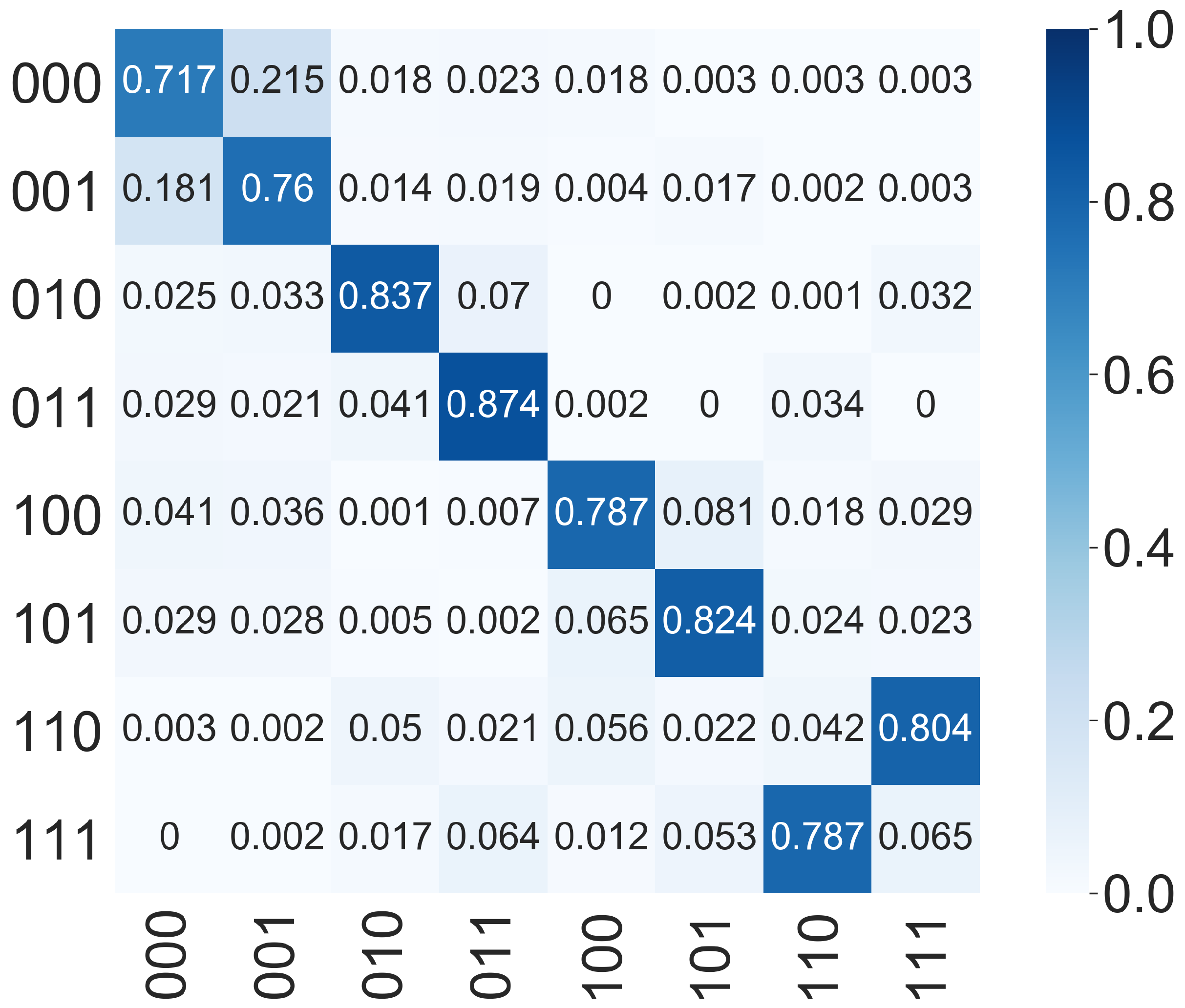}
        \subcaption{{\bf The performance of \RTM \\on {\bf ibmq\_ourense}.}}
        \label{fig:rtofm_ourense_prob}
    \end{minipage}
    
    \begin{minipage}[t]{.33\hsize}
        \centering
        \includegraphics[keepaspectratio, width=60mm, angle=0]{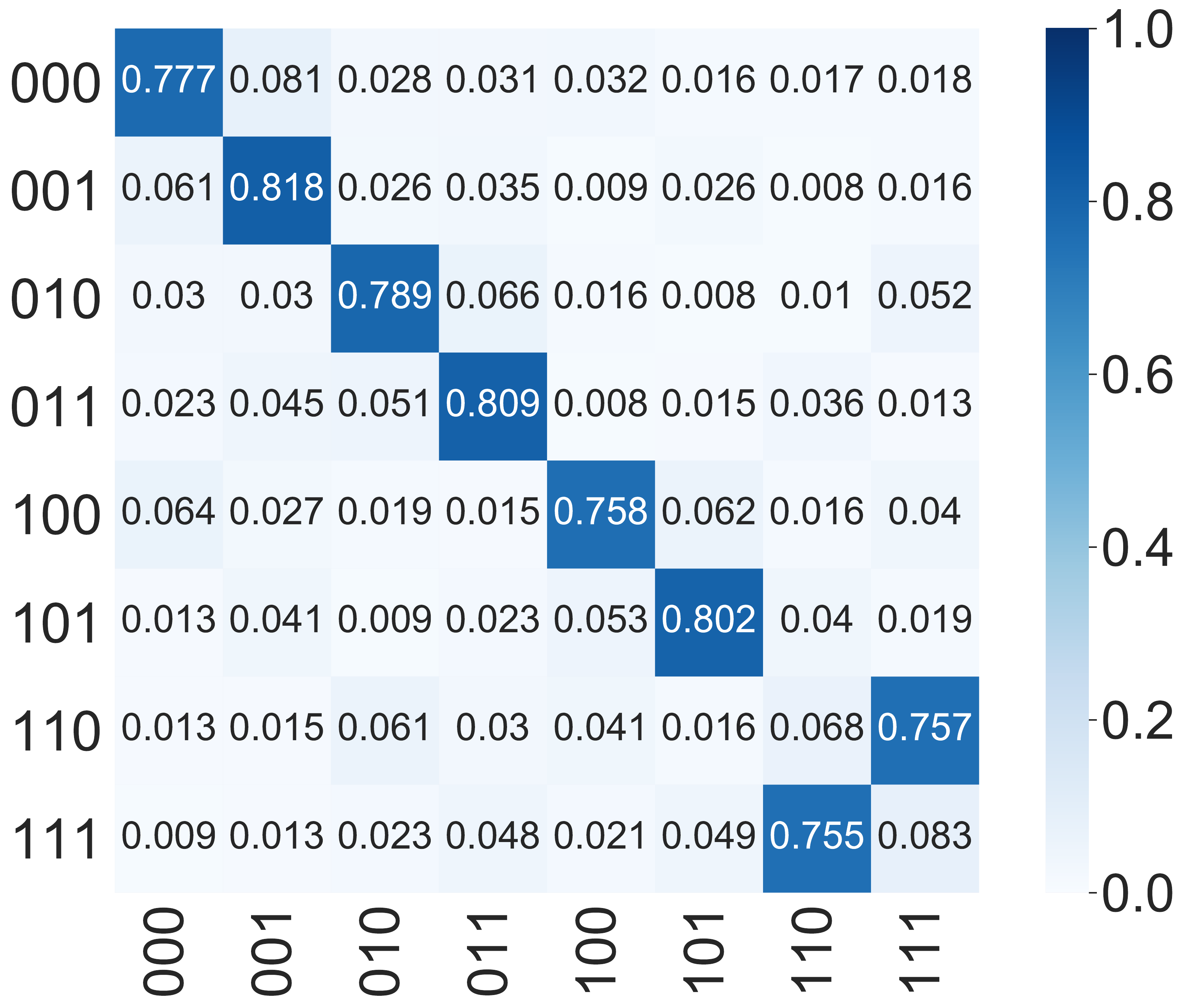}
        \subcaption{{\bf The performance of Toffoli with SWAP \\on {\bf ibmq\_ourense}.}}
        \label{fig:swaptof_ourense_prob}
    \end{minipage} \\ 
    
    \begin{minipage}[t]{.33\hsize}
    \centering
        \includegraphics[keepaspectratio, width=60mm, angle=0]{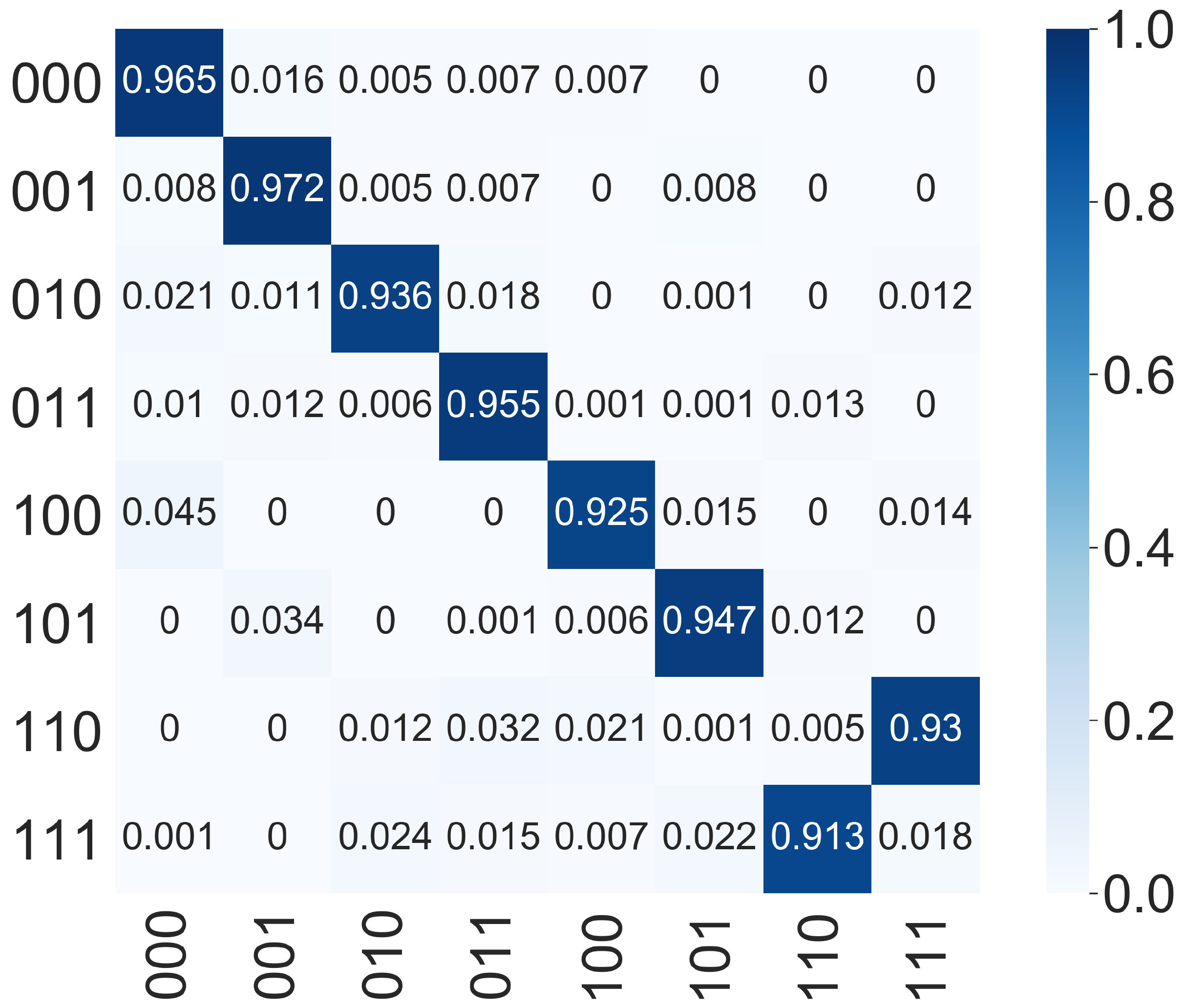}
        \subcaption{{\bf The performance of \RTX \\on {\bf ibmq\_valencia}.}}
        \label{fig:rtofix_valencia_prob}
    \end{minipage}
    
    \begin{minipage}[t]{.33\hsize}
        \centering
        \includegraphics[keepaspectratio, width=60mm, angle=0]{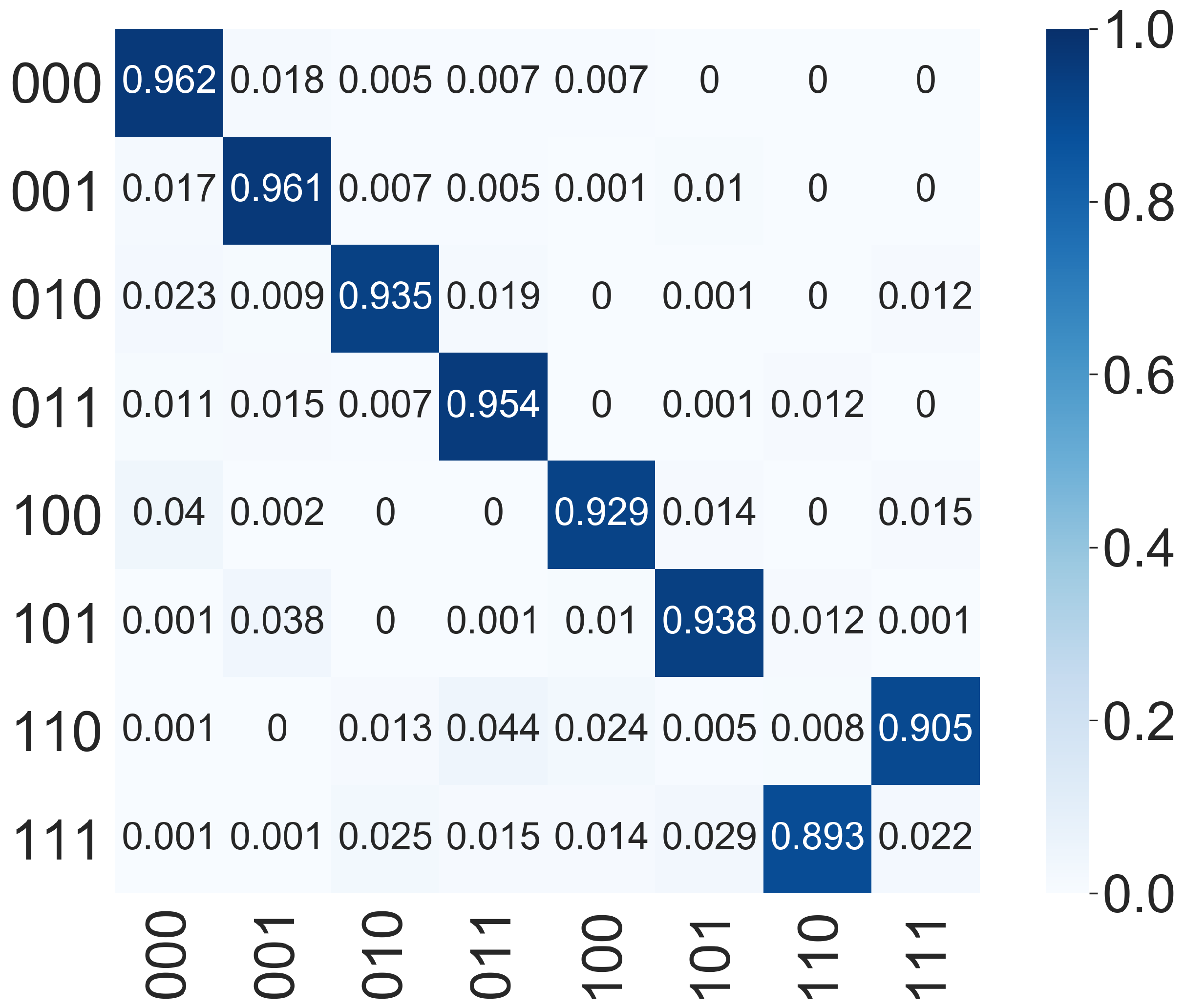}
        \subcaption{{\bf The performance of \RTM \\on {\bf ibmq\_valencia}.}}
        \label{fig:rtofm_valencia_prob}
    \end{minipage}
    
    \begin{minipage}[t]{.33\hsize}
        \centering
        \includegraphics[keepaspectratio, width=60mm, angle=0]{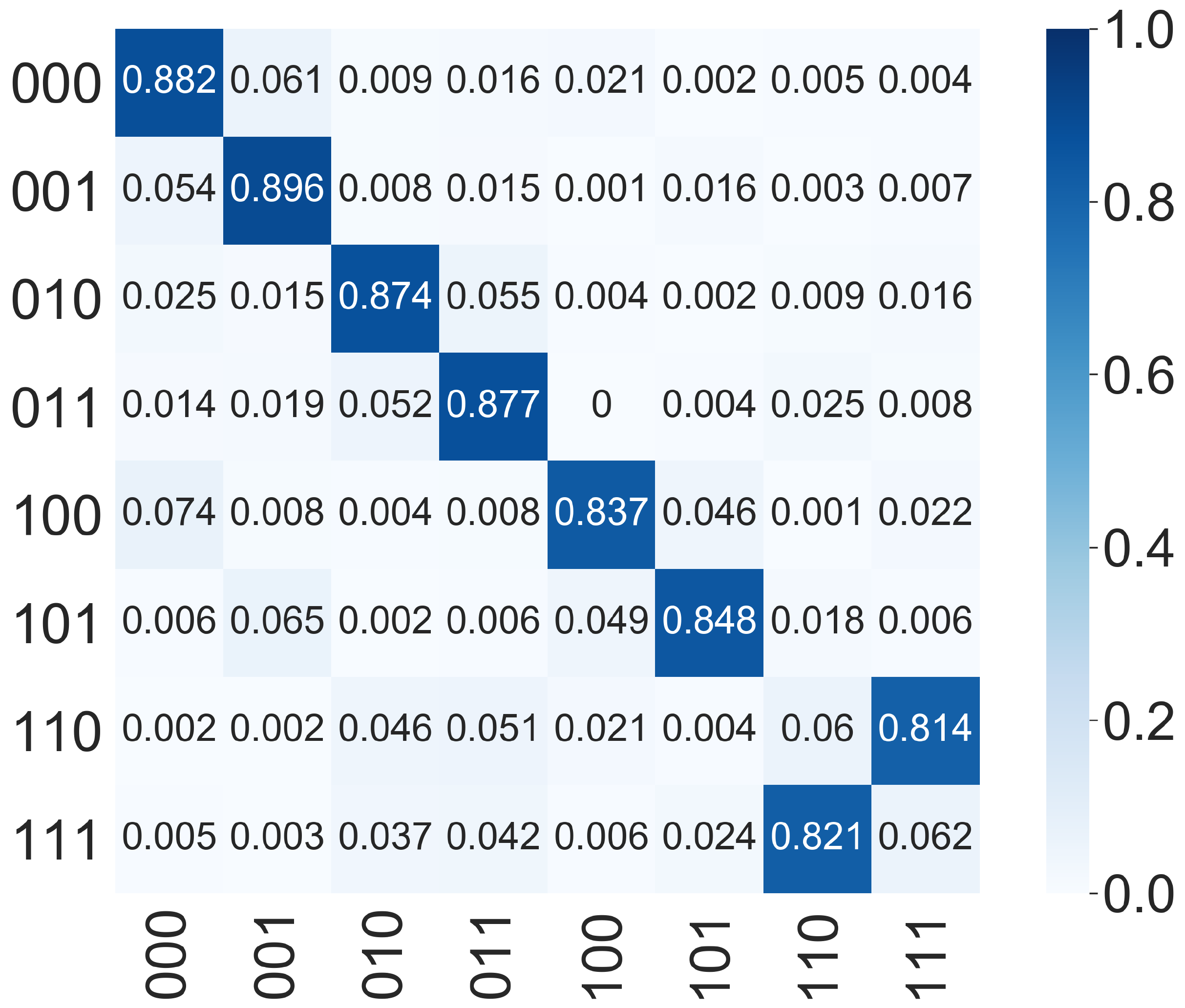}
        \subcaption{{\bf The performance of Toffoli with SWAP \\on {\bf ibmq\_valencia}.}}
        \label{fig:swaptof_valencia_prob}
    \end{minipage}
    
    \end{tabular}
    \caption{{\bf Execution of three types of Toffoli gate on the real devices}. To generate the results for each row, 8192 trials were performed for each input. Entries are output probabilities, with each row summing to approximately 1. Each row denotes the input value, and each column the output value.
    }
\label{fig:tof_prob}
\end{figure*}

\subsection{Performance of composite gates}
\label{appendix:performance_cnx}

We prepared different input states and measured using the computational basis after applying three types of Toffoli gate.
FIG.~\ref{fig:tof_prob} shows the experimental results using two processors.

\begin{figure*}[tp]
    \begin{tabular}{c}
    \begin{minipage}[t]{.50\hsize}
        \includegraphics[keepaspectratio, width=75mm, angle=0]{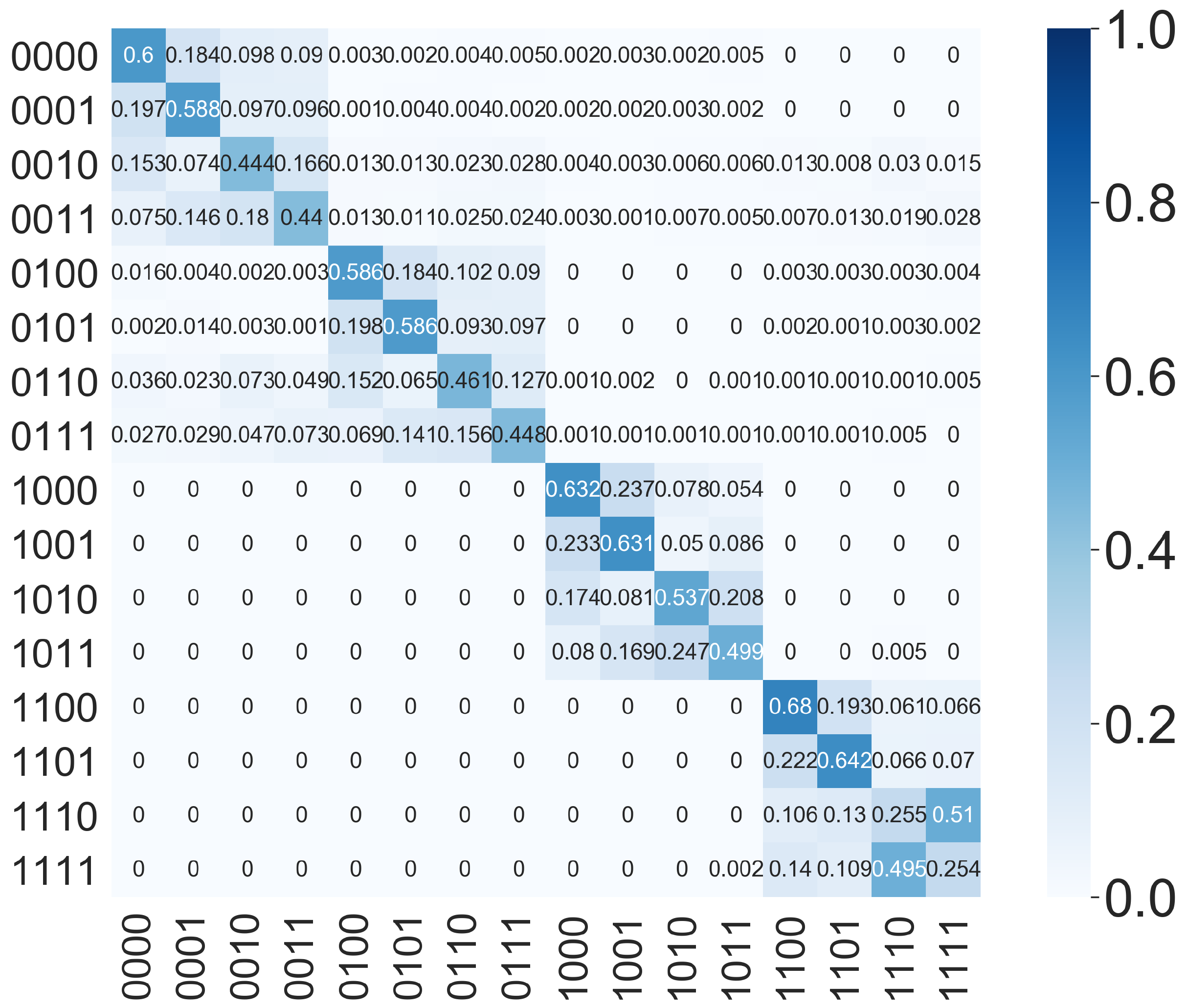}
        \subcaption{{\bf The performance of $C^{\otimes 3}X$ gate with \RTX \\on {\bf ibmq\_ourense}.}}
        \label{fig:rc3xix_ourense_prob}
    \end{minipage}
    
    \begin{minipage}[t]{.50\hsize}
        \centering
        \includegraphics[keepaspectratio, width=75mm, angle=0]{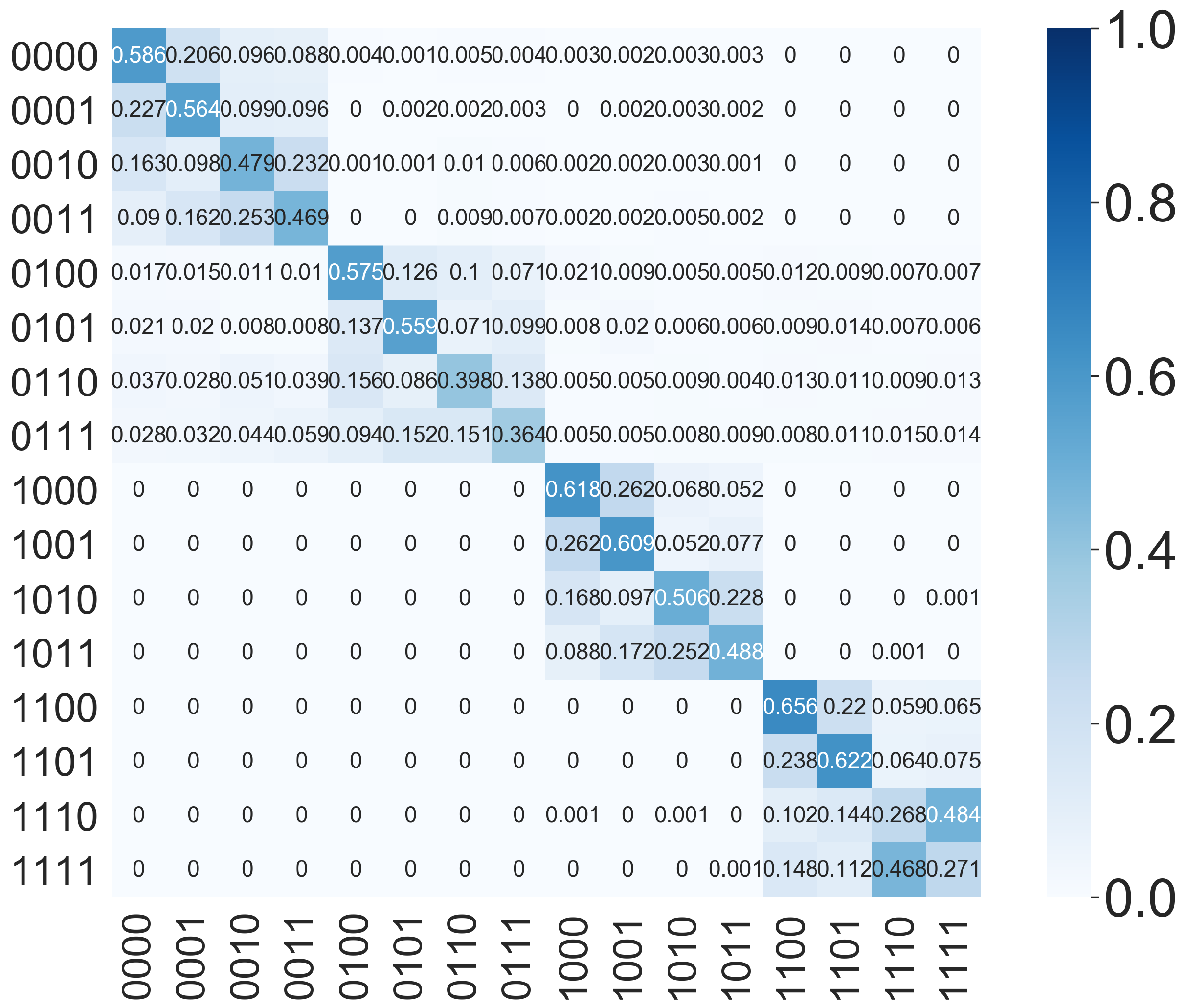}
        \subcaption{{\bf The performance of $C^{\otimes 3}X$ gate with \RTM \\on {\bf ibmq\_ourense}.}}
        \label{fig:rc3xm_ourense_prob}
    \end{minipage} \\ 
    
    \begin{minipage}[t]{.50\hsize}
    \centering
        \includegraphics[keepaspectratio, width=75mm, angle=0]{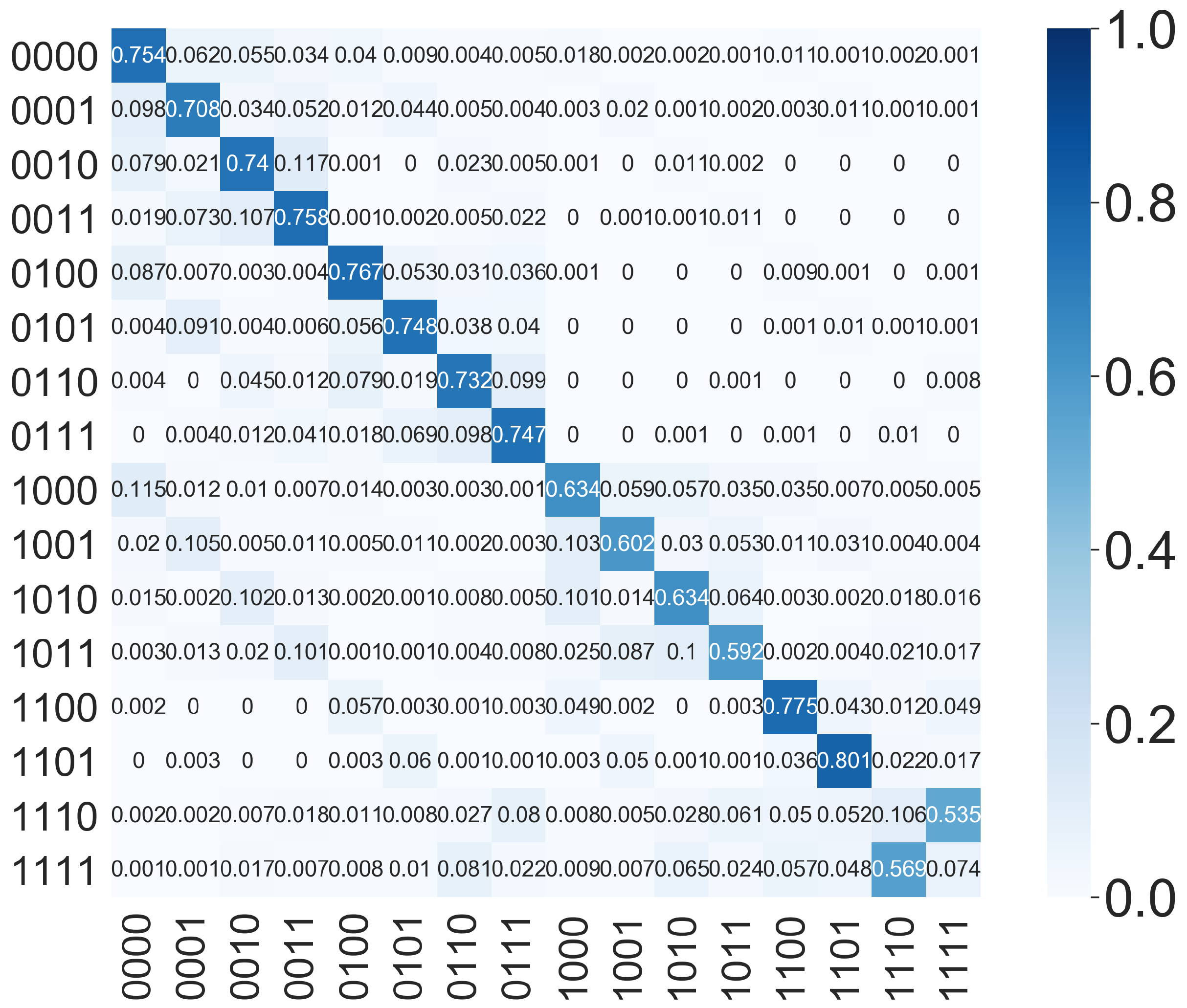}
        \subcaption{{\bf The performance of $C^{\otimes 3}X$ gate with \RTX \\on {\bf ibmq\_valencia}.}}
        \label{fig:rc3xix_valencia_prob}
    \end{minipage}
    
    \begin{minipage}[t]{.50\hsize}
        \centering
        \includegraphics[keepaspectratio, width=75mm, angle=0]{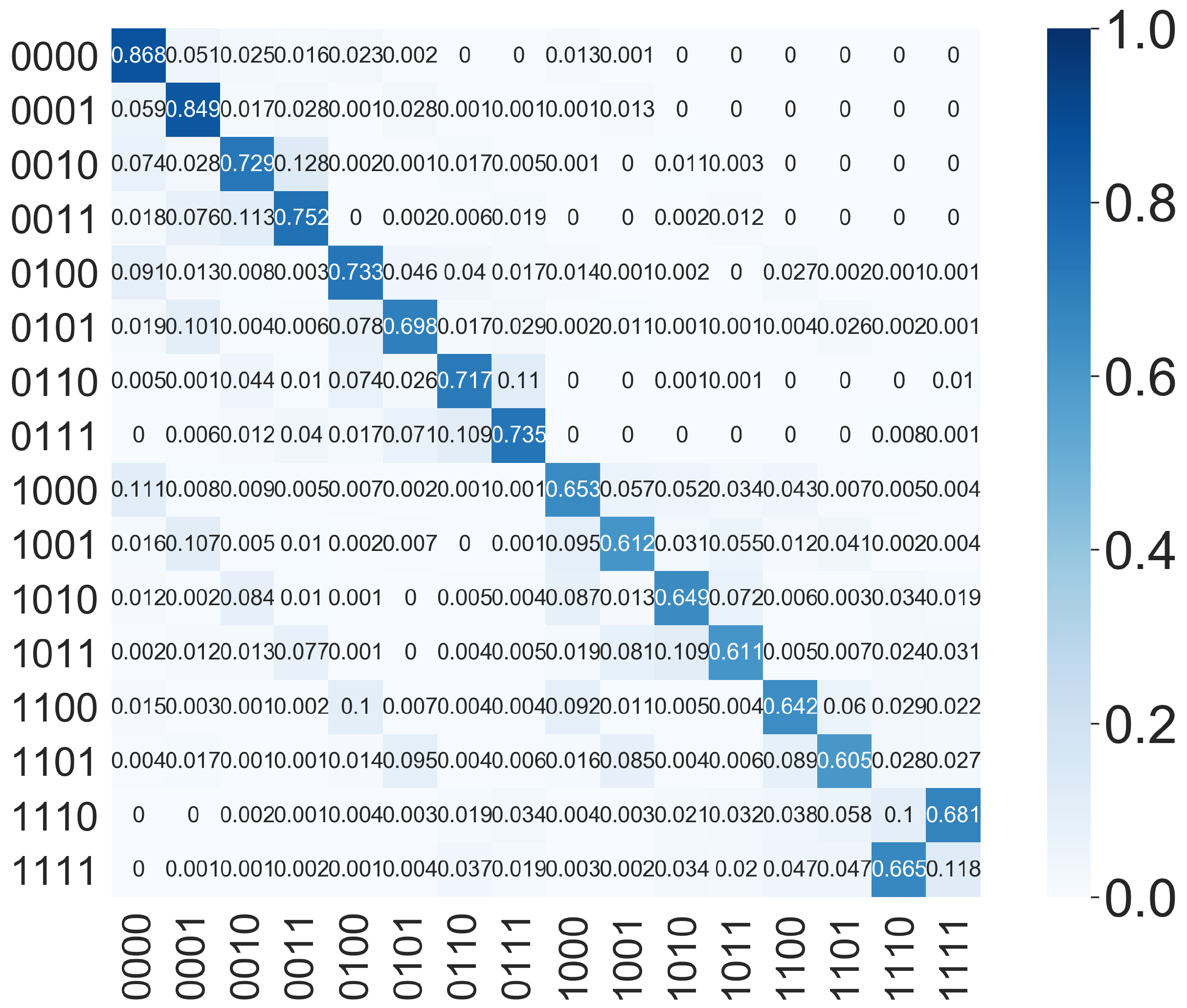}
        \subcaption{{\bf The performance of $C^{\otimes 3}X$ gate with \RTM \\on {\bf ibmq\_valencia}.}}
        \label{fig:rc3xm_valencia_prob}
    \end{minipage}
    \end{tabular}
    \caption{{\bf Execution of $C^{\otimes 3}X$ gate on the real devices}. To generate the results for each row, 8192 trials were performed for each input. Entries are output counts, with each row summing to approximately 1. Each row denotes the input value, and each column the output value.
    }
    \label{fig:c3x_prob}
\end{figure*}

We also perform $C^{\otimes 3}X$ gate to different input states and measured using the computational basis.
FIG.~\ref{fig:c3x_prob} shows the experimental results using two processors.


When testing small circuits such as these complex gates on real systems, state preparation and measurement (SPAM) errors will distort the results compared to the circuit itself. Therefore, we adopted the standard measurement error mitigation approach recommended for use with Qiskit, utilizing the library functions {\tt CompleteMeasFitter()} and {\tt complete\_meas\_cal()} in Qiskit~\cite{qiskit_MEM}.
First we execute the set of circuits created by {\tt complete\_meas\_cal()} to take measurements for each of the $2^5$ basis states for five qubits on {\bf ibm\_ourense} or {\bf ibm\_valencia}, and collect the results into a matrix $C_{\textrm{noisy}}$. 
We then use {\tt CompleteMeasFitter()} to find $M$ that satisfies the following equation:
\begin{align}
    C_{\textrm{noisy}} = M C_{\textrm{ideal}} 
\end{align}
where $C_{\textrm{ideal}}$ denotes ideal result matrix not containing noise. If $M$ is invertible, we can mitigate the measurement errors by applying the inverse of $M$ to the raw data matrix $R$ from the actual circuit (e.g., FIG.~\ref{fig:tof_prob}):
\begin{align}
    R_{\textrm{mitigated}} = M^{-1} R_{\textrm{noisy}}.
\end{align}
However, in general, $M$ is not invertible; instead, the corresponding Qiskit filter object derived from $M$ applies a least-squares fit.  All of the real-device data figures in this paper utilize this approach.

\section{The circuits for the MAX-CUT problem}
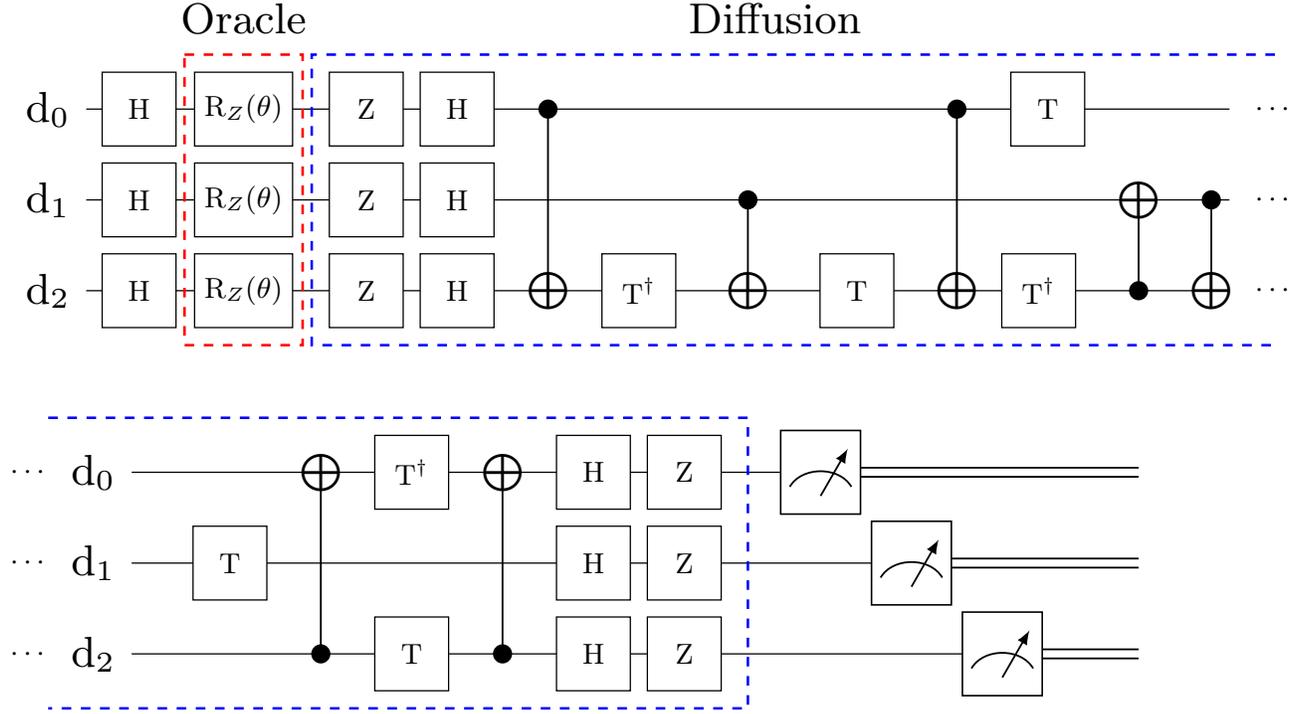
\begin{figure*}[p!]
    \centering
    \resizebox{1\textwidth}{!}{
    \begin{tikzpicture}[background rectangle/.style={fill=white},
                    show background rectangle]
                    
    \tikzset{meter/.append style={draw,fill=white, 
        inner sep=10, rectangle, font=\vphantom{A}, 
        minimum width=25, 
        line width=.5,
        path picture={\draw[line width=0.2mm][black] ([shift={(.1,.3)}]
        path picture bounding box.south west) to[bend left=60] ([shift={(-.1,.3)}]path picture bounding box.south east);
        \draw[line width=0.2mm][black,-latex] ([shift={(0,.2)}]path picture bounding box.south) -- ([shift={(.3,-.2)}]path picture bounding box.north);}
        }
    }
    
    \tikzstyle{operator} = [draw, fill=white,minimum size=2.5em] 
    \tikzstyle{ctrl} = [fill,shape=circle,minimum size=5pt,inner sep=0pt]

    \node[scale=1.5] at (0, 0) (q0-0) {$\rm d_0$};
    \node[scale=1.5] at (0,-1) (q1-0) {$\rm d_1$};
    \node[scale=1.5] at (0, -2) (q2-0) {$\rm d_2$};
    \draw(q0-0) -- (13,  0);
    \draw(q1-0) -- (13, -1);
    \draw(q2-0) -- (13, -2);
    
    \node [operator] (H00) at (1, 0) {H};
    \node [operator] (H10) at (1, -1){H};
    \node [operator] (H20) at (1, -2){H};
    
    \node [operator] (RZ01) at (2.15, 0) {R$_Z$($\rm{\theta}$)};
    \node [operator] (RZ11) at (2.15, -1){R$_Z$($\theta$)};
    \node [operator] (RZ21) at (2.15, -2){R$_Z$($\theta$)};

    \draw[red, thick, dashed] (1.5, 0.6) -- (2.8, 0.6) -- (2.8, -2.6) -- (1.5, -2.6) -- (1.5, 0.6);
    \node[scale=1.5] (vs1) at (2.15, 1) {Oracle};
    
    \node [operator] (Z03) at (3.5, 0) {Z};
    \node [operator] (Z13) at (3.5, -1){Z};
    \node [operator] (Z23) at (3.5, -2){Z};
    
    \node [operator] (H04) at (4.5, 0) {H};
    \node [operator] (H14) at (4.5, -1){H};
    \node [operator] (H24) at (4.5, -2){H};
    
    \node [ctrl, scale=1.2] (ctrl05) at (5.5, 0) {};
    \node [scale=1.3] at (5.5, -2){$\bigoplus$};
    \draw[line width=0.2mm](5.5, 0) -- (5.5, -2);
    
    \node [operator] at (6.5, -2) {T$^\dagger$};
    
    \node [ctrl, scale=1.2] (ctrl07) at (7.7, -1) {};
    \node [scale=1.3] at (7.7, -2){$\bigoplus$};
    \draw[line width=0.2mm](7.7, -1) -- (7.7, -2);
    
    \node [operator] (T28) at (8.9, -2) {T};
    
    \node [ctrl, scale=1.2] (ctrl09) at (10, 0) {};
    \node [scale=1.3] at (10, -2) {$\bigoplus$};
    \draw[line width=0.2mm](10, 0) -- (10, -2);
    
    \node [operator] (T010) at (11, 0) {T};
    \node [operator] at (10.9, -2) {T$^\dagger$};
    
    \node [ctrl, scale=1.2] (ctrl211) at (12, -2) {};
    \node [scale=1.3] at (12, -1) {$\bigoplus$};
    \draw[line width=0.2mm](12, -1) -- (12, -2);
    
    x12 = 12.8;
    \node [ctrl, scale=1.2] (ctrl112) at (12.8, -1) {};
    \node [scale=1.3] at (12.8, -2){$\bigoplus$};
    \draw[line width=0.2mm](12.8, -1) -- (12.8, -2);
    
    \node at (13.5, 0) {$\cdots$};
    \node at (13.5,-1) {$\cdots$};
    \node at (13.5, -2) {$\cdots$};

    \draw[blue, thick, dashed] (13.5, 0.6) -- (2.9, 0.6) -- (2.9, -2.6) -- (13.5, -2.6);
    \node[scale=1.5] (vs1) at (8, 1) {Diffusion};

    \node at (-0.2, -4) {$\cdots$};
    \node at (-0.2,-5) {$\cdots$};
    \node at (-0.2, -6) {$\cdots$};

    \node[scale=1.5] at (0.5, -4) (q0-1) {$\rm d_0$};
    \node[scale=1.5] at (0.5,-5) (q1-1) {$\rm d_1$};
    \node[scale=1.5] at (0.5, -6) (q2-1) {$\rm d_2$};
    \draw (q0-1) -- (8.5,  -4);
    \draw (q1-1) -- (9.5, -5);
    \draw (q2-1) -- (10.5, -6);
    
    \node[operator] (T52) at (2, -5) {T};
    
    \node [ctrl, scale=1.2] (ctrl63) at (3, -6) {};
    \node [scale=1.3] at (3, -4){$\bigoplus$};
    \draw[line width=0.2mm](3, -4) -- (3, -6);
    
    \node [operator] (Tag44) at (4, -4) {T$^\dagger$};
    \node [operator] (T64) at (4, -6) {T};
    
    \node [ctrl, scale=1.2] (ctrl65) at (5, -6) {};
    \node [scale=1.3] at (5, -4){$\bigoplus$};
    \draw[line width=0.2mm](5, -4) -- (5, -6);

    \node [operator] (H46) at (6, -4) {H};
    \node [operator] (H56) at (6, -5){H};
    \node [operator] (H66) at (6, -6){H};
    
    \node [operator] (H47) at (7, -4) {Z};
    \node [operator] (H57) at (7, -5){Z};
    \node [operator] (H67) at (7, -6){Z};
    
    \draw[line width=0.2mm][blue, thick, dashed] (0, -3.4) -- (7.7, -3.4) -- (7.7, -6.6) -- (0, -6.6);
    
    \draw[line width=0.2mm] (8.5, -3.95) -- (12, -3.95);
    \draw[line width=0.2mm] (8.5, -4.05) -- (12, -4.05);
    
    \draw[line width=0.2mm] (9.5, -4.95) -- (12, -4.95);
    \draw[line width=0.2mm] (9.5, -5.05) -- (12, -5.05);
    
    \draw[line width=0.2mm] (10.5, -5.95) -- (12, -5.95);
    \draw[line width=0.2mm] (10.5, -6.05) -- (12, -6.05);
    
    \node [meter] (meter)  at (8.5, -4) {};
    \node [meter] (meter)  at (9.5, -5) {};
    \node [meter] (meter)  at (10.5, -6) {};
    
    \end{tikzpicture}
}
    \caption{{\bf The circuit of MAX-CUT solver for $K_{1,3}$.} The $\theta$ of $R_Z$ changes the amplification rate for correct answer.}
    \label{fig:3grover_circuit}
\end{figure*}

We show the circuit to find MAX-CUT of $K_{1,3}$ in Fig.~\ref{fig:3grover_circuit}.
Unlike Eq.~(\ref{eq:diff}), we adopted $ZH (HZ)$ for $HX (XH)$ and Toffoli with SWAP gate for Toffoli gate.
The former change allows us to reduce the number of $U3$ gates, thereby reducing gate errors (in the case where a series of single-qubit gates are not integrated into one $U3$ gate).
The latter change avoids connectivity constraints with minimal overhead.

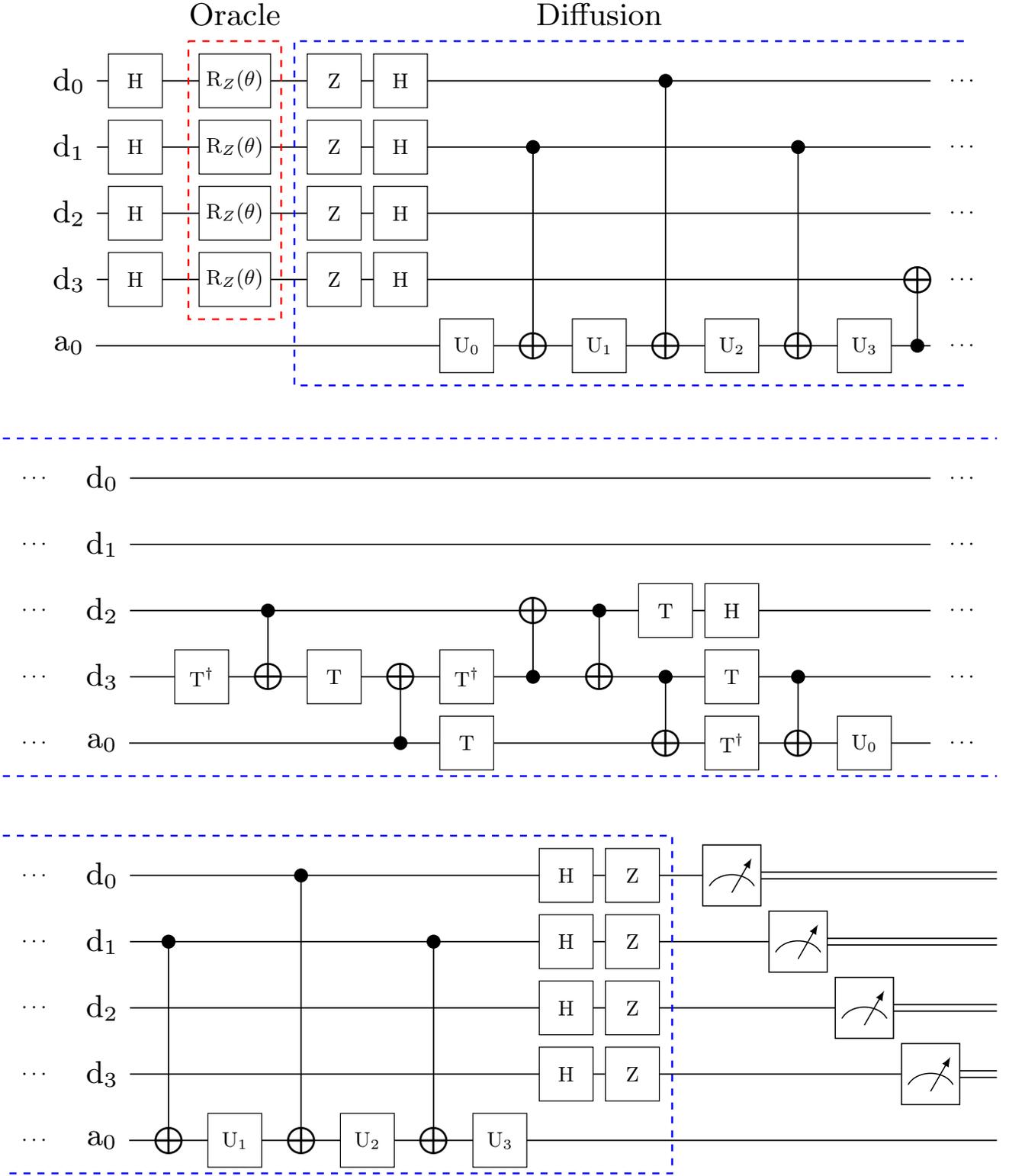
\begin{figure*}[p!]
    \centering
    \resizebox{1\textwidth}{!}{
    \begin{tikzpicture}[background rectangle/.style={fill=white},
                    show background rectangle]
              
    \tikzset{meter/.append style={draw,fill=white, 
        inner sep=10, rectangle, font=\vphantom{A}, 
        minimum width=25, 
        line width=.5,
        path picture={\draw[black] ([shift={(.1,.3)}]
        path picture bounding box.south west) to[bend left=60] ([shift={(-.1,.3)}]path picture bounding box.south east);
        \draw[black,-latex] ([shift={(0,.2)}]path picture bounding box.south) -- ([shift={(.3,-.2)}]path picture bounding box.north);}
        }
    }
    
    \tikzstyle{operator} = [draw, fill=white,minimum size=2.5em] 
    \tikzstyle{ctrl} = [fill,shape=circle,minimum size=5pt,inner sep=0pt]

    \node[scale=1.5] at (0, 0) (q0-0) {$\rm d_0$};
    \node[scale=1.5] at (0,-1) (q1-0) {$\rm d_1$};
    \node[scale=1.5] at (0,-2) (q2-0) {$\rm d_2$};
    \node[scale=1.5] at (0,-3) (q3-0) {$\rm d_3$};
    \node[scale=1.5] at (0,-4) (q4-0) {$\rm a_0$};
    
    \draw[line width=0.2mm] (q0-0) -- (13,  0);
    \draw[line width=0.2mm] (q1-0) -- (13, -1);
    \draw[line width=0.2mm] (q2-0) -- (13, -2);
    \draw[line width=0.2mm] (q3-0) -- (13, -3);
    \draw[line width=0.2mm] (q4-0) -- (13, -4);
    
    \node [operator] (H00) at (1, 0) {H};
    \node [operator] (H10) at (1, -1){H};
    \node [operator] (H20) at (1, -2){H};
    \node [operator] (H30) at (1, -3){H};
    
    \node [operator] (RZ01) at (2.5, 0) {R$_Z$($\theta$)};
    \node [operator] (RZ11) at (2.5, -1){R$_Z$($\theta$)};
    \node [operator] (RZ21) at (2.5, -2){R$_Z$($\theta$)};
    \node [operator] (RZ31) at (2.5, -3){R$_Z$($\theta$)};

    \draw[line width=0.2mm][red, thick, dashed] (1.8, 0.6) -- (3.2, 0.6) -- (3.2, -3.6) -- (1.8, -3.6) -- (1.8, 0.6);
    \node[scale=1.5] (vs1) at (2.5, 1) {Oracle};
    
    \node [operator] (Z03) at (4, 0) {Z};
    \node [operator] (Z13) at (4, -1){Z};
    \node [operator] (Z23) at (4, -2){Z};
    \node [operator] (Z33) at (4, -3){Z};
    
    \node [operator] (H04) at (5, 0) {H};
    \node [operator] (H14) at (5, -1){H};
    \node [operator] (H24) at (5, -2){H};
    \node [operator] (H34) at (5, -3){H};
    
    \node [operator] (H05) at (6, -4) {U$_0$};
    
    \node [ctrl, scale=1.2] (ctrl06) at (7, -1){};
    \node [scale=1.3] at (7, -4){$\bigoplus$};
    \draw[line width=0.2mm](7, -1) -- (7, -4);
    
    \node [operator] (H07) at (8, -4) {U$_1$};
    
    \node [ctrl, scale=1.2] (ctrl08) at (9, 0){};
    \node [scale=1.3] at (9, -4){$\bigoplus$};
    \draw[line width=0.2mm](9, 0) -- (9, -4);
    
    \node [operator] (H09) at (10, -4) {U$_2$};
    
    \node [ctrl, scale=1.2] (ctrl010) at (11, -1){};
    \node [scale=1.3] at (11, -4){$\bigoplus$};
    \draw[line width=0.2mm](11, -1) -- (11, -4);
    
    \node [operator] (H011) at (12, -4) {U$_3$};
    
    \node [ctrl, scale=1.2] (ctrl012) at (12.8, -4){};
    \node [scale=1.3] at (12.8, -3){$\bigoplus$};
    \draw[line width=0.2mm](12.8, -3) -- (12.8, -4);

    \node at (13.5, 0) {$\cdots$};
    \node at (13.5,-1) {$\cdots$};
    \node at (13.5,-2) {$\cdots$};
    \node at (13.5,-3) {$\cdots$};
    \node at (13.5,-4) {$\cdots$};
    
    \draw[blue, thick, dashed] (13.5, 0.6) -- (3.4, 0.6) -- (3.4, -4.6) -- (13.5, -4.6);
    \node[scale=1.5] (vs1) at (8, 1) {Diffusion};

    \node at (-0.5, -6) {$\cdots$};
    \node at (-0.5, -7) {$\cdots$};
    \node at (-0.5, -8) {$\cdots$};
    \node at (-0.5, -9) {$\cdots$};
    \node at (-0.5, -10) {$\cdots$};
    
    \node[scale=1.5] at (0.5, -6) (q0-1) {$\rm d_0$};
    \node[scale=1.5] at (0.5, -7) (q1-1) {$\rm d_1$};
    \node[scale=1.5] at (0.5, -8) (q2-1) {$\rm d_2$};
    \node[scale=1.5] at (0.5, -9) (q3-1) {$\rm d_3$};
    \node[scale=1.5] at (0.5, -10) (q4-1) {$\rm a_0$};
    
    \draw[line width=0.2mm] (q0-1) -- (13, -6);
    \draw[line width=0.2mm] (q1-1) -- (13, -7);
    \draw[line width=0.2mm] (q2-1) -- (13, -8);
    \draw[line width=0.2mm] (q3-1) -- (13, -9);
    \draw[line width=0.2mm] (q4-1) -- (13, -10);
    
    \node [operator] (Tdg92) at (2, -9) {T$^\dagger$};
    
    \node [ctrl, scale=1.2] (ctrl103) at (3, -8){};
    \node [scale=1.3] at (3, -9){$\bigoplus$};
    \draw[line width=0.2mm](3, -8) -- (3, -9);
    
    \node [operator] (T94) at (4, -9) {T};
    
    \node [ctrl, scale=1.2] (ctrl103) at (5, -10){};
    \node [scale=1.3] at (5, -9){$\bigoplus$};
    \draw[line width=0.2mm](5, -9) -- (5, -10);
    
    \node [operator] (Tdg96) at (6, -9) {T$^\dagger$};
    \node [operator] (T10x6) at (6, -10) {T};
    
    \node [ctrl, scale=1.2] (ctrl97) at (7, -9){};
    \node [scale=1.3] at (7, -8){$\bigoplus$};
    \draw[line width=0.2mm](7, -8) -- (7, -9);
    
    \node [ctrl, scale=1.2] (ctrl108) at (8, -8){};
    \node [scale=1.3] at (8, -9){$\bigoplus$};
    \draw[line width=0.2mm](8, -8) -- (8, -9);
    
    \node [operator] (T106) at (9, -8) {T};
    \node [ctrl, scale=1.2] (ctrl99) at (9, -9){};
    \node [scale=1.3] at (9, -10){$\bigoplus$};
    \draw[line width=0.2mm](9, -9) -- (9, -10);
    
    \node [operator] (H108) at (10, -8) {H};
    \node [operator] (T109) at (10, -9) {T};
    \node [operator] (T1010) at (10, -10) {T$^\dagger$};
  
    \node [ctrl, scale=1.2] (ctrl119) at (11, -9){};
    \node [scale=1.3] at (11, -10){$\bigoplus$};
    \draw[line width=0.2mm](11, -9) -- (11, -10);

    \node [operator] (U1110) at (12, -10) {U$_0$};
    
    \node at (13.5,-6) {$\cdots$};
    \node at (13.5,-7) {$\cdots$};
    \node at (13.5,-8) {$\cdots$};
    \node at (13.5,-9) {$\cdots$};
    \node at (13.5,-10) {$\cdots$};
    
    \draw[blue, thick, dashed] (-1, -5.4) -- (14, -5.4);
    \draw[blue, thick, dashed] (-1, -10.5) -- (14, -10.5);

    \node at (-0.5, -12) {$\cdots$};
    \node at (-0.5, -13) {$\cdots$};
    \node at (-0.5, -14) {$\cdots$};
    \node at (-0.5, -15) {$\cdots$};
    \node at (-0.5, -16) {$\cdots$};
    
    \node[scale=1.5] at (0.5, -12) (q0-2) {$\rm d_0$};
    \node[scale=1.5] at (0.5, -13) (q1-2) {$\rm d_1$};
    \node[scale=1.5] at (0.5, -14) (q2-2) {$\rm d_2$};
    \node[scale=1.5] at (0.5, -15) (q3-2) {$\rm d_3$};
    \node[scale=1.5] at (0.5, -16) (q4-2) {$\rm a_0$};
    
    \draw[line width=0.2mm] (q0-2) -- (10, -12);
    \draw[line width=0.2mm] (q1-2) -- (11, -13);
    \draw[line width=0.2mm] (q2-2) -- (12, -14);
    \draw[line width=0.2mm] (q3-2) -- (13, -15);
    \draw[line width=0.2mm] (q4-2) -- (14, -16);
    
    \node [ctrl, scale=1.2] (ctrl212) at (1.5, -13){};
    \node [scale=1.3] at (1.5, -16){$\bigoplus$};
    \draw[line width=0.2mm](1.5, -13) -- (1.5, -16);
    
    \node [operator] (H07) at (2.5, -16) {U$_1$};
    
    \node [ctrl, scale=1.2] (ctrl08) at (3.5, -12){};
    \node [scale=1.3] at (3.5, -16){$\bigoplus$};
    \draw[line width=0.2mm](3.5, -12) -- (3.5, -16);
    
    \node [operator] (H09) at (4.5, -16) {U$_2$};
    
    \node [ctrl, scale=1.2] (ctrl010) at (5.5, -13){};
    \node [scale=1.3] at (5.5, -16){$\bigoplus$};
    \draw[line width=0.2mm](5.5, -13) -- (5.5, -16);
    
    \node [operator] (716) at (6.5, -16) {U$_3$};

    \node [operator] (Z03) at (7.5, -12) {H};
    \node [operator] (Z13) at (7.5, -13){H};
    \node [operator] (Z23) at (7.5, -14){H};
    \node [operator] (Z33) at (7.5, -15){H};
    
    \node [operator] (H04) at (8.5, -12) {Z};
    \node [operator] (H14) at (8.5, -13){Z};
    \node [operator] (H24) at (8.5, -14){Z};
    \node [operator] (H34) at (8.5, -15){Z};
    
    \draw[blue, thick, dashed] (-1, -11.4) -- (9.1, -11.4) -- (9.1, -16.5) --  (-1, -16.5);

    
    \draw[line width=0.2mm] (10, -11.95) -- (14, -11.95);
    \draw[line width=0.2mm] (10, -12.05) -- (14, -12.05);
    
    \draw[line width=0.2mm] (11, -12.95) -- (14, -12.95);
    \draw[line width=0.2mm] (11, -13.05) -- (14, -13.05);
    
    \draw[line width=0.2mm] (12, -13.95) -- (14, -13.95);
    \draw[line width=0.2mm] (12, -14.05) -- (14, -14.05);
    
    \draw[line width=0.2mm] (13, -14.95) -- (14, -14.95);
    \draw[line width=0.2mm][line width=0.2mm] (13, -15.05) -- (14, -15.05);

    \node [meter] (meter)  at (10, -12) {};
    \node [meter] (meter)  at (11, -13) {};
    \node [meter] (meter)  at (12, -14) {};
    \node [meter] (meter)  at (13, -15) {};
    
    \end{tikzpicture}
}
    \caption{{\bf The circuit of MAX-CUT solver for $K_{1,4}$.} Each $U_{k}$ gate is determined by the type of adopted $RTOF$ gate.}
    \label{fig:4grover_circuit}
\end{figure*}

We show the circuit to find MAX-CUT of $K_{1,4}$ in Fig.~\ref{fig:4grover_circuit}.
The gate set of the diffusion part except $ZH$ and $HZ$ constitutes one $C^{\otimes 3}X$ gate.

\section{Qiskit Versions}

The version of Qiskit packages we use are listed in Table \ref{table:QiskitVer}.

\begin{table}[ht]
    \centering
    \begin{tabular}{c|c}
        \hline
        name & version \\ \hline
        qiskit & 0.14.0 \\
        qiskit-terra & 0.11.0 \\
        qiskit-aer & 0.3.4 \\
        qiskit-ignis & 0.2.0 \\
        qiskit-aqua & 0.6.1 \\   
        qiskit-chemistry & 0.5.0 \\
        qiskit-ibmq-provider & 0.4.4 \\ \hline
    \end{tabular}
    \caption{{\bf Qiskit packages version}}
    \label{table:QiskitVer}
\end{table}

\section{Date-time}

Each experiment was performed on the dates listed in Table \ref{table:experiments}.

\begin{table}[ht]
    \centering
    \begin{tabular}{l|c}
        \hline
        Experiment & Date-time \\ \hline \hline
        
        \begin{tabular}{l}
            Performance of $RTOF_{iX}$ gate, $RTOF_{M}$ gate and 
            \\Toffoli with SWAP gate on {\bf ibmq\_ourense}
        \end{tabular} & 2019/12/24 \\ \hline

        \begin{tabular}{l}
            Performance of $C^{\otimes 3}X$ with $RTOF_{iX}$ gate and \\ $C^{\otimes 3}X$ with $RTOF_{M}$ gate on {\bf ibmq\_ourense}
        \end{tabular} & 2019/12/24 \\ \hline

        \begin{tabular}{l}
             MAX-CUT solver on {\bf ibmq\_valencia} 
        \end{tabular} & 2020/1/1 \\ \hline
        \begin{tabular}{l}
             MAX-CUT solver on {\bf ibmq\_ourense} 
        \end{tabular} & 2020/1/1 \\ \hline
        
        \begin{tabular}{l}
             Subdivided phase Oracle Grover algorithm\\on {\bf ibmq\_ourense} 
        \end{tabular} & 2020/1/1 \\ \hline
        Simulated gate fidelities of various Toffoli gates & 2020/1/1 \\ \hline

        \begin{tabular}{l}
            Performance of $RTOF_{iX}$ gate, $RTOF_{M}$ gate and 
            \\Toffoli with SWAP gate on {\bf ibmq\_valencia}
        \end{tabular} & 2020/1/6 \\ \hline
        
        \begin{tabular}{l}
            Performance of $C^{\otimes 3}X$ with $RTOF_{iX}$ gate and \\ $C^{\otimes 3}X$ with $RTOF_{M}$ gate on {\bf ibmq\_valencia}
        \end{tabular} & 2020/1/6 \\ \hline
        
        Performance of $R_Y$ gate on {\bf ibmq\_ourense} & 2020/1/8 \\ \hline
        Performance of $R_Y$ gate on {\bf ibmq\_valencia} & 2020/1/8 \\ \hline
        
    \end{tabular}
    \caption{{\bf Date and time when experimental data have been taken}}
    \label{table:experiments}
\end{table}

\section{Performance of IBM Q processors}
\label{appendix:performance_device}
We show single-qubit gate and readout performance of IBM Q processors in TAB.~\ref{table:device_qubit}.
We also show two-qubit gates performance in TAB.~\ref{table:device_cx}.

\begin{table}[H]
    \centering
    \begin{tabular}{c|c|c|c}
    \hline
        & U2 gate error & U3 gate error & Readout error \\ \hline
        \multicolumn{4}{c}{ibmq\_ourense} \\ \hline
        $Q_0$ & $3.04E-4$ & $6.09E-4$ & $1.80E-2$ \\
        $Q_1$ & $3.32E-4$ & $6.63E-4$ & $2.80E-2$ \\
        $Q_2$ & $3.67E-4$ & $7.33E-4$ & $2.80E-2$ \\
        $Q_3$ & $3.79E-4$ & $7.58E-4$ & $3.40E-2$ \\
        $Q_4$ & $3.77E-4$ & $7.53E-4$ & $4.90E-2$ \\ \hline
        \multicolumn{4}{c}{ibmq\_valencia}\\ \hline
        $Q_0$ & $5.31E-4$ & $1.06E-3$ & $2.75E-2$ \\
        $Q_1$ & $3.35E-4$ & $6.70E-4$ & $4.13E-2$ \\
        $Q_2$ & $5.51E-4$ & $1.10E-3$ & $2.50E-2$ \\
        $Q_3$ & $3.22E-4$ & $6.45E-4$ & $2.50E-2$ \\
        $Q_4$ & $4.26E-4$ & $8.52E-4$ & $4.00E-2$ \\ \hline
    \end{tabular}
    
    \caption{Qubit performance on Jan 1 2020.}
    \label{table:device_qubit}
\end{table}

\begin{table}[htb]
    \centering
    \begin{tabular}{c|c|c}
    \hline
        &  {\bf ibmq\_ourense} &  {\bf ibmq\_valencia} \\ \hline
        $CX~(0, 1)$ & $7.22E-3$ & $7.67E-3$ \\
        $CX~(1, 2)$ & $9.55E-3$ & $9.62E-3$ \\
        $CX~(1, 3)$ & $1.34E-2$ & $1.13E-2$ \\
        $CX~(3, 4)$ & $7.35E-3$ & $7.71E-3$ \\
        \hline
    \end{tabular}
    \caption{$CX$ gate performance on Jan 1 2020.}
    \label{table:device_cx}
\end{table}


\end{document}